\documentclass[12pt]{article}
\input{definition}

\begin{document}

\begin{center}
{\bf\Large Kernel Angle Dependence Measures for Complex Objects}
\\\vskip 0.3cm
Yilin Zhang and Songshan Yang
\\\vskip0.3cm
\textit{ Center of Applied Statistics and Institute of Statistics and Big Data, Renmin University of China}
\begin{singlespace}
   \footnotetext[1]{Yilin Zhang (Email: yzhang97@ruc.edu.cn) is a Ph.D student, and Songshan Yang (Email: yangss@ruc.edu.cn) is Assistant Professor and the corresponding author of the article, Center of Applied Statistics and Institute of Statistics and Big Data,  Renmin University of China, % 59 Zhongguancun Avenue, Haidian District,
   Beijing 100872, P. R. China. 
   Zhang's research  is supported by the Fundamental Research Fund  for Central Universities  and  the Research Fund  of Renmin University of China (21XNH157).
}
\end{singlespace}
\today
\end{center}

\begin{singlespace}
\begin{abstract}

Measuring and testing dependence between complex objects is of great importance in modern statistics. Most existing work relied on the distance between random variables, which inevitably required the moment conditions to guarantee the distance is well-defined. Based on the geometry element ``angle", we develop a novel class of nonlinear dependence measures for data in metric space that can avoid such conditions. Specifically, by making use of the reproducing kernel Hilbert space equipped with Gaussian measure, we introduce kernel angle covariances that can be applied to complex objects such as random vectors or matrices. We estimate kernel angle covariances based on $U$-statistic and establish the corresponding independence tests via gamma approximation. Our kernel angle independence tests, imposing no-moment conditions on kernels, are robust with heavy-tailed random variables. 
We conduct comprehensive simulation studies and apply our proposed methods to a facial recognition task. Our kernel angle covariances-based tests show remarkable performances in dealing with image data.
%We conduct comprehensive simulation studies and a real data application to demonstrate the performance of the proposed independence testing methods for complex objects. {\color{red} real data}

\end{abstract}
\end{singlespace}

\newpage

\csection{Introduction}
Measuring and testing the dependence between random variables is an important problem in statistical research and has a wide range of applications. 
With the development of modern technology, it is of great importance to detect the dependence for complex objects.
For example, data usually come as images in computer vision research, and it is of great interest to explore the relationship between the images of human faces and facial expressions \citep{ying2022frechet}.
%data are often high-dimensional and over-dispersed in the microbiome and metagenomic studies, and it is challenging to measure and detect the association between the composition of microbial communities and the host factors \citep{zhang2020distance};
Other complex objects are also routinely collected in various applications such as medical imaging, computational biology, and geological analysis \citep{pan2020ball,moon2022interpoint}.
Most existing independence tests are able to measure the dependence between random vectors and cannot be directly applied to complex objects such as random matrices. Researchers tend to vectorize the random matrices to make the testing methods feasible but such transformation ignores intrinsic features of the data.
The above limitation of the current methods motivates us to develop new dependence measures that can be adapted to data in metric spaces.

Let $X$ and $Y$ be random variables on $\calX\times\calY$ with joint measure $\mu$ and marginal measures $\mu_{X}$ and $\mu_{Y}$, where $\calX$ and $\calY$ are two separable metric spaces. In this article, we focus on testing 
\beqrs
H_{0}: \text{X and Y are independent } \textrm{  versus  } H_{1}: \text{ otherwise}. 
\eeqrs
The recent surge of interest in measuring and testing  nonlinear dependence mainly focuses on developing dependence measures based on the distances between random variables \citep{moon2022interpoint}.
The leading example is the distance correlation \citep{szekely2007measuring,szekely2014partial}, which built dependence metrics between random vectors based on the distances in Euclidean spaces. 
Motivated by \cite{szekely2007measuring}, a series of Euclidean distance-based metrics are proposed such as \citet{yao2018testing},\citet{shi2020distribution} and \citet{deb2021multivariate}.
To make these dependence measures adapted to random variables in metric spaces,
\citet{gretton2005kernel} and  \citet{gretton2007kernel} introduced the Hilbert-Schmidt independence criterion by applying kernels to quantify the distance between random objects. The Hilbert-Schmidt independence criterion received considerable attention since it can be applied to complex data  such as  
fMRI signals \citep{gretton2005kernel}, compositional data \citep{rudra2022compositional} and image data \citep{damodaran2017sparse,greenfeld2020robust}. Inspired by \cite{gretton2005kernel}, \cite{pfister2018kernel,ke2019expected,deb2020measuring} further put forward kernel-based measures that can be widely used in different research areas.

The aforementioned methods are focusing on measuring the ``distance" between random variables.
To guarantee the distance-based measures are well-defined, it is required to impose  moment conditions on kernel functions.
Instead of measuring the distance between random variables,  some new dependence measures focus on quantifying the ``angle" between random variables. \cite{zhu2017projection} proposed the projection correlation based on measuring the angle between random vectors in Euclidean spaces.
\cite{zhu2017projection} imposed no moment conditions, so the independence tests based on projection correlation are robust with  heavy-tailed random variables.
There is a rich literature motivated by \cite{zhu2017projection}, and the references include, but not limited to \cite{kim2020robust, li2020projective, lai2021testing, xu2022power, liu2022new,zhang2022projective}.
Although these methods possess solid statistical properties, they failed to measure the dependence for complex objects such as compositional data and random matrices. 
In addition, the aforementioned methods can only detect the linear dependence structure when the random vectors are high dimensional \citep{zhu2020distance}. 

To make the angle-based independence tests applicable to complex objects and able to detect the nonlinear dependence for high dimensional data, we introduce the kernel angle dependence measures using reproducing Hilbert kernel spaces.
We derive the kernel angle covariances directly from the integration by making use of the good properties of reproducing Hilbert kernel spaces equipped with Gaussian measure  \citep{van2008reproducing}.
Since different kernels can be selected to catch the intrinsic features of different objects, our kernel angle covariances  can be adapted to different complex objects in metric spaces.
When distance kernel is used, the independence tests based on kernel angle covariances are equivalent to the existing angel-based independence tests using Euclidean distances \citep{zhu2017projection,kim2020robust,zhang2022projective}.
Compared to Hilbert-Schmidt information criterion \citep{gretton2007kernel}, our angle-based dependence measures impose no moment conditions on the kernel functions so they are more robust to heavy-tailed  random variables.
In addition, our framework can induce  generalized distance correlation \citep{sejdinovic2013equivalence} by the integration of the covariance using a certain weight function.
In the testing procedure, the existing angle-based dependence tests involve random permutations in approximating the null distributions, which is computationally expensive \citep{zhu2017projection,xu2022power,kim2020robust}.
We overcome such challenges using gamma approximation to estimate the null distribution of test statistics, which highly accelerates the testing procedure.
Using our framework, other test statistics on univariate or multivariate random variables can also be extended to metric spaces.

The remaining article is organized as follows. 
In Section 2, 
we put forward kernel angle dependence measures by making use of reproducing kernel Hilbert space equipped with Gaussian measures.
Then we propose estimates of angle dependence measures based on U-statistics and conduct independence tests based on the corresponding measures.
We introduce gamma approximation to estimate the null distributions to avoid random permutations.
To demonstrate the performance of the independence tests on complex data, we conduct comprehensive simulation studies and an application to microbiome data in Section 3. 
In Section 4, we further induce  generalized distance covariance by the integration of our framework. 
Section 5 summarizes our contributions and discusses some potential extensions. 
We provide the technical proofs in the Appendix.
%%%%%%%%%%%%%%%%%%%%%%%%%%%%%%%%%%%%%%%%%%%%%%%%%%%%%%%%%%%
%%%%%%%%%%%%%%%%%%%%%%%%%%%%%%%%%%%%%%%%%%%%%%%%%%%%%%%%%%%

\csection{Kernel Angle Based Dependence Measures}

\csubsection{Preliminaries}

In this section, we give a brief review of the reproducing kernel Hilbert spaces and Gaussian measures, and then establish crucial integral results for the reproducing kernel Hilbert spaces equipped with Gaussian measures.

Suppose $\calZ$ is a compact metric space equipped with a finite Borel measure $\mu$, $\calH$ is a separable Hilbert space, and  $Q:\calH\to \calH$ is the continuous, symmetric, positive definite, and linear bounded operator.

{\defi{\label{definition:RKHS}}
	Suppose $\calH$ is a Hilbert space of real-valued functions defined on $\calZ$. A function $K:\calZ\times\calZ\to\mR$  is the reproducing kernel of $\calH$ if 
	\begin{itemize}
		\item[(i)] for any $z\in\calZ$, $K(\cdot,z)\in \calH$; 
		\item[(ii)] for any $z\in\calZ$ and $f\in\calH$, $\langle f, K(\cdot,z)\rangle_{\calH} = f(z)$. 
	\end{itemize}
	$\calH$ is called a reproducing kernel Hilbert space if it possesses a reproducing kernel.
}

Moore-Aronszajn Theorem \citep[Theorem 3]{berlinet2011reproducing} claimed that, 
%given any positive semi-definite kernel $K:\calZ\times\calZ\to\mR$, 
there exists a unique Hilbert space $\calH_{K}$ of real-valued functions on $\calZ$ with positive semi-definite kernel $K$ as reproducing kernel, for any given positive semi-definite kernel $K:\calZ\times\calZ\to\mR$. $\phi(z)\defby K(\cdot,z)$ is called the canonical map of $K$ and satisfies that $\langle f,\phi(z)\rangle_{\calH_{K}} = f(z)$.

{\defi{\label{definition:GaussianMeasure}\citep[Theorem 1.11]{da2014introduction}}
	 A Gaussian measure $\mu$ on $\calH$, with mean $s\in\calH$ and covariance operator $Q$, has the characteristic function
	\beqrs
	\int_{h\in\calH}\exp(i\langle h, f\rangle_{\calH}) \mu(dh) = \exp(i\langle f,s\rangle_{\calH} - 2^{-1}\langle Qf,f\rangle_{\calH}),
	\eeqrs	
where $h, f \in\calH$. 
}

The reproducing kernel Hilbert space arises because it determines the ``geometry" of the concentration of the  Gaussian measure \citep{van2008reproducing}.
 The spectral view of reproducing kernel Hilbert space provides us the eigenfunctions to represent Gaussian measure, which facilitates us to derive the explicit form of integration results in Lemma \ref{lemma:normal_hilbertspace}.
We include the detailed derivation in the Appendix.

{\lemm{\label{lemma:normal_hilbertspace}} Suppose $s_1, s_2$ and  $h$ are in the separable reproducing kernel  Hilbert space $\calH_{K}$, which has the continuous reproducing kernel $K:\calZ\times\calZ\rightarrow\mR$. 
Let $\mu$ be the Gaussian measure on $\calH_{K}$ with mean zero and  covariance identity operator. Then
\beqr\label{equation:angle1}
	&&\int_{h\in\calH_{K}}1(\langle s_{1},h\rangle_{\calH_{K}}\leq 0) 1(\langle s_{2}, h\rangle_{\calH_{K}}\leq 0) \mu(dh)\\
	&=& 2^{-1} - (2\pi)^{-1}\arccos\{ \langle s_1,s_2\rangle_{\calH_{K}}/(\|s_1\|_{\calH_{K}}\|s_2\|_{\calH_{K}})\}. \nonumber
 \eeqr
In addition, if $U\in\mR$ follows the standard normal distribution, then
	\beqr\label{equation:angle2}
    &&\int_{h\in\calH_{K}}E\big\{1(\langle s_{1},h\rangle_{\calH_{K}}\leq U) 1(\langle s_{2}, h\rangle_{\calH_{K}}\leq U) \big\} \mu(dh)\\
    &=& 2^{-1} - (2\pi)^{-1}\arccos\{ (1+\langle s_1,s_2\rangle_{\calH_{K}})(1+\|s_1\|_{\calH_{K}}^{2})^{-1/2}(1+\|s_2\|_{\calH_{K}}^{2})^{-1/2}\}.\nonumber
    \eeqr	
}
The integration results are crucial for the 
derivation of kernel angle covariances in Section 2.2. 

\csubsection{Kernel Angle Covariances}

Suppose $\calH_{1}$ and $\calH_{2}$ are separable reproducing kernel Hilbert spaces generated from two universal kernels $K_1:\calX\times\calX\to\mR$ and $K_2:\calY\times\calY\to\mR$, respectively, 
and $\phi_{1}:\calX\to\calH_{1}$ and $\phi_{2}:\calY\to\calH_{2}$ are two canonical maps of $K_{1}$ and $K_{2}$.

{\lemm{\label{lemma:independence_equivalence}} 
	$X$ and $Y$ are independent  if and only if	$\cov\{1(f(X)\leq u), 1(g(Y)\leq v)\} = 0$ for any $f\in \calH_1$, $g\in \calH_2$ and $u,v\in\mR$.
}
\\
\noindent\cite{gretton2005kernel} put forward constraint covariance satisfying that $X$ and $Y$ are independent if and only if $\sup_{f\in\calH_{1}, g\in\calH_{2}}\cov\{f(X), g(Y)\} = 0$.
To make \\$\sup_{f\in\calH_{1}, g\in\calH_{2}}\cov\{f(X), g(Y)\}<\infty$, 
\cite{gretton2005kernel} implicitly required that both $K_{1}$ and $K_{2}$ are bounded kernels.
Inspired by \cite{zhu2017projection,kim2020robust}, we adopt indicator functions in covariance term and $\cov\{1(f(X)\leq u), 1(g(Y)\leq v)\}<\infty$ always holds. This avoids imposing further conditions about the bounded kernel   assumptions or bounded moment assumptions $E\{K_{1}(X_{1}, X_{2})\}<\infty$ and \\$E\{K_{2}(Y_{1}, Y_{2})\}<\infty$.

Based on the Lemma \ref{lemma:independence_equivalence}, we propose kernel angle  covariance 
\beqr\label{equation:integration}
\KAcov(X,Y) = c\int_{\calH_1}\int_{\calH_2}\int_{\mR^2}\cov^{2}\{1( f(X)\leq u), 1(g(Y)\leq v)\}
d \omega(u,v) \mu_1(df) \mu_2(dg).
\eeqr
To derive the explicit form of the kernel angle covariance, we set scale parameter $c=  4\pi^2$, and $\mu_{1}, \mu_{2}$ to be Gaussian measures with means zero  and covariances identity operators for $\calH_{1}$ and $\calH_{2}$. 
Let $U\defby f(X)$ and $V\defby g(Y)$.
We consider three different choices of weight functions  $\omega(u,v)$:
(1)  $d\omega_{1}(u,v) = d \Phi_{2}(u,v)$; 
(2) $d\omega_{2}(u,v) = dF_{U,V}(u,v)$;  
(3) $d\omega_{3}(u,v) = dF_{U}(u)F_{V}(v)$,
where $\Phi_{p}$ represents cumulative distribution function for $p$-dimensional standard normal variable,  $F_{U}(u)$, $F_{V}(v)$ and $F_{U,V}(u,v)$ are the cumulative distribution functions for $U$, $V$ and $(U,V)$, respectively.

We denote $\KAcov_{m}(X,Y)$ using corresponding weight functions $\omega_{m}(u,v), m = 1,2,3$. 
By applying integral results in Lemma \ref{lemma:normal_hilbertspace}, we can derive the expressions for $\KAcov_{m}(X,Y)$ in Theorem \ref{theorem:angel_correlation}.
Suppose that $\{(X_{i}, Y_{i}), i=1,\ldots, n\}$ is a random sample of $(X,Y)$.  $\KAcov_{m}(X,Y), m =1,2,3$ can be represented using the first six independent copies of $(X,Y)$.

{\theo{\label{theorem:angel_correlation}} 
	We represent $\KAcov_{m}(X,Y), m = 1,2,3$  as follows:
	\begin{itemize}
	\item[(1)] $\KAcov_{1}(X,Y)= E\{\ang_{1}^{\prime}(X_{1}, X_{2})\ang_{2}^{\prime}(Y_{1}, Y_{2}) - 2 \ang_{1}^{\prime}(X_{1}, X_{2})\ang_{2}^{\prime}(Y_{1}, Y_{3}) $\\$+ \ang_{1}^{\prime}(X_{1}, X_{2})\ang_{2}^{\prime}(Y_{3}, Y_{4})\}$;
	\item[(2)] $\KAcov_{2}(X,Y) =  E\{\ang_{1}(X_{1}, X_{2};X_{5})\ang_{2}(Y_{1}, Y_{2};Y_{5}) - 2 \ang_{1}(X_{1}, X_{2};X_{5})$\\$\ang_{2}(Y_{1}, Y_{3};Y_{5}) + \ang_{2}(X_{1}, X_{2};X_{5})\ang_{2}(Y_{3}, Y_{4};Y_{5})\}$;
	\item[(3)] 
	$\KAcov_{3}(X,Y) =E\{\ang_{1}(X_{1}, X_{2};X_{5})\ang_{2}(Y_{1}, Y_{2};Y_{6}) - 2 \ang_{1}(X_{1}, X_{2};X_{5})$\\$\ang_{2}(Y_{1}, Y_{3};Y_{6}) + \ang_{1}(X_{1}, X_{2};X_{5})\ang_{2}(Y_{3}, Y_{4};Y_{6})\}$,
    \end{itemize}
where
$\ang_{1}^{\prime}(X_{i}, X_{j}) \defby \arccos\big[\{K_{1}(X_{i}, X_{j}) + 1\}\{K_{1}(X_{i}, X_{i})+1\}^{-1/2}\{K_{1}(X_{j}, X_{j})+$\\
$1\}^{-1/2}\big]$ and
$\ang_{1}(X_{i}, X_{j};X_{k}) \defby \arccos\big[\{K_{1}(X_{i}, X_{j}) - K_{1}(X_{i}, X_{k})  - K_{1}(X_{j}, X_{k}) +$\\
$ K_{1}(X_{k}, X_{k}) \}\{K_{1}(X_{i}, X_{i}) - 2K_{1}(X_{i}, X_{k}) +K_{1}(X_{k}, X_{k})\}^{-1/2}\{K_{1}(X_{j}, X_{j}) - 2K_{1}(X_{j}, $\\$X_{k}) +K_{1}(X_{k}, X_{k})\}^{-1/2}\big]$. $\ang_{2}^{\prime}(Y_{i}, Y_{j})$ and $\ang_{2}(Y_{i}, Y_{j};Y_{k})$  are expressed in an analogous manner. If $j=k$ or $i=k$, $\ang_{2}(X_{i}, X_{j};X_{k})=\ang_{2}(Y_{i}, Y_{j};Y_{k})=0$. 
}

\begin{figure}[h]
\begin{tikzpicture}
	\draw[black, thick] (0,0) -- (2.8,0);
	\draw[black, thick] (0,0) -- (2,2);
	\draw (0.5,0) arc (0:45:0.5cm);
	\filldraw[black] (0,0) circle (1.3pt) node[anchor=east]{$(\phi_{1}(X_k), 0)$};
	\filldraw[black] (2.8,0) circle (1.3pt) node[anchor=west]{$(\phi_{1}(X_i), 0)$};
	\filldraw[black] (2,2) circle (1.3pt) node[anchor=west]{$(\phi_{1}(X_j), 0)$};
	\draw[black, thick] (7,0) -- (9.8,0);
	\draw[black, thick] (7,0) -- (9,2);
	\draw (7.5,0) arc (0:45:0.5cm);
	\filldraw[black] (7,0) circle (1.3pt) node[anchor=east]{$(\vz_{\calH_{1}}, 1)$};
	\filldraw[black] (9.8,0) circle (1.3pt) node[anchor=west]{$(\phi_{1}(X_i), 0)$};
	\filldraw[black] (9,2) circle (1.3pt) node[anchor=west]{$(\phi_{1}(X_j), 0)$};
	\draw[-latex] (0.6,0.25) -- (1.1,0.6) node[anchor=west] {$\ang(X_{i},X_{j};X_{k})$};
	\draw[-latex] (7.6,0.25) -- (8.1,0.6) node[anchor=west] {$\ang^{\prime}(X_{i},X_{j})$};
\end{tikzpicture}
    \caption{The left panel is the angle for $\ang(X_{i}, X_{j};X_{k})$, and the right panel is the angle for $\ang^{\prime}(X_{i}, X_{j})$.}
    \label{figure:twoangle}
\end{figure}
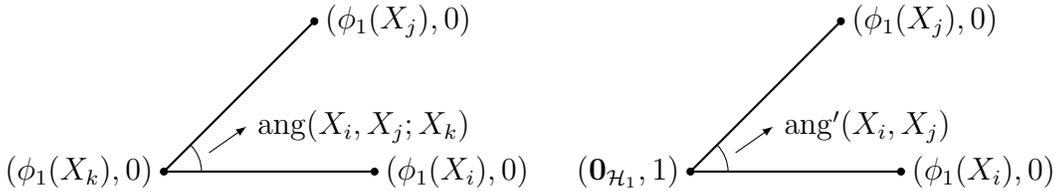

Suppose that the Hilbert space $\calH^\prime_{1}\defby\calH_{1}\times\mR$, which has the induced inner product $\langle(f_1,a_1),(f_2,a_2)\rangle_{\calH_{1}^{\prime}} = \langle f_1,f_2 \rangle_{\calH_{1}} + a_{1}a_{2}$ for $f_1,f_2\in\calH_{1}$ and $a_1,a_2\in\mR$. 
$\calH_{2}^{\prime}\defby\calH_{2}\times\mR$ has analogous induced inner product.
$\vz_{\calH_{m}}$ are zero elements in $\calH_{m}, m=1,2$.
Actually, both $\ang_{1}^{\prime}(X_{i}, X_{j})$ and $\ang_{1}(X_{i}, X_{j}; X_{k})$ can be represented as the angle formed by three points in Hilbert space $\calH_{1}^{\prime}$.
$\ang_{1}^{\prime}(X_{i}, X_{j})$ represents the angle formed by three points $(\phi_{1}(X_i), 0)$, $(\phi_{1}(X_j), 0)$ and $(\vz_{\calH_1}, 1)$, and $(\vz_{\calH_1}, 1)$ is the vertex. $\ang_{1}(X_{i}, X_{j};X_{k})$ represents the angle formed by $(\phi_{1}(X_i), 0)$, $(\phi_{1}(X_j), 0)$ and $(\phi_{1}(X_k), 0)$, and $(\phi_{1}(X_k), 0)$ is the vertex. The two angles are illustrated in Figure \ref{figure:twoangle}.
$\ang_{1}^{\prime}(Y_{i}, Y_{j})$ and $\ang_{1}(Y_{i}, Y_{j};Y_{k})$ can also be seen as angles in $\calH_{2}^{\prime}$.

Specifically, $\KAcov_{1}(X,Y)$ uses $(\vz_{\calH_{1}},1)$ and $(\vz_{\calH_{2}},1)$ as vertexes, $\KAcov_{2}(X,Y)$ uses $(\phi_{1}(X_{5}),0)$ and $(\phi_{2}(Y_{5}),0)$ as vertexes, and  $\KAcov_{3}(X,Y)$ uses $(\phi_{1}(X_{5}),0)$ and $(\phi_{2}(Y_{6}),0)$ as vertexes. 
Then, kernel angle correlations can be defined by $\KAC_{m}(X,Y)\defby\KAcov_{m}(X,Y)\{\KAcov_{m}(X,X)\KAcov_{m}(Y,Y)\}^{-1/2}$ for $m = 1,2,3$.

Our kernel angle dependence measures can be adapted to complex objects with different choices of kernels.
In particular, when $\calX = \mR^{p_{1}}$, $\calY= \mR^{p_{2}}$, $K_{1}(X_{1}, X_{2}) = X_{1}\trans X_{2}$ and $K_{2}(Y_{1}, Y_{2}) = Y_{1}\trans Y_{2}$,
$\KAcov_{1}(X,Y)$, $\KAcov_{2}(X,Y)$ and $\KAcov_{3}(X,Y)$ are equivalent to  the improved projection covariance \citep{zhang2022projective}, projection covariance \citep{zhu2017projection} and multivariate Blum-Kiefer-Rosenblatt coefficient \citep[Theoerm 7.2]{kim2020robust}, respectively.

\csubsection{$U$-statistic Estimates and the Asymptotics}
We build estimates for $\KAcov_{m}(X,Y)$, $m=1,2,3$ upon $U$-statistics \citep{serfling1980approximation}. To simplify the notations, we define
$a_{ij}^{\prime}\defby\ang_{1}^{\prime}(X_{i}, X_{j})$, $b_{ij}^{\prime}\defby\ang_{2}^{\prime}(Y_{i}, Y_{j})$,
$a_{ijk}\defby\ang_{1}(X_{i}, X_{j}; X_{k})$ and $b_{ijk}\defby\ang_{2}(Y_{i}, Y_{j}; Y_{k})$ for $i,j,k=1,\ldots,n$.
The estimates for $\KAcov_{m}(X,Y), m=1,2,3$ are
\beqr\label{equation:definitionkacov}
\wh\KAcov_{1}(X,Y) &\defby& \{(n)_{4}\}^{-1} \sum_{(i,j,k,l)}^{n}  \left(a_{ij}^{\prime}b_{ij}^{\prime} - 2a_{ij}^{\prime}b_{ik}^{\prime} + a_{ij}^{\prime}b_{kl}^{\prime}\right),\\
\wh\KAcov_{2}(X,Y) &\defby&  \{(n)_{5}\}^{-1} \sum_{(i,j,k,l,r)}^{n}  \left(a_{ijr}b_{ijr} - 2a_{ijr}b_{ikr} + a_{ijr}b_{klr}\right),\nonumber\\
\wh\KAcov_{3}(X,Y) &\defby& \{(n)_{6}\}^{-1}  \sum_{(i,j,k,l,r,t)}^{n}  \left(a_{ijr}b_{ijt} - 2a_{ijr}b_{ikt} + a_{ijr}b_{klt}\right), \nonumber
\eeqr
where $(n)_m\defby n(n-1)\cdots(n-m+1)$, and $(i,j,k,l)$, $(i,j,k,l,r)$ and $(i,j,k,l,r,t)$ are taken over the indexes from $\{1,\ldots,n\}$ that are different from each other.
We provide an equivalent representation of equation \eqref{equation:definitionkacov} as follows.
\beqr\label{equation:calculationkacov}
\wh\KAcov_{1}(X,Y) &=&\{n(n-3)\}^{-1}\Big[\tr(\A^{\prime}\B^{\prime})\\
&&\hspace{2.5cm}-2(n-2)^{-1}\vo_{n}\trans\A^{\prime}\B^{\prime}\vo_{n}+\{(n-1)_2\}^{-1}\vo_{n}\trans\A^{\prime}\vo_{n}\vo_{n}\trans\B^{\prime}\vo_{n}\Big],\nonumber\\
\wh\KAcov_{2}(X,Y) &=& \{n(n-1)(n-4)\}^{-1}\sum\limits_{r=1}^{n}\Big[\tr(\A_{r}\B_{r})\nonumber\\
&&-2(n-3)^{-1}\vo_{n-1}\trans\A_{r}\B_{r}\vo_{n-1}+\{(n-2)_2\}^{-1}\vo_{n-1}\trans\A_{r}\vo_{n-1}\vo_{n-1}\trans\B_{r}\vo_{n-1}\Big],\nonumber\\
\wh\KAcov_{3}(X,Y) &=& \{n(n-1)(n-2)(n-5)\}^{-1}\sum\limits_{(r,t)}^{n}\Big[\tr(\A_{rt}\B_{tr}) \nonumber\\
&&-2(n-4)^{-1}\vo_{n-2}\trans\A_{rt}\B_{tr}\vo_{n-2}+\{(n-3)_2\}^{-1}\vo_{n-2}\trans\A_{rt}\vo_{n-2}\vo_{n-2}\trans\B_{tr}\vo_{n-2}\Big], \nonumber
\eeqr
where $\textbf{1}_{n}\in\mR^{n}$ with all elements equal to one, $\A^{\prime},\B^{\prime} \in \mR^{n\times n}$ with $\A^{\prime} \defby (a^{\prime}_{ij}$) and $\B^{\prime} \defby (b^{\prime}_{ij})$, for $i,j= 1,\ldots,n$,
$\A_{k},\B_{k} \in \mR^{(n-1)\times (n-1)}$ with  $\A_{k} \defby (a_{ijk})$ and $\B_{k} \defby (b_{ijk})$ for $i,j \neq k$,
$\A_{rt},\B_{tr} \in \mR^{(n-2)\times (n-2)}$ with  $\A_{rt} \defby (a_{ijr})$ and $\B_{tr} \defby (b_{ijt})$ for $i,j \neq t$, $i,j \neq r$.
Equations in \eqref{equation:calculationkacov} indicate that the computation complexies of $\wh\KAcov_{m}(X,Y), m = 1,2,3$ are $O(n^{2})$, $O(n^{3})$ and $O(n^{4})$ respectively.
The estimates for kernel angle correlations are  $\wh\KAC_{m}(X,Y)\defby\wh\KAcov_{m}(X,Y)\{\wh\KAcov_{m}(X,X)\wh\KAcov_{m}(Y,Y)\}^{-1/2}$ for $m=1,2,3$, correspondingly.

{\theo{\label{theorem:angel_correlation_ustat}} 
	 For $m = 1,2,3$, we have
	\begin{itemize}
		\item[(1)] when $X$ and $Y$ are independent,
		\beqrs
		n\wh\KAcov_{m}(X,Y) \stackrel{d}{\longrightarrow}\sum_{j=1}^{\infty} \eta_{m,j} (\zeta^{2}_{1,m,j}-1);
		\eeqrs
		\item[(2)] when $X$ and $Y$ are not independent,
		\beqrs
		n^{1/2}\big\{\wh\KAcov_{m}(X,Y)-\KAcov_{m}(X,Y)\big\} \stackrel{d}{\longrightarrow}\calN(0, \sigma_{m}^{2}),
		\eeqrs 
	\end{itemize}
 as $n$ diverges to infinity.
 Here, $\{\zeta_{1,m,j}, m=1,2,3, j = 1,\ldots,\infty \}$ are independent  and follow the standard normal distribution, and $\{\eta_{m,j},m=1,2,3,  j = 1,\ldots,\infty\}$ are the eigenvalues of the corresponding Hilbert-Schmidt integral operator depending on the distributions of $X$ and $Y$. $\sigma_{m}, m=1,2,3$ are given in \eqref{equation:sigma1},\eqref{equation:sigma2} and   \eqref{equation:sigma3} in the Appendix.
}

Theorem \ref{theorem:angel_correlation_ustat} establishes the asymptotics for $\wh\KAcov_{m}(X, Y)$ under both the null and alternative hypothesis.
Under the null hypothesis, $n\wh\KAcov_{m}(X,Y)$ converges in distribution to $I_{1,m} - I_{2,m}$, where
\beqrs
I_{1,m}\defby \sum_{j=1}^{\infty} \eta_{m,j} \zeta^{2}_{1,m,j} \textrm{ and } 
I_{2,m}\defby \sum_{j=1}^{\infty} \eta_{m,j} .
\eeqrs
The distributions of $I_{1,m} - I_{2,m}, m=1,2,3$ are not intractable, since parameters $\{\eta_{m,j},m=1,2,3,  j = 1,\ldots,\infty\}$ are unknown.
To make kernel angle independence tests into practice, the researchers often implement random permutations to approximate the distribution of $I_{1,m} - I_{2,m}$, which results in heavy computational burdens.
To accelerate the testing procedure,  we introduce gamma approximations \citep{welch1938significance,satterthwaite1946approximate} to estimate the distribution of the test statistics, which approximate the distribution of $I_{1,m}$ using Gamma distribution by matching the first two moments of $I_{1,m}$.
By defining $a_{1} \defby E\big\{\ang_{1}^{\prime}(X_1,X_2)\big\}E\big\{\ang_{2}^{\prime}(Y_1,Y_2)\big\}$,  $b_{1} \defby E\big\{\KAcov_{1}(X_1,X_2)\big\}E\big\{\KAcov_{1}(Y_1,Y_2)\big\}$,
$a_{2} \defby E\big\{\ang_{1}(X_1,X_2;X_3)\big\}E\big\{\ang_{1}(Y_1,Y_2;$\\$Y_3)\big\}$ and $b_{2} \defby E\big\{\ang_{1}(X_1,X_2;X_3)\big\}E\big\{\ang_{1}(Y_1,Y_2;Y_3)\big\}$, 
we derive the shape $\alpha_m$ and rate parameters $\beta_m$ of Gamma distributions in the proof of Theorem \ref{theorem:angel_correlation_ustat}, where $\alpha_{1} = a_{1}^2/(2b_{1})$, $\beta_{1} = a_{1}/(2b_{1})$, $\alpha_{2} = a_{2}^2/(2b_{2})$, $\beta_{2} = a_{2}/(2b_{2})$,
$\alpha_{3} = \alpha_{2}$ and $\beta_{3} = \beta_{2}$.
Given the fact that $I_{2,m} = E(I_{1,m}) = \alpha_{m}\beta_{m}^{-1}$, the distributions $I_{1,m} - I_{2,m}, m=1,2,3$ can be approximated by $\textrm{Gamma}(\alpha_{m},\beta_{m}) - \alpha_{m}\beta_{m}^{-1}$ respectively.

Suppose that $\z\defby(Z_1,\ldots,Z_{p})\trans\in\mR^{p}$  and $\Z\in\mR^{p\times p}$. 
In practice,  we should choose different kernels for different types of data to improve the performance of the tests.
For low dimension random vector, we suggest Gaussian kernel $K(\z_1, \z_2) = \exp\{-\|\z_1-\z_2\|_{2}^{2}/\gamma^2\}$ and Laplacian kernel $K(\z_1, \z_2) = \exp\{-\|\z_1-\z_2\|_{2}/\gamma\}$ \citep{sriperumbudur2011universality}, where $\gamma$ is the median of pairwise distances of random vectors \citep{scholkopf2002learning}. 
Distance kernel $K(\z_1, \z_2) = (\|\z_1\|_{2}^{\alpha}+ \|\z_2\|_{2}^{\alpha}- \|\z_1 - \z_2\|_{2}^{\alpha})/2$, $0<\alpha\leq 2$ \citep[Example 15]{sejdinovic2013equivalence} is another possible choice. 
When $\alpha=2$, $K(\z_1, \z_2) = \z_{1}\trans\z_{2}$ corresponds to inner product in Euclidean space.
For high dimensional random vector, \cite{zhu2020distance,yan2021kernel} pointed out that  if the kernel $K(\z_{1},\z_{2})$ is a smooth function of $\|\z_{1}-\z_{2}\|_{2}^{2}$ (i.e. Gaussian kernel, Laplacian kernel, and distance kernel), the dependence measures with such kernels can only detect the linear dependence.
To overcome such difficulty, we  suggest $L^1$-norm based kernel \citep{sarkar2020some,sarkar2018some}
\beqrs
K(\z_1, \z_2) = 2^{-1}\left\{\left(\sum_{u=1}^{p}\abs{Z_{1u}}\right)^{2}+ \left(\sum_{u=1}^{p}\abs{Z_{2u}}\right)^2 - \left(\sum_{u=1}^{p}\abs{Z_{1u} - Z_{2u}}\right)^{2}\right\}.
\eeqrs
For the symmetric positive definite matrix,
\cite{arsigny2007geometric} proposed log-Euclidean distance, which is the geodesic distance on the manifold.
We denote the corresponding log-Euclidean kernel as
\beqrs
K(\Z_1, \Z_2) =  2^{-1}\left\{\|\log (\Z_{1})\|_{F}^{2} + \|\log (\Z_{2})\|_{F}^{2} - \|\log (\Z_{1}) - \log (\Z_{2}) \|_{F}^{2}\right\},
\eeqrs
where $\log(\Z)$ represents the logarithm for matrix $\Z$ and $\|\cdot\|_{F}$ denotes the Frobenius matrix norm.

%%%%%%%%%%%%%%%%%%%%%%%%%%%%%%%%%%%%%%%%%%%%%%%%%%%%%%%%%%%
%%%%%%%%%%%%%%%%%%%%%%%%%%%%%%%%%%%%%%%%%%%%%%%%%%%%%%%%%%%

%%%%%%%%%%%%%%%%%%%%%%%%%%%%%%%%%%%%%%%%%%%%%%%%%%%%%%%%%%%
%%%%%%%%%%%%%%%%%%%%%%%%%%%%%%%%%%%%%%%%%%%%%%%%%%%%%%%%%%%

\csection{Numerical Studies}

In this section, we conduct extensive simulations to examine the performance of our proposed independence tests based on $\wh\KAcov_{m}(X, Y)$ for $m=1,2,3$. 
We consider three different objects and select suitable kernels for them.
We examine the estimation accuracy of gamma approximation for the critical value and compare the power of the proposed tests with other existing methods.
We include  four existing distance-based independence tests in the simulation studies, which are the distance correlation test \citep{szekely2007measuring}, Ball covariance test \citep{pan2020ball}, Hilbert-Schmidt information criterion
\citep{gretton2007kernel} and the multivariate test of \citep{heller2013consistent}. $199$ random permutations are used to approximate the distributions under the null hypothesis for these four existing tests.
We set sample size $n=100$ and report empirical size and power based on 500 replicates.
Throughout the simulations, $X$ and $Y\in\mR$ are univariate random variables, $\x$ and $\y$ are random vectors and $\X$, and $\Y$ are random matrices.

\begin{figure}[htbp!]
	%\addtocounter{subfigure}{0}
	\captionsetup{font = footnotesize}
	\centering
	\subfigure[Linear Normal]{
		\begin{minipage}[t]{0.45\linewidth}
			\centering
			\includegraphics[width=2.6in]{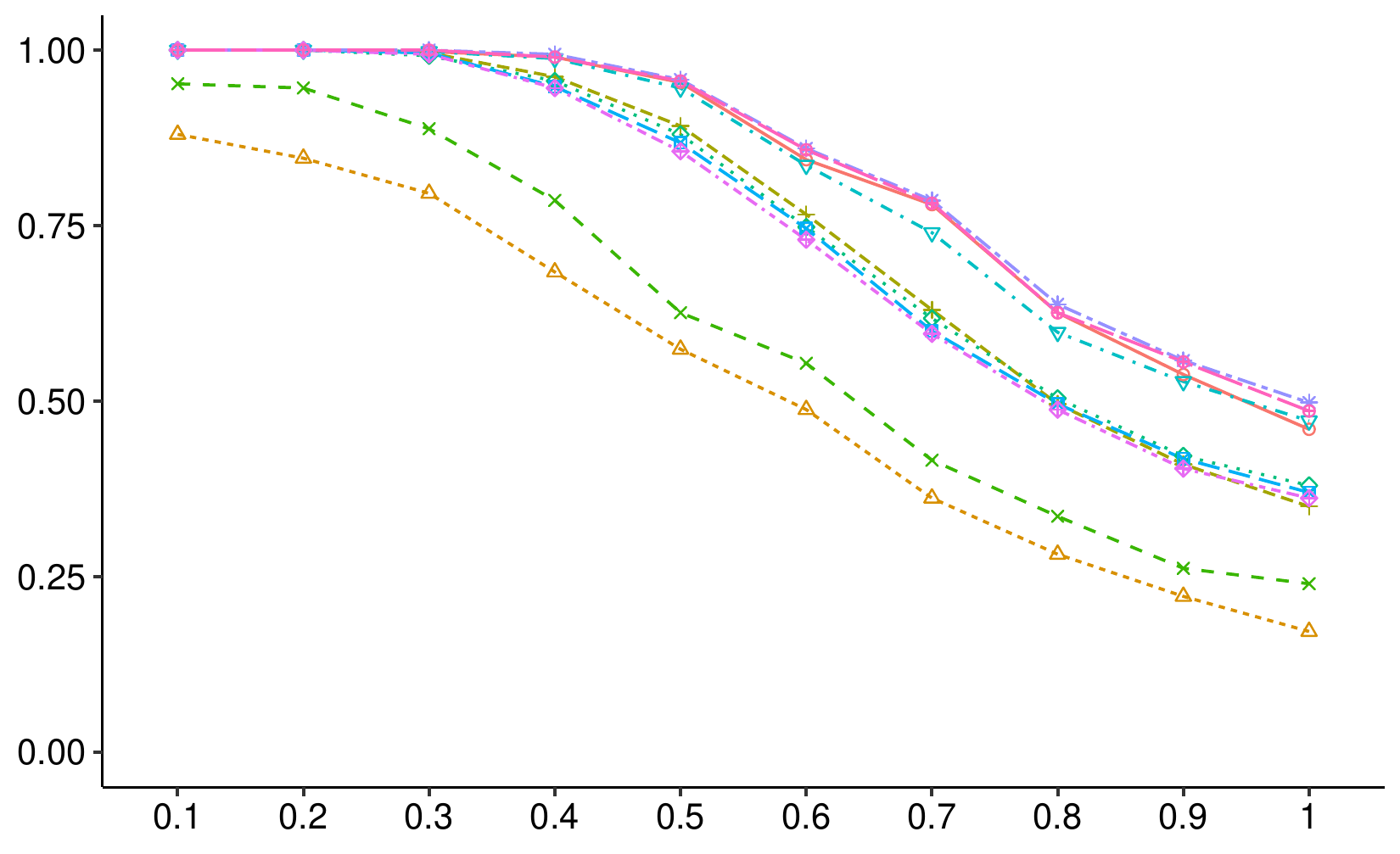}
			%\caption{fig1}
		\end{minipage}%
	}%
	\subfigure[Linear $t_3$]{
	\begin{minipage}[t]{0.45\linewidth}
		\centering
		\includegraphics[width=2.6in]{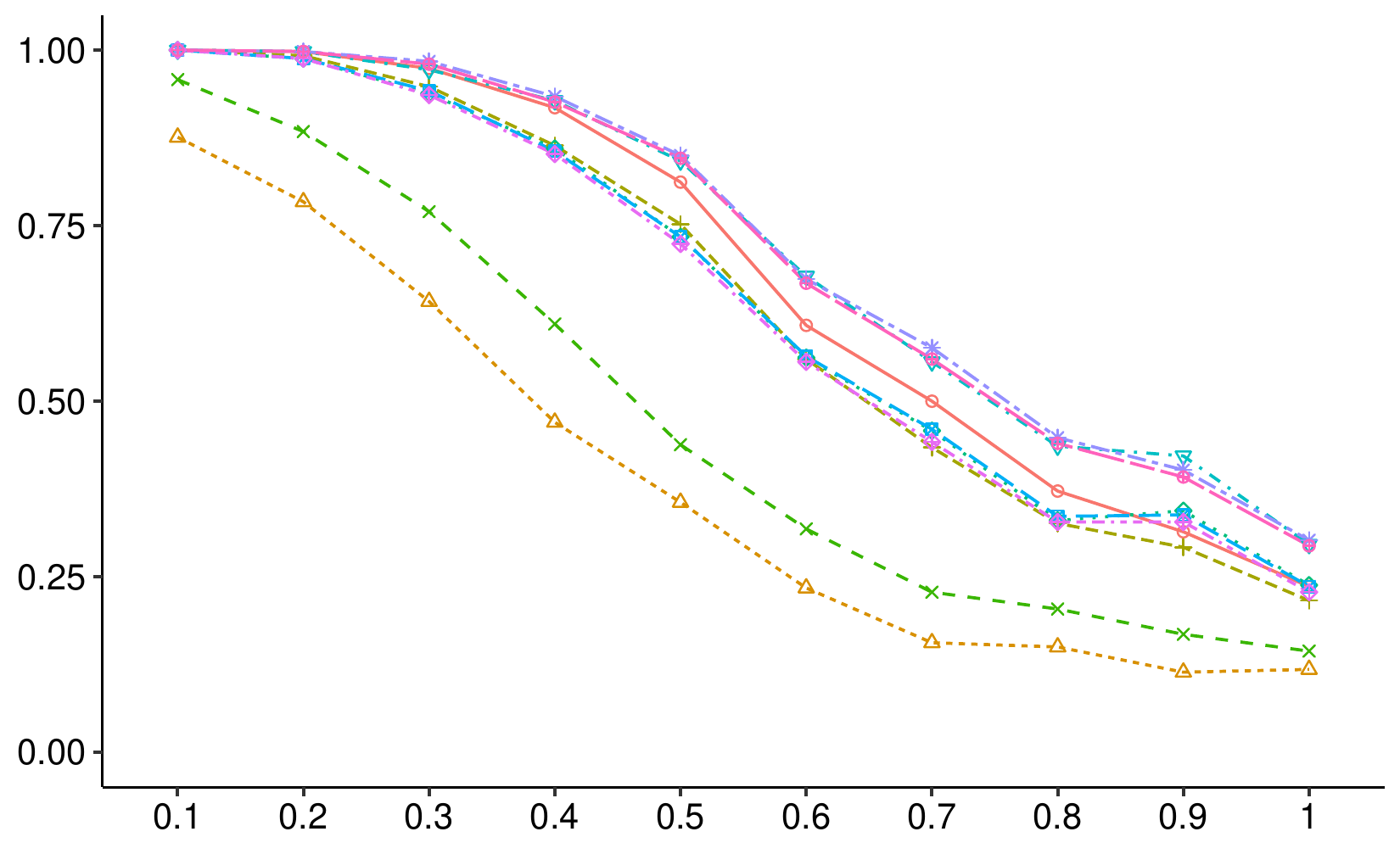}
		%\caption{fig1}
	\end{minipage}%
   }%

	\subfigure[Log Normal]{
		\begin{minipage}[t]{0.45\linewidth}
			\centering
			\includegraphics[width=2.6in]{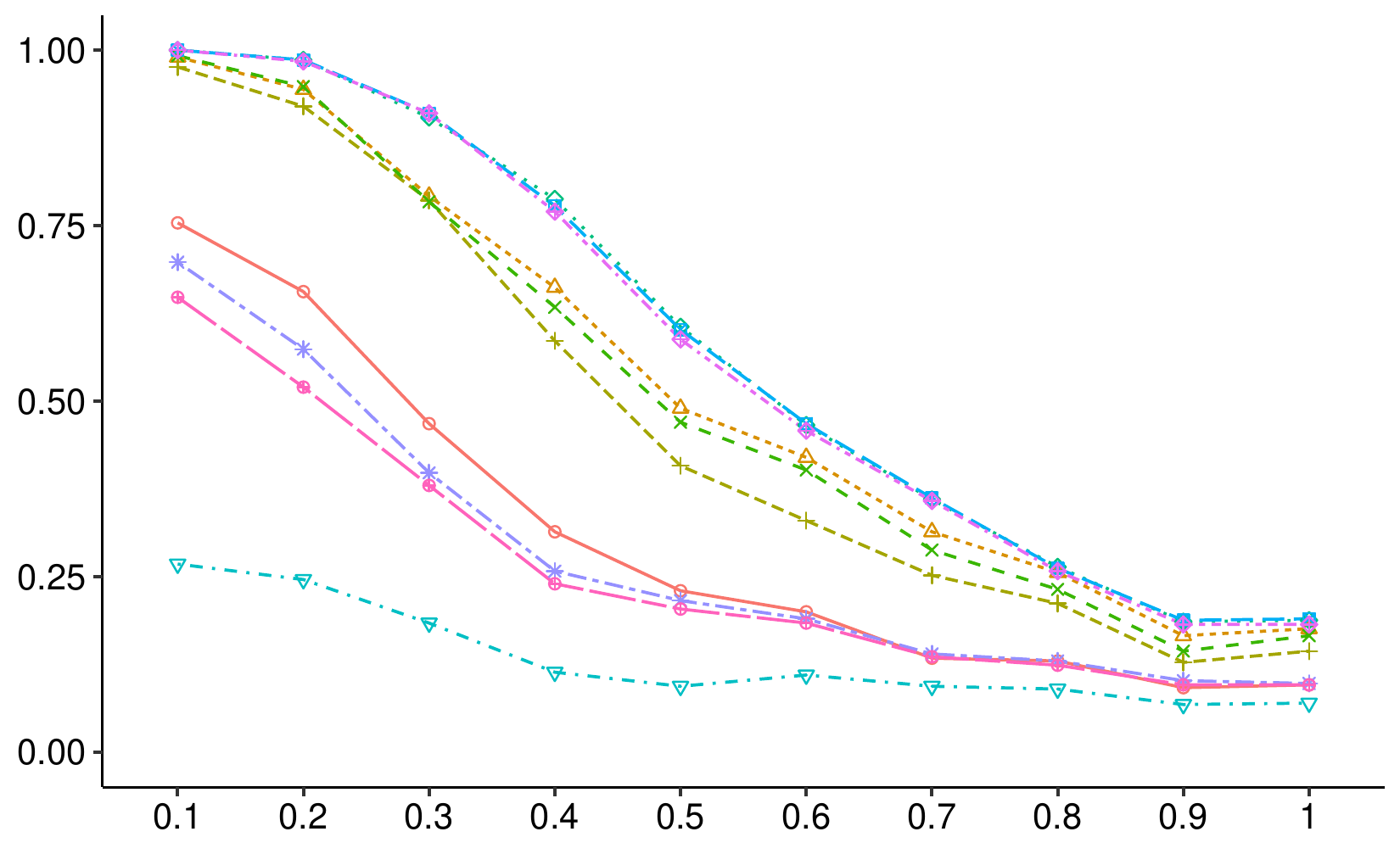}
			%\caption{fig1}
		\end{minipage}%
	}%
    \subfigure[Log $t_{3}$]{
    	\begin{minipage}[t]{0.45\linewidth}
    		\centering
    		\includegraphics[width=2.6in]{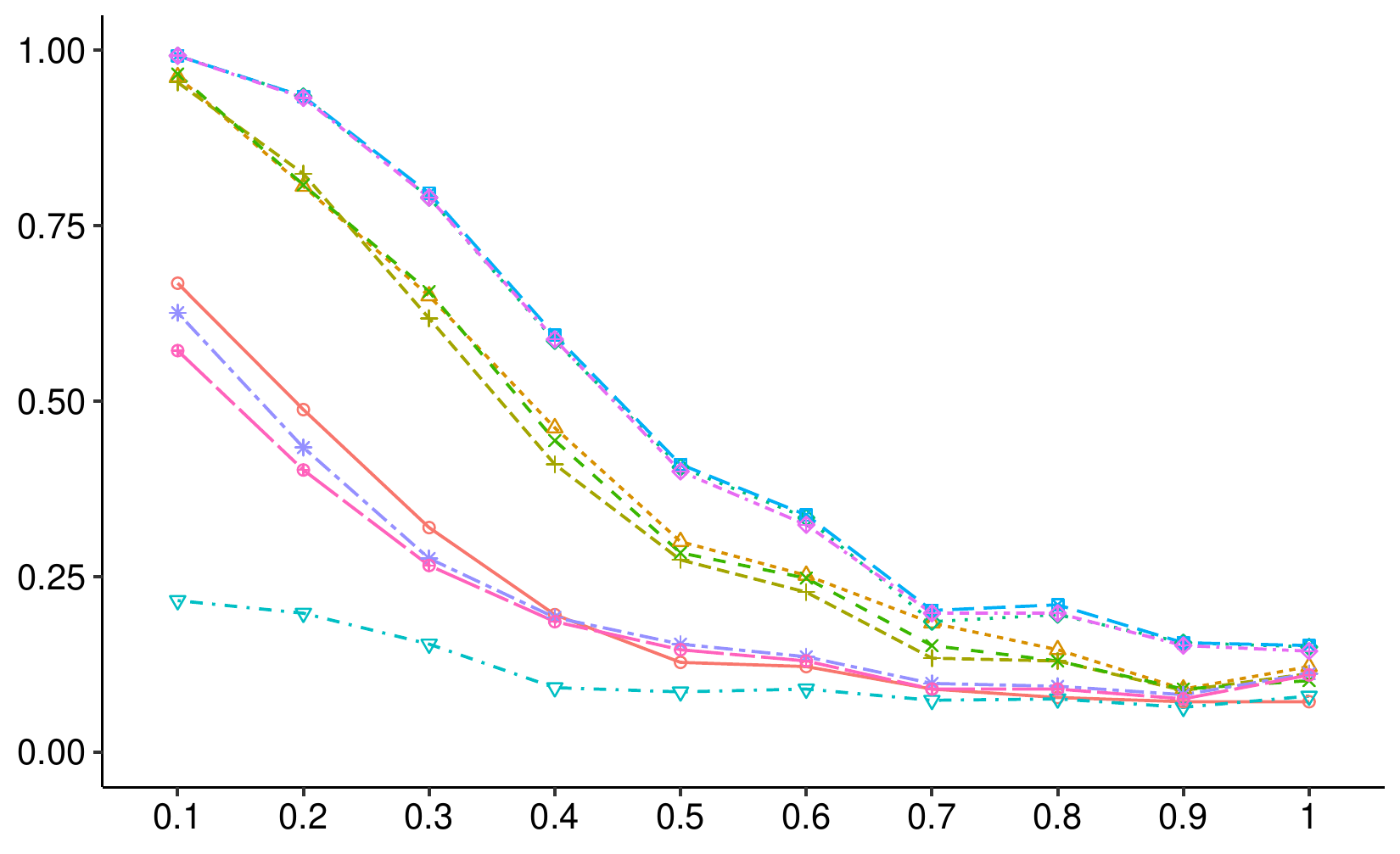}
    		%\caption{fig1}
    	\end{minipage}%
    }%

	\subfigure[Quaratic Normal]{
		\begin{minipage}[t]{0.45\linewidth}
			\centering
			\includegraphics[width=2.6in]{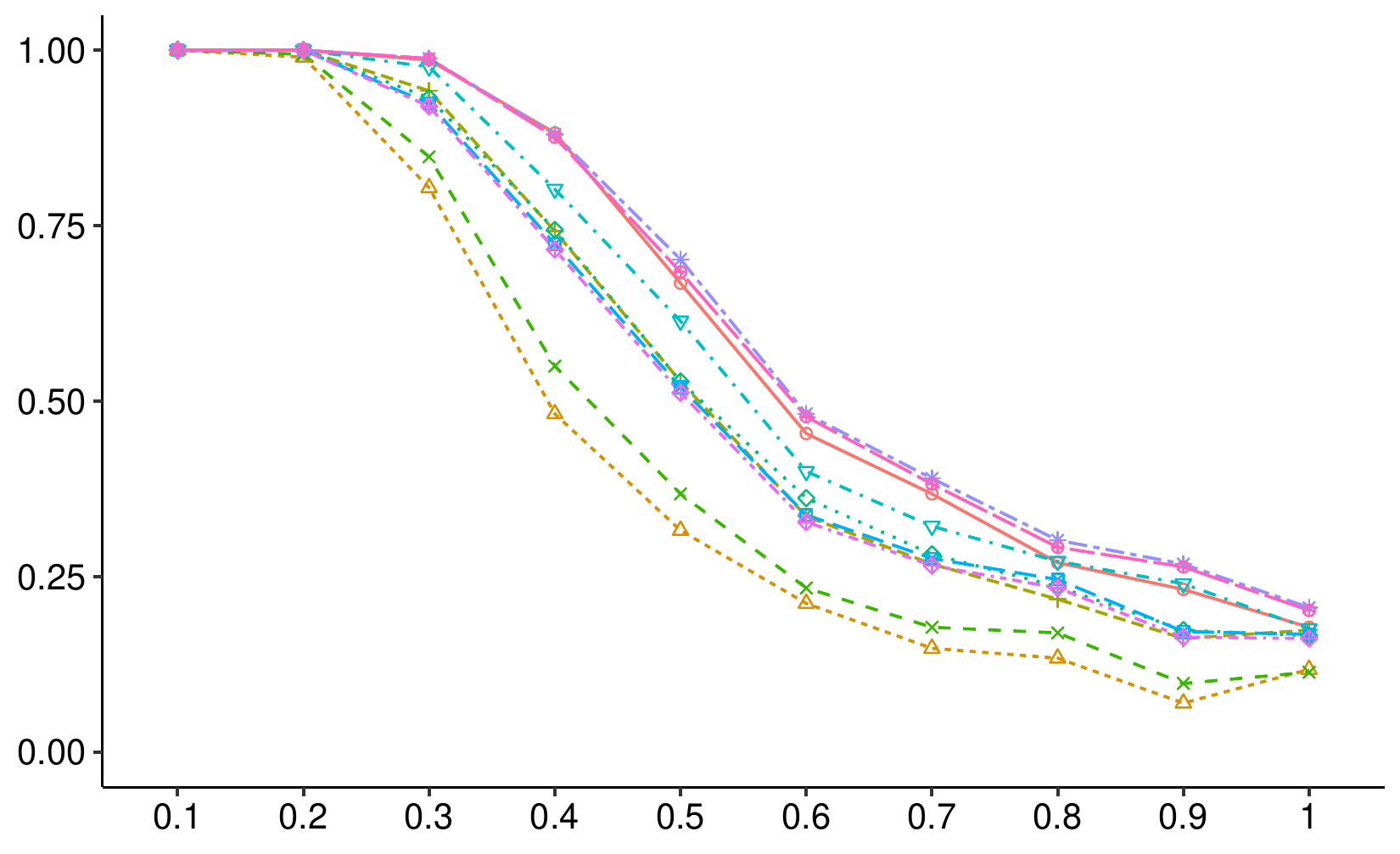}
			%\caption{fig2}
		\end{minipage}%
	}%
	\subfigure[Quadratic $t_{3}$]{
		\begin{minipage}[t]{0.45\linewidth}
			\centering
			\includegraphics[width=2.6in]{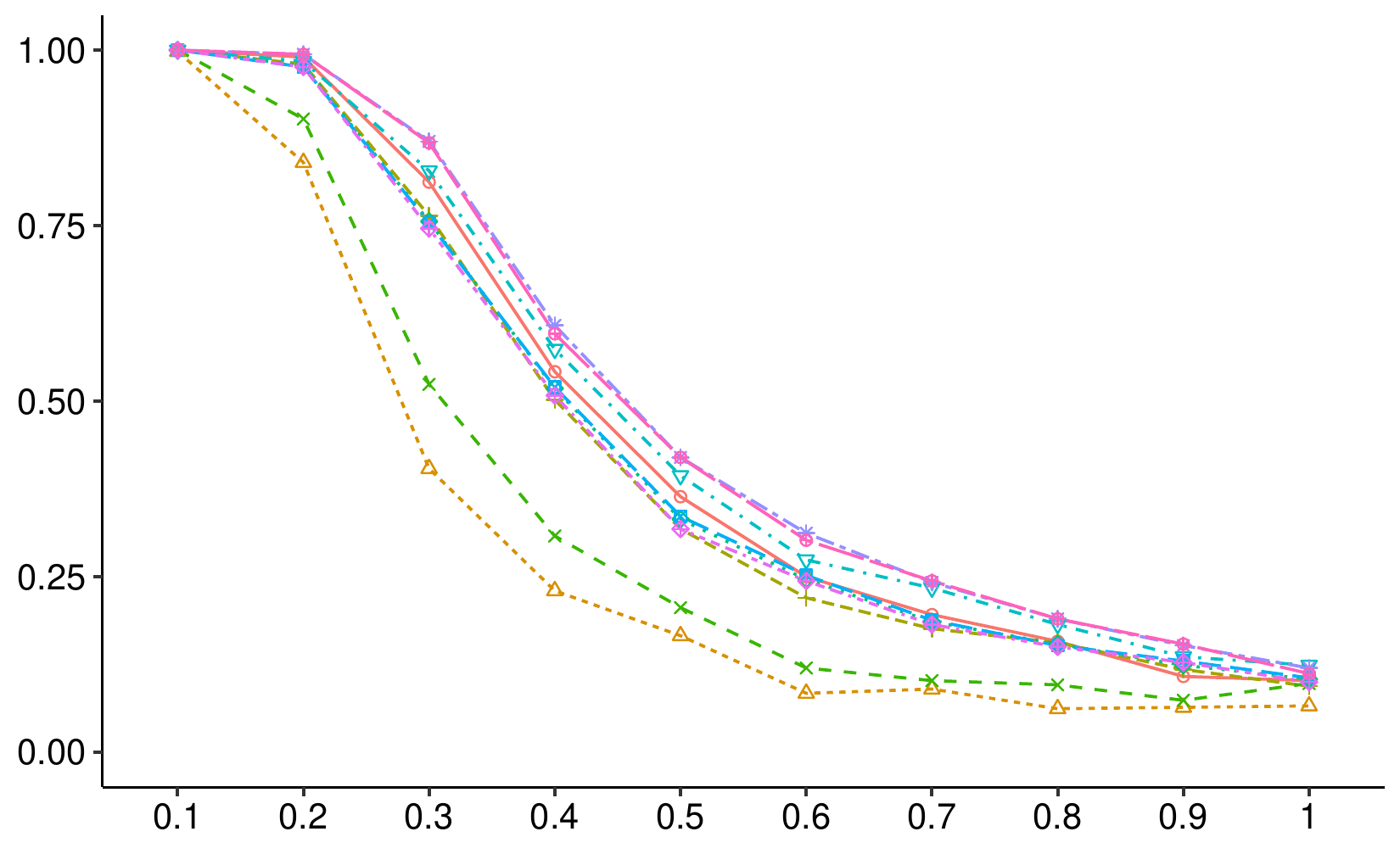}
			%\caption{fig2}
		\end{minipage}%
	}%
	\centering
	\caption{The empirical powers of tests under three scenarios with noise of standard normal distribution ($z$) and $t$ distribution with 3 degrees of freedom ($t_3$). The horizontal axis represents $\lambda$ which controls the size of noise and the vertical axis represents the empirical power. We use different marks to represents different tests, distance correlation test: $\circ$; the multivariate test of \cite{heller2013consistent}: $\triangle$; the Hilbert-Schmidt independence criterion with Gaussian kernel: $+$; Ball covariance test: $\times$; $\KAcov_1(\x,\y)$ using Laplacian kernel: $\Diamond$; $\KAcov_2(\x,\y)$ using Laplacian kernel: $\boxtimes$;  $\KAcov_3(\x,\y)$ using Laplacian kernel: $\protect\diamondplus$; $\KAcov_1(\x,\y)$ using distance kernel: $\bigtriangledown$; $\KAcov_2(\x,\y)$ using distance kernel: $\ast$;  $\KAcov_3(\x,\y)$ using distance kernel: $\oplus$.}
	\label{fig:powervector}
\end{figure}

\noindent\textbf{Study 1.}
We first test dependence between low-dimensional random vectors $\x$ and $\y$, where $\x \defby (X_1, \ldots, X_5)\trans$ and $\y \defby (Y_1,\ldots, Y_5)\trans$.
We choose distance kernel and Laplacian kernel for  $\KAcov_m(\x,\y), m=1,2,3$.
Three scenarios are considered as follows{\color{red}:}
\begin{enumerate}
	\item[(1)] Linear: $Y_{j} = 0.4X_{j} + 0.2X_{j+1} + 2\lambda\varepsilon_{j}$ for $j = 1,2$;
	\item[(2)] Log: $Y_{j} = 1.2\log(X_{j}^2) + 6\lambda\varepsilon_{j}$ for $j = 1,2$;
	\item[(3)] Quadratic: $Y_{j} = 0.2(X_j-2)^2 + 6\lambda\varepsilon_{j}$ for $j = 1,2$.
\end{enumerate}
We generate $X_j, j=1,\ldots,5$ and $Y_j, j=3,4,5$ independently from the standard normal distribution, or $\varepsilon_{j}, j=1,2$ from two different distributions, standard normal distribution, and $t$ distribution with $3$ degrees of freedom. 
We first examine whether these tests can control type I error. 
To generate data when $\x$ and $\y$ are independent, we generate two independent pairs $(\x_{1}, \y_{1})$ and $(\x_{2}, \y_{2})$. We test the dependence between $\x_{1}$ and $\y_{2}$.
We fix $\lambda=1$ and report the empirical size at significance level $\alpha=0.05$ in Table \ref{table:sizevector}.
Then, we compare power performances by testing dependence between $\x_{1}$ and $\y_{1}$.
Figure \ref{fig:powervector} depicts the empirical powers over $\lambda\in\{0.1,\ldots,1\}$.

\begin{table}[htbp!]
	\captionsetup{font = footnotesize}
 \caption{\footnotesize The empirical sizes of tests under three scenarios with noises of standard normal distribution ($z$) and $t$ distribution with 3 degrees of freedom ($t_3$). The tests include the distance correlation test (DC), the multivariate test of \cite{heller2013consistent} (HHG), the Hilbert-Schmidt independence criterion with Gaussian kernel (HSIC), Ball covariance test (Ball), $\KAcov_{m}(\x,\y), m=1,2,3$ using distance kernel and Laplacian kernel.}
\label{table:sizevector}   
\centering
 \footnotesize{ 
  \begin{tabular}{r|r|ccccc}
    \hline
     \hline
	\multirow{6}{*}{(1)} 	& & DC & HHG  &  HSIC & Ball & $\KAcov_{1,\textrm{Laplacian}}$ \\
	\cline{2-7} 
     & $z$ & 0.052 & 0.044 & 0.058 & 0.050 & 0.054 \\
     & $t_3$ & 0.050 & 0.058 &  0.056 &  0.044 & 0.064 \\
     \cline{2-7} 
      &  & $\KAcov_{1,\textrm{distance}}$ & $\KAcov_{2,\textrm{Laplacian}}$  &  $\KAcov_{2,\textrm{distance}}$ & $\KAcov_{3,\textrm{Laplacian}}$ & $\KAcov_{3,\textrm{distance}}$ \\
     \cline{2-7} 
     & $z$ & 0.048 & 0.056 & 0.052 & 0.048 & 0.050 \\
     & $t_3$ & 0.062& 0.064& 0.064 &0.064 & 0.062 \\
		\hline
			\multirow{6}{*}{(2)} 	& & DC & HHG  &  HSIC & Ball & $\KAcov_{1,\textrm{Laplacian}}$ \\
		\cline{2-7} 
		& $z$ & 0.048  & 0.056 & 0.044 & 0.052 & 0.054 \\
		& $t_3$  &0.038  &0.036 & 0.050 & 0.046 & 0.038 \\
		\cline{2-7} 
		&  & $\KAcov_{1,\textrm{distance}}$ & $\KAcov_{2,\textrm{Laplacian}}$  &  $\KAcov_{2,\textrm{distance}}$ & $\KAcov_{3,\textrm{Laplacian}}$ & $\KAcov_{3,\textrm{distance}}$ \\
		\cline{2-7} 
		& $z$ & 0.044 & 0.062 & 0.052 & 0.050 & 0.054 \\
		& $t_3$ & 0.046 & 0.044 & 0.052 & 0.040 & 0.046 \\
		\hline
			\multirow{6}{*}{(3)} & & DC & HHG  &  HSIC & Ball & $\KAcov_{1,\textrm{Laplacian}}$ \\
		\cline{2-7} 
		& $z$ & 0.046 &0.052& 0.040 &0.050& 0.064 \\
		& $t_3$ &  0.046 &0.050& 0.044 &0.054 &0.068\\
		\cline{2-7} 
		&  & $\KAcov_{1,\textrm{distance}}$ & $\KAcov_{2,\textrm{Laplacian}}$  &  $\KAcov_{2,\textrm{distance}}$ & $\KAcov_{3,\textrm{Laplacian}}$ & $\KAcov_{3,\textrm{distance}}$ \\
		\cline{2-7} 
		& $z$ & 0.054 &0.064 &0.054& 0.064& 0.048 \\
		& $t_3$ & 0.056 &0.068 &0.060& 0.066& 0.060\\
		\hline
		\hline
	\end{tabular}}
\end{table}

Table \ref{table:sizevector} confirms that all methods control the type I error very well at significance level $\alpha = 0.05$.
Figure \ref{fig:powervector} illustrates that  if we use the same kernel for $\KAcov_{m}(\x, \y), m=1,2,3$, the corresponding tests based on them have similar performance. 
We also find that our tests using distance kernel  outperform other tests in the linear dependent case, and our tests using Laplacian kernel outperform others in log and quadratic dependent cases.  
The result indicates that different kernels are suitable for detecting different dependent structures.

\begin{figure}[htbp!]
	%\addtocounter{subfigure}{0}
	\captionsetup{font = footnotesize}
	\centering
	\subfigure[Circle $z$]{
		\begin{minipage}[t]{0.45\linewidth}
			\centering
			\includegraphics[width=2.6in]{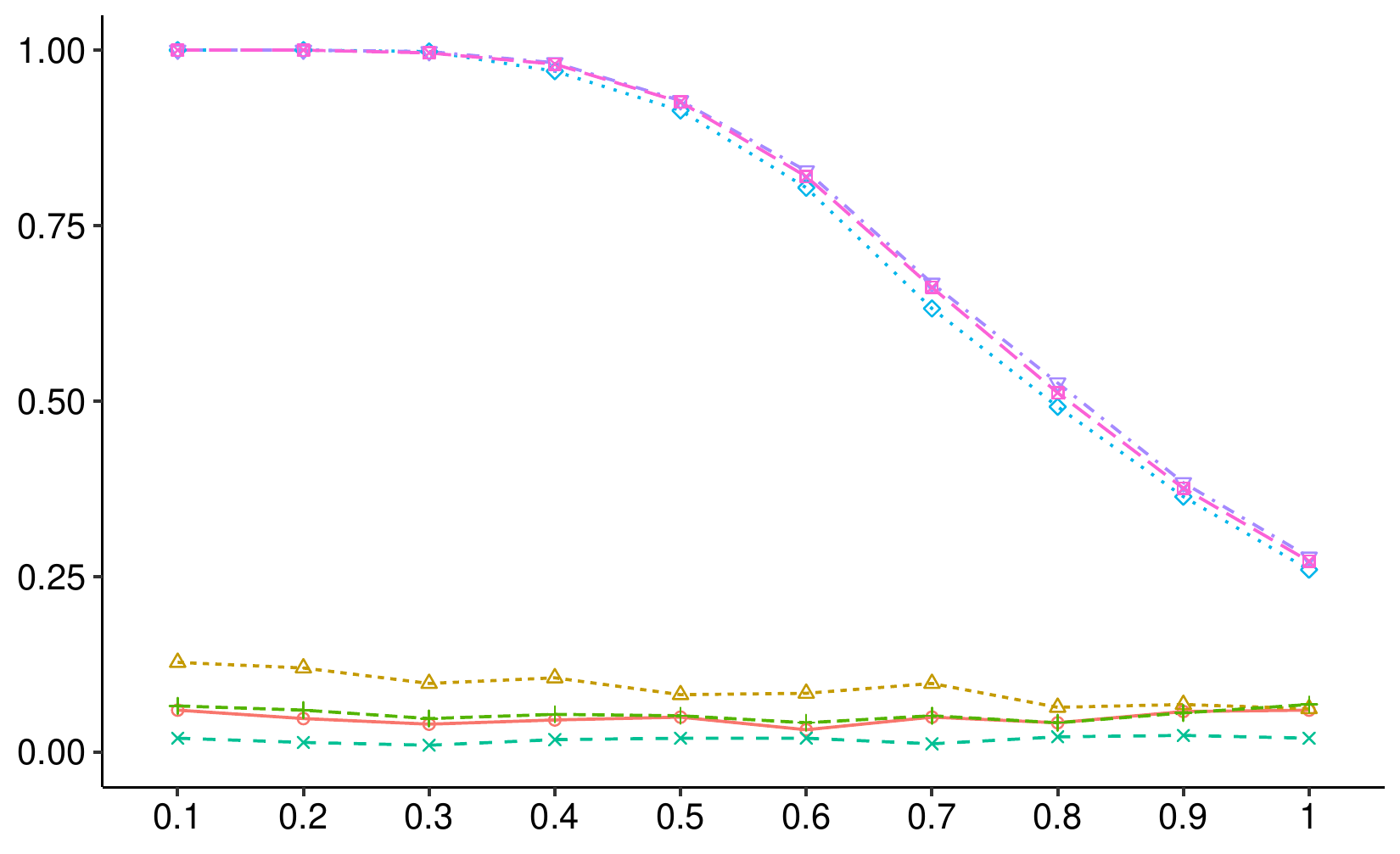}
			%\caption{fig1}
		\end{minipage}%
	}%
	\subfigure[Circle $t_{3}$]{
		\begin{minipage}[t]{0.45\linewidth}
			\centering
			\includegraphics[width=2.6in]{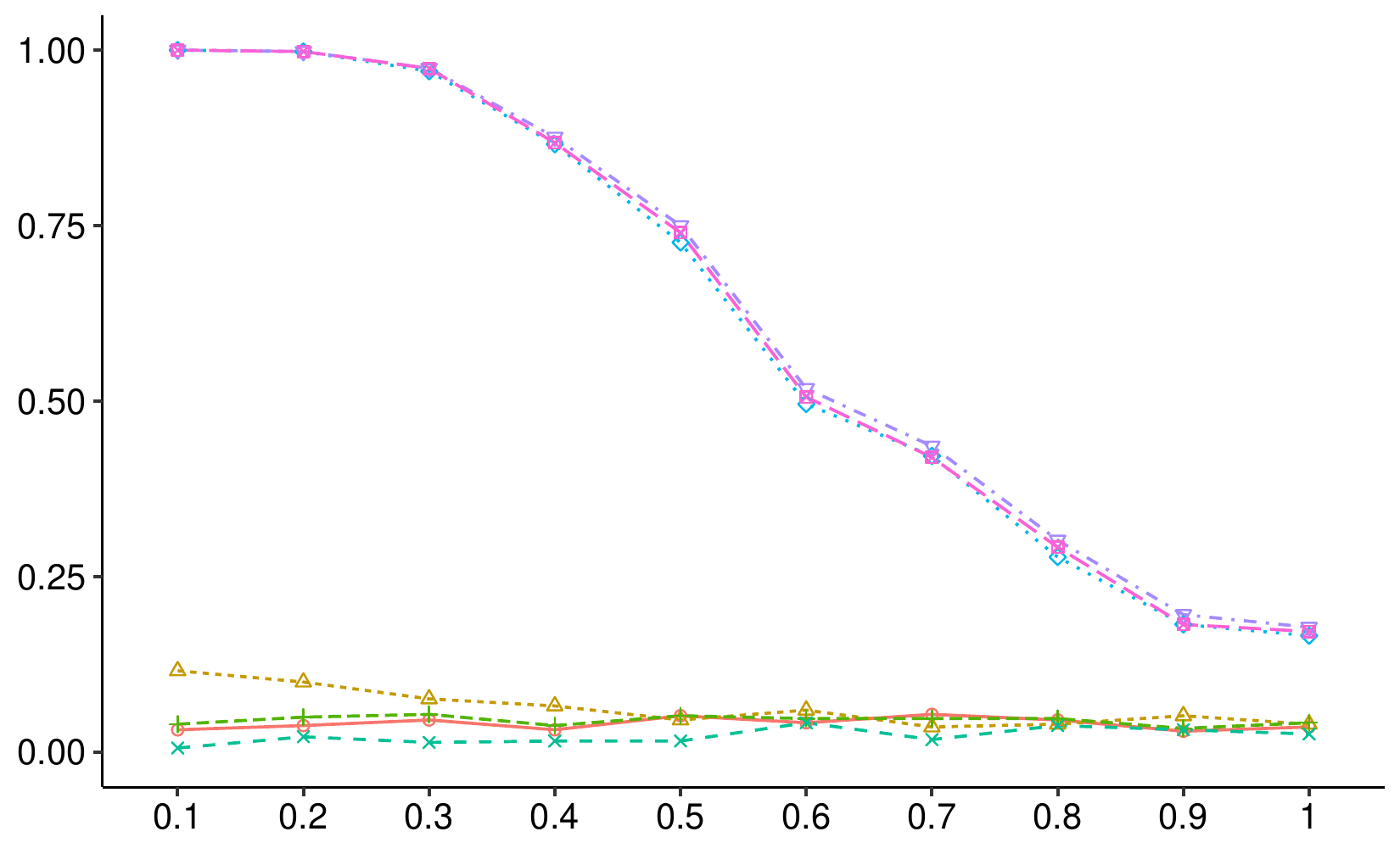}
			%\caption{fig1}
		\end{minipage}%
	}%
	
	\subfigure[Two Parabola $z$]{
		\begin{minipage}[t]{0.45\linewidth}
			\centering
			\includegraphics[width=2.6in]{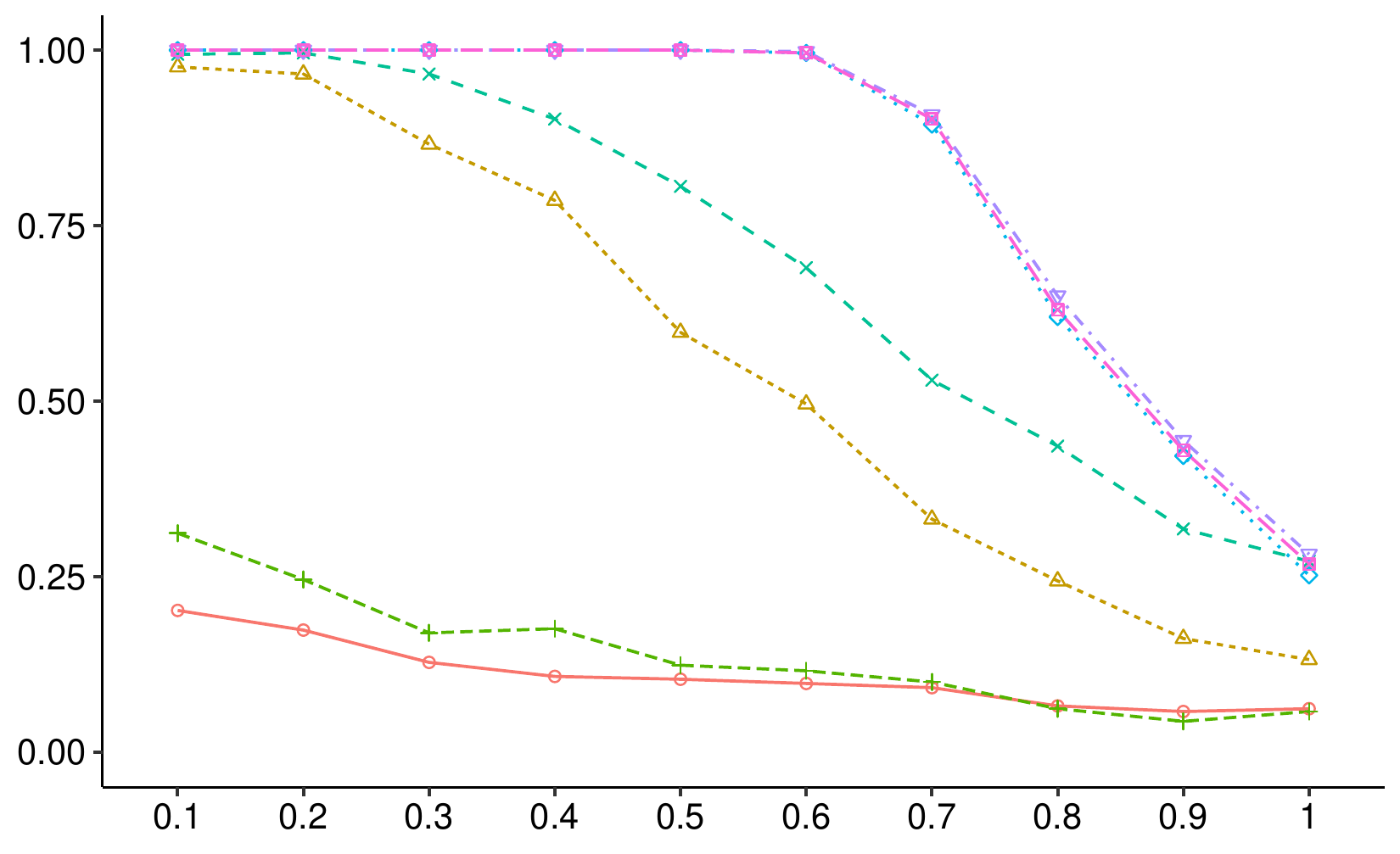}
			%\caption{fig1}
		\end{minipage}%
	}%
	\subfigure[Two Parabola $t_{3}$]{
		\begin{minipage}[t]{0.45\linewidth}
			\centering
			\includegraphics[width=2.6in]{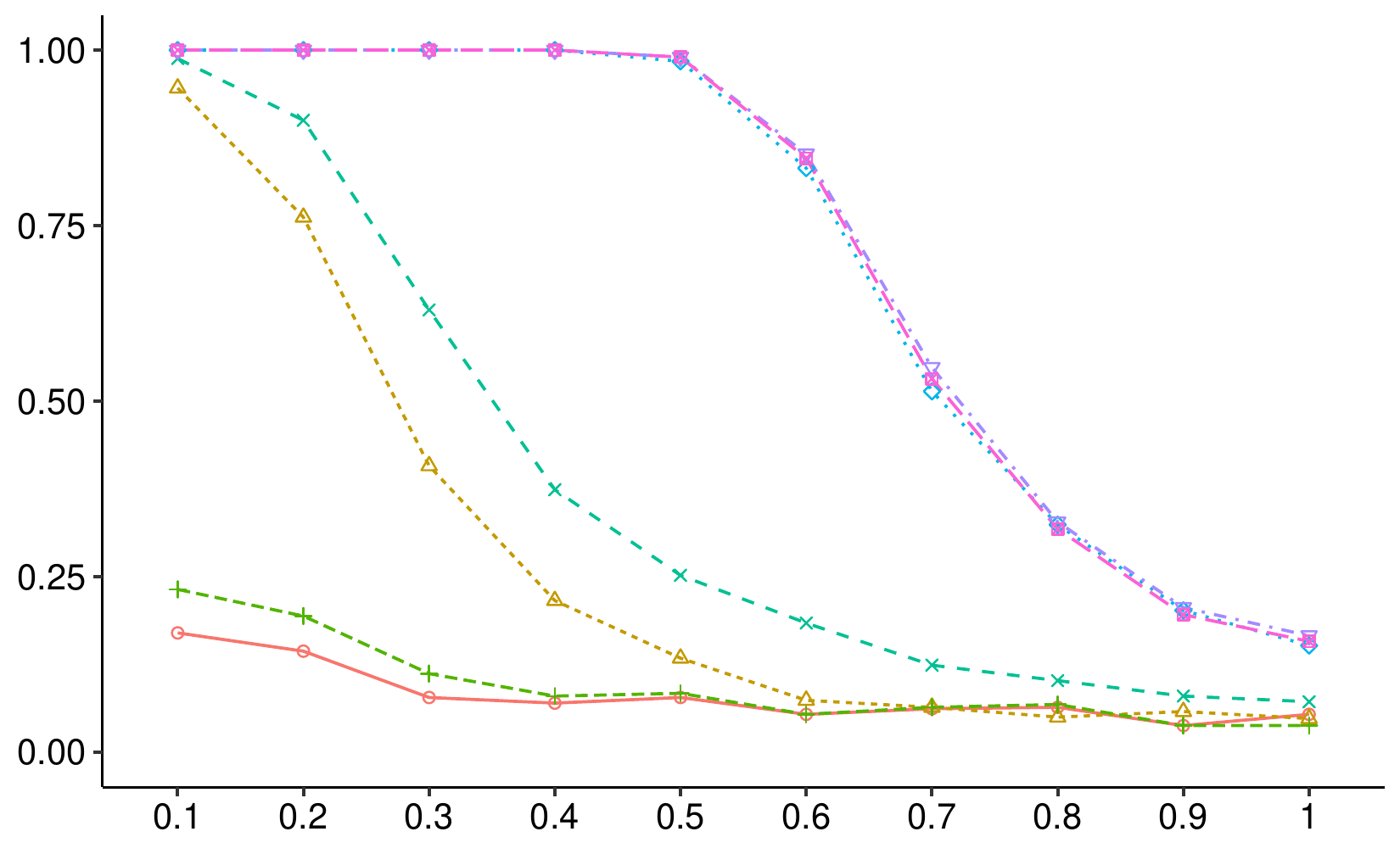}
			%\caption{fig1}
		\end{minipage}%
	}%
	
	\subfigure[Sinusoidal $z$]{
		\begin{minipage}[t]{0.45\linewidth}
			\centering
			\includegraphics[width=2.6in]{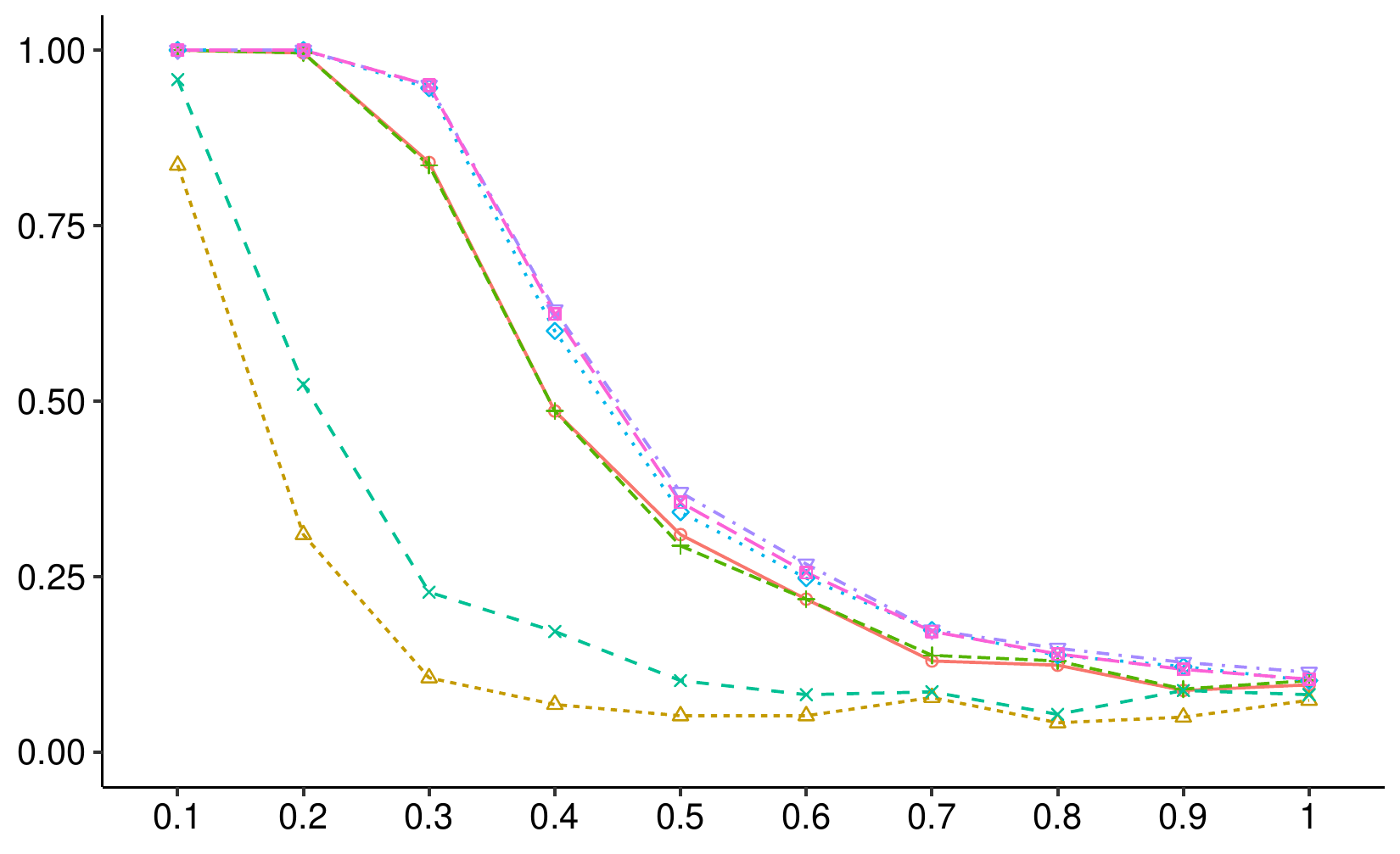}
			%\caption{fig2}
		\end{minipage}%
	}%
	\subfigure[Sinusoidal $t_{3}$]{
		\begin{minipage}[t]{0.45\linewidth}
			\centering
			\includegraphics[width=2.6in]{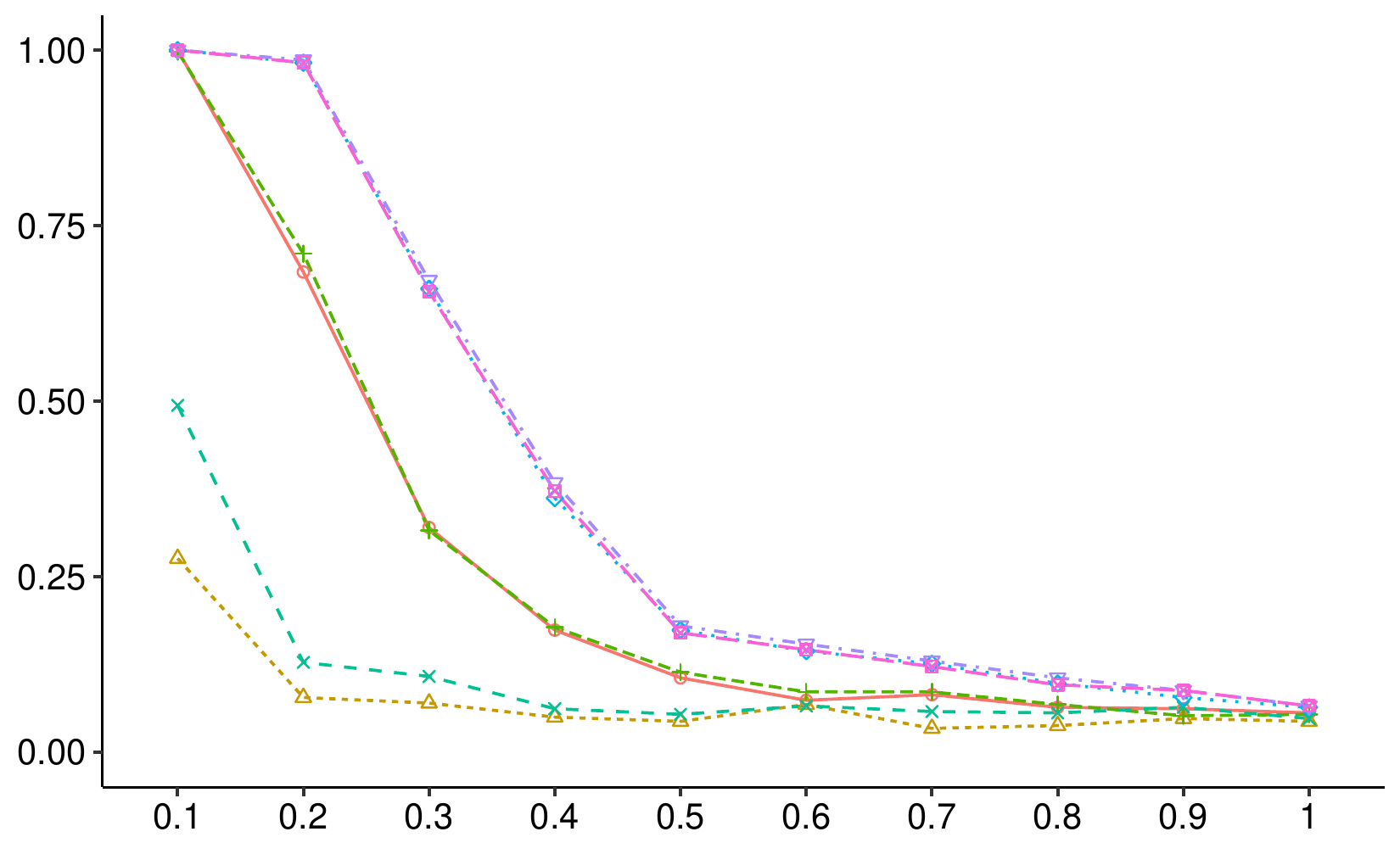}
			%\caption{fig2}
		\end{minipage}%
	}%
	\centering
	\caption{The empirical power of tests under three scenarios with noise of standard normal distribution ($z$) and $t$ distribution with 3 degrees of freedom ($t_3$). The horizontal axis represents $\lambda$ which controls the size of noise and the vertical axis represents the empirical power. We use different marks to represents different tests, distance correlation test: $\circ$;  the multivariate test of \cite{heller2013consistent}: $\triangle$; the Hilbert-Schmidt independence criterion with Gaussian kernel: $+$; Ball covariance test: $\times$; $\KAcov_1(\x,\y)$ using $L^1$ norm kernel: $\Diamond$; $\KAcov_2(\x,\y)$ using  $L^1$ norm kernel: $\bigtriangledown$;   $\KAcov_3(\x,\y)$ using $L^1$ norm kernel: $\boxtimes$.}
	\label{fig:powerhigh}
\end{figure}

\noindent\textbf{Study 2.}
We test dependence between high-dimensional random vectors $\x$ and $\y$, where $\x \defby (X_1, \ldots, X_{100})\trans$ and $\y \defby (Y_1,\ldots, Y_{100})\trans$.
We choose $L^1$-norm based kernel for both $\x$ and $\y$ in $\KAcov_m(\x,\y), m=1,2,3$, and consider the following three scenarios {\color{red}:} 
\begin{enumerate}
	\item[(4)] Circle: $Y_{j} =  1.5(1-X_{j}^2)^{1/2}Z_{j} +  \lambda\varepsilon_{j}$;  $\pr(Z_{j}=\pm1) = 1/2$, for $j = 1,\ldots,100$.
	\item[(5)] Two Parabola: $Y_{j} = X_{j}^2Z_{j} +  0.7\lambda\varepsilon_{j}$,  $\pr(Z_{j}=\pm1) = 1/2$, for $j = 1,\ldots,100$;
	\item[(6)] Sinusodial: $Y_{j} = \sin(4\pi X_{j})+  4\lambda\varepsilon_{j}$, for $j = 1,\ldots,100$.
\end{enumerate}
We generate $X_j, j=1,\ldots,100$ independently from uniform distribution on $[0,1]$, and  noise $\varepsilon_{j}, j=1,\ldots,100$ from the standard normal distribution and $t$ distribution with $3$ degrees of freedom.
Similarly to Study 1, we conduct the independent tests for $(\x_{1},\y_{2})$ generated from the above three scenarios. We report the empirical size in Table \ref{table:sizehigh} at significance level $\alpha=0.05$. 
To compare the power performances, we test dependence between $\x_{1}$ and $\y_{1}$, and depict the empirical powers over $\lambda\in\{0.1,\ldots,1\}$ in Figure \ref{fig:powerhigh}.

\begin{table}[htbp!]
	\captionsetup{font = footnotesize}
	\caption{\footnotesize The empirical sizes of tests under three scenarios with noises of standard normal distribution ($z$) and $t$ distribution with 3 degrees of freedom ($t_3$). The tests include the distance correlation test (DC), the multivariate test of \cite{heller2013consistent} (HHG), the Hilbert-Schmidt independence criterion with Gaussian kernel (HSIC), Ball covariance test (Ball), $\KAcov_{m}(\x,\y), m=1,2,3$ using  $L^1$-norm based kernel.}
	\label{table:sizehigh}    
	\centering
	\begin{tabular}{r|r|ccccccc}
		\hline
		\hline
	&	& DC & HHG  &  HSIC & Ball & $\KAcov_{1}$ &  $\KAcov_{2}$  &   $\KAcov_{3}$ \\
		\hline
		\multirow{2}{*}{(4)} & $z$ & 0.042 & 0.068 & 0.044 & 0.056& 0.044& 0.050& 0.044 \\
		& $t_3$ & 0.038& 0.034 & 0.040 & 0.046 & 0.052 & 0.052 & 0.050 \\
		\hline
		\multirow{2}{*}{(5)}  & $z$ & 0.032 & 0.056 & 0.046 & 0.068 & 0.046 &0.048 &0.044 \\
		& $t_3$ & 0.060 & 0.054 & 0.058 & 0.058 & 0.050 & 0.048 & 0.040 \\
		\hline
		\multirow{2}{*}{(6)}  & $z$ & 0.064 & 0.024 & 0.068 &  0.050 & 0.050 & 0.056 & 0.052\\
		& $t_3$     & 0.052 & 0.040 & 0.050 & 0.046 & 0.052 & 0.044 & 0.042 \\
		\hline
		\hline
	\end{tabular}
\end{table}

Table \ref{table:sizehigh} confirms that when data are high dimensional random vectors, all methods work well in controlling the empirical sizes. Figure \ref{fig:powerhigh}  illustrates the superiority of our tests with $L^1$ norm-based kernels in terms of power performances.
For example, other tests have empirical powers less than $0.2$ and the proposed tests $\KAcov_{m}(\x,\y), m=1,2,3$ have power close to $1$ in the circle-dependent scenario.
Additionally, the proposed tests have similar performance with different noises, which indicates that the proposed tests are robust to heavy-tailed noise.

\begin{figure}[htbp!]
	%\addtocounter{subfigure}{0}
	\captionsetup{font = footnotesize}
	\centering
	\subfigure[Matrix-matrix $z$]{
		\begin{minipage}[t]{0.45\linewidth}
			\centering
			\includegraphics[width=2.6in]{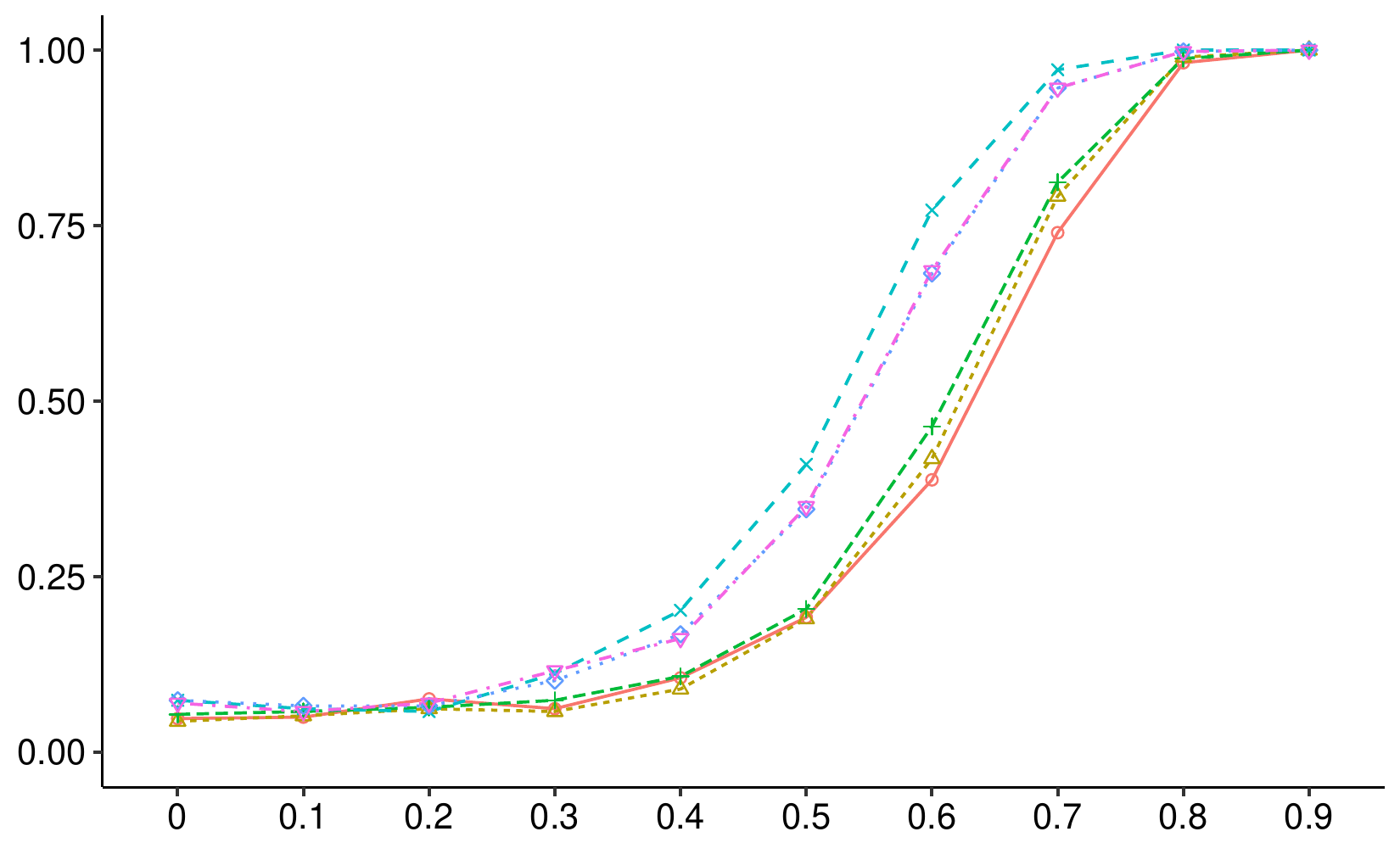}
			%\caption{fig1}
		\end{minipage}%
	}%
	\subfigure[Matrix-matrix $t_3$]{
		\begin{minipage}[t]{0.45\linewidth}
			\centering
			\includegraphics[width=2.6in]{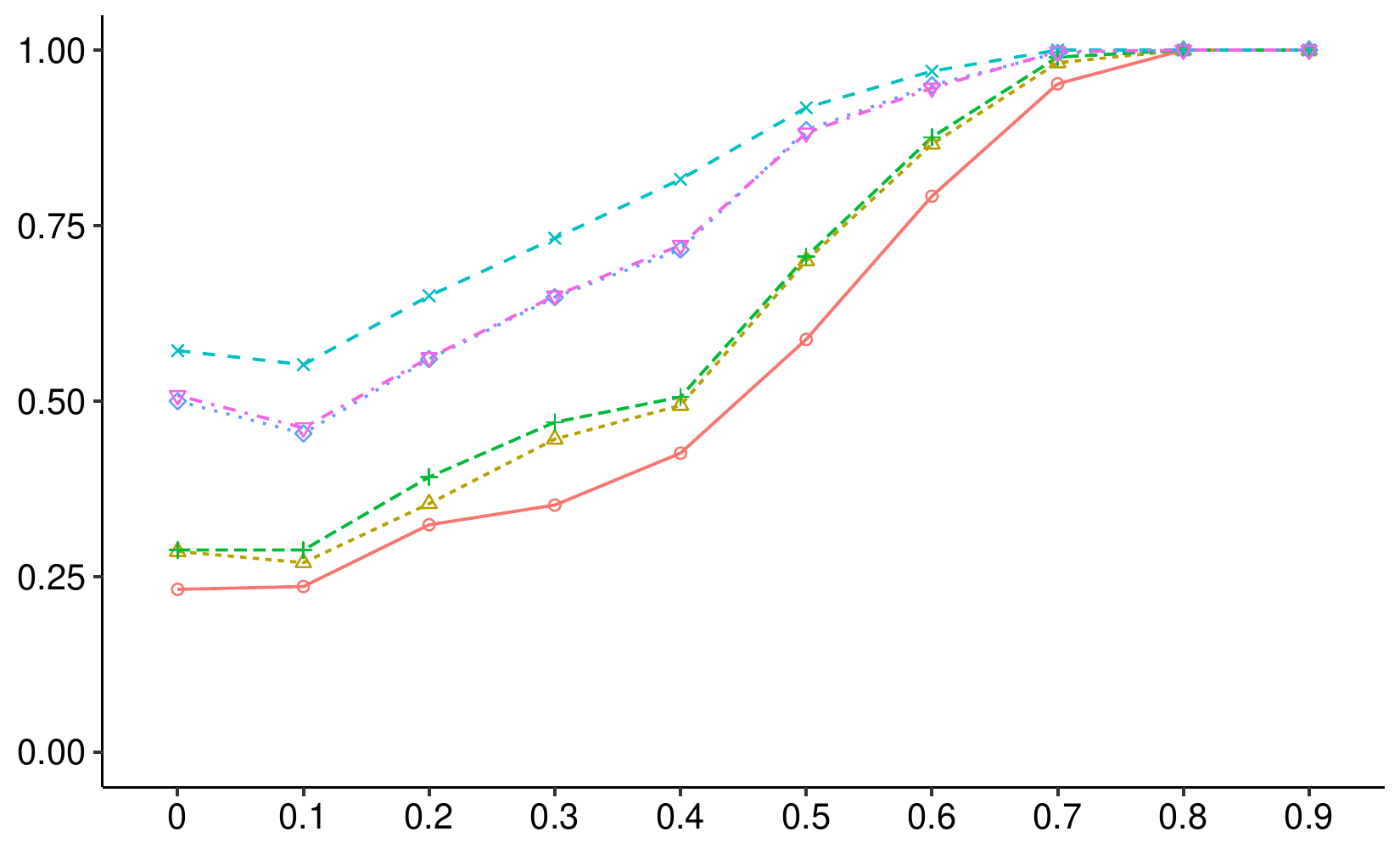}
			%\caption{fig1}
		\end{minipage}%
	}%
	
	\subfigure[Block matrix $z$]{
		\begin{minipage}[t]{0.45\linewidth}
			\centering
			\includegraphics[width=2.6in]{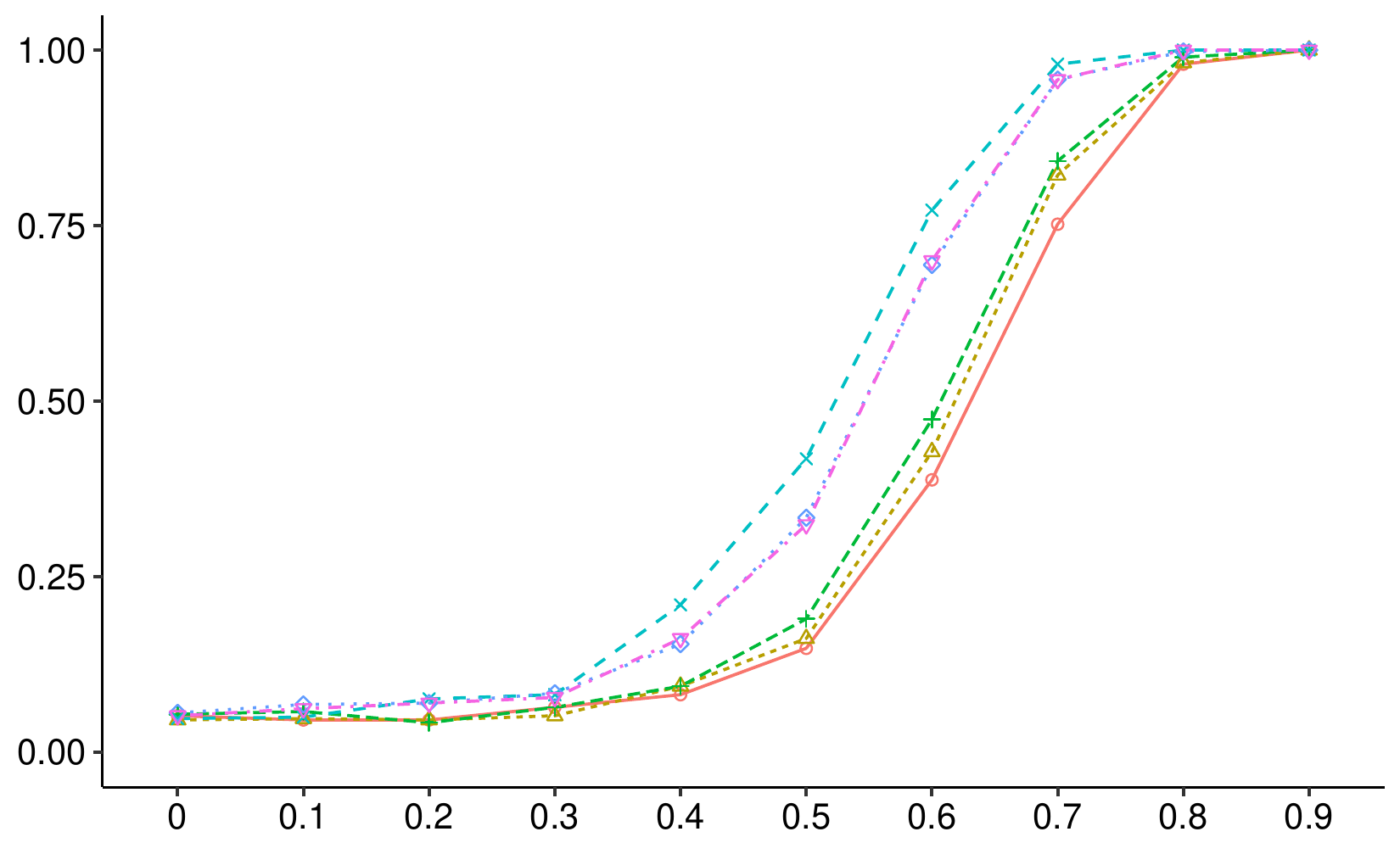}
			%\caption{fig1}
		\end{minipage}%
	}%
	\subfigure[Block matrix $t_3$]{
		\begin{minipage}[t]{0.45\linewidth}
			\centering
			\includegraphics[width=2.6in]{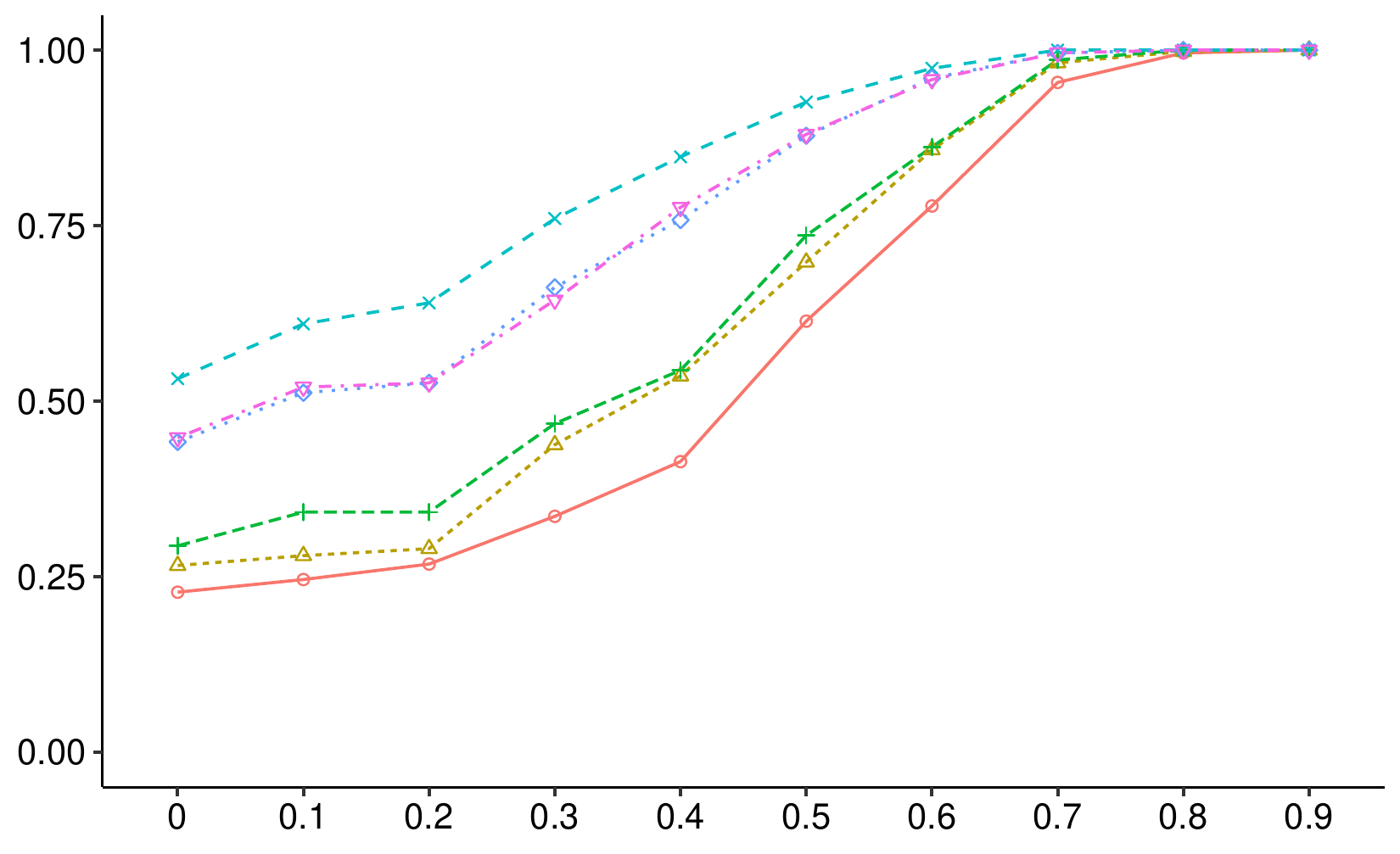}
			%\caption{fig1}
		\end{minipage}%
	}%
	
	\subfigure[Matrix-vector $z$]{
		\begin{minipage}[t]{0.45\linewidth}
			\centering
			\includegraphics[width=2.6in]{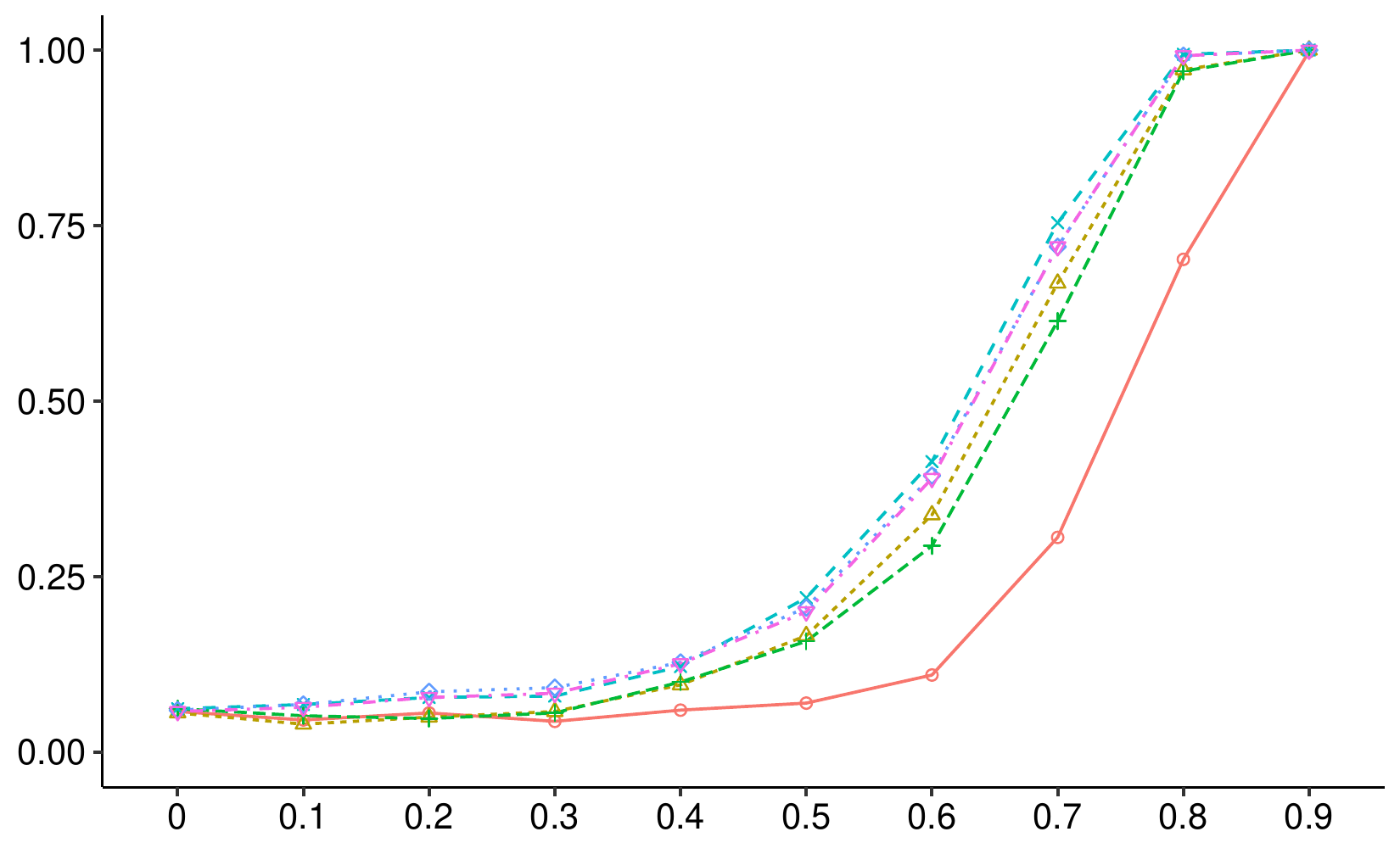}
			%\caption{fig2}
		\end{minipage}%
	}%
	\subfigure[Matrix-vector $t_3$]{
		\begin{minipage}[t]{0.45\linewidth}
			\centering
			\includegraphics[width=2.6in]{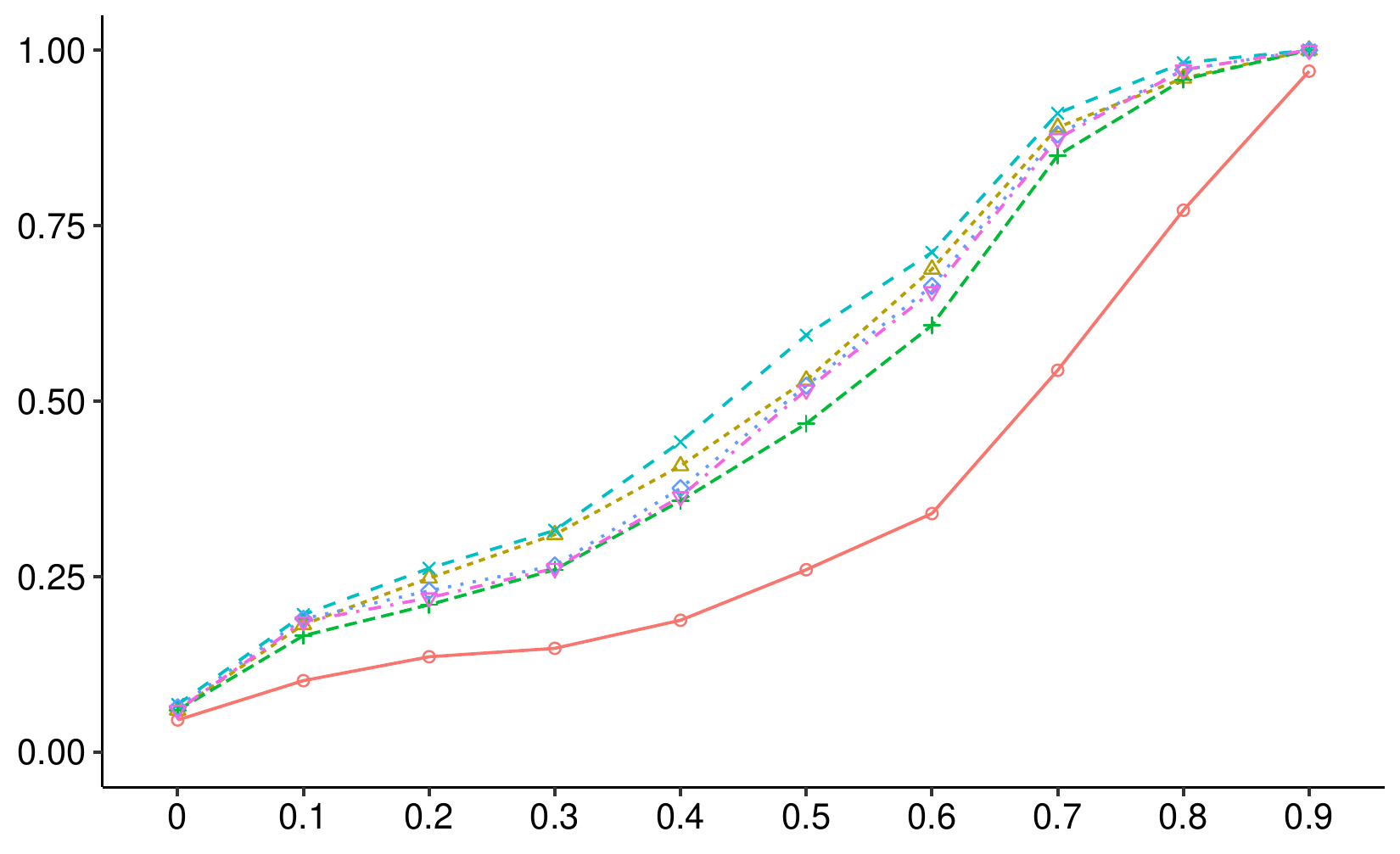}
			%\caption{fig2}
		\end{minipage}%
	}%
	\centering
	\caption{The empirical powers of tests under three scenarios with  $(Z_{1}, Z_{2})$ from standard normal distribution ($z$) and $t$ distribution with 3 degrees of freedom ($t_3$). The horizontal axis represents $\rho$ which controls the strength of dependence and the vertical axis represents the empirical power. We use different marks to represent different tests: the generalized distance correlation test: $\circ$;  the multivariate test of \cite{heller2013consistent}: $\triangle$; Ball covariance test: $+$; $\KAcov_1$:  $\times$; $\KAcov_2$: $\Diamond$;   $\KAcov_3$: $\bigtriangledown$.}
	\label{fig:powermatrix}
\end{figure}

\noindent\textbf{Study 3.} We implement dependence tests between random matrices.
We choose the log-Euclidean kernel for the symmetric positive definite matrix and the Laplacian kernel for univariate random variables in our proposed tests.
We compare the proposed tests with the generalized distance correlation test \citep{sejdinovic2013equivalence}, the multivariate test of \cite{heller2013consistent}, and the Ball covariance test \citep{pan2020ball}. Since these three methods are distance based, we can also calculate the distance between matrices based on the log-Euclidean kernel, where the distance between matrices $\Z_{1}$ and $\Z_{2}$ is $d(\Z_{1},\Z_{2})\defby \left\{K(\Z_{1}, \Z_{1}) + K(\Z_{2}, \Z_{2}) - 2K(\Z_{1}, \Z_{2})  \right\}^{1/2}$.
We consider the following three scenarios:
\begin{enumerate}
	\item[(7)] Matrix-matrix : $\X,\Y\in \mR^{3\times 3}$. $(\X)_{rl} = 1/(1+Z_{1}^2)$ and $(\Y)_{rl} = 1/(1+Z_{2}^2)$ for $r\neq l$, $(\X)_{rr} = 1$ and $(\Y)_{rr} = 1$  for $r=1,2,3$;
	\item[(8)] Block matrix: $\Z_1,\Z_2\in \mR^{2\times 2}$, where 
	$(\Z_1)_{11} = (\Z_1)_{22} = (\Z_2)_{11} = (\Z_2)_{22}=  1$, $(\Z_1)_{12} = (\Z_1)_{21} = 1/(1+Z_{1}^2)$ and $(\Z_2)_{12} = (\Z_2)_{21} = 1/(1+Z_{2}^2)$. $\X = diag(\Z_1, \Z_1)$ and $\Y = diag(\Z_2, \Z_2)$;
	\item[(9)] Matrix-vector: $\X\in \mR^{3\times 3}$, $Y\in\mR$. $(\X)_{rl} = 1/(1+Z_{1}^2)$ for $r\neq l$, $(\X)_{rr} = 1$ for $r=1,2,3$. $Y = (Z_{2}-2)^2$.
\end{enumerate}
We generate $Z_1$ and $Z_2$ from two the  binomial normal distribution and the binomial $t$ distribution with $3$ degrees of freedom. Both binomial normal distribution and binomial $t$ distribution have zero mean and covariance matrix or scale matrix as $\bSig$, where $(\bSig)_{11} = (\bSig)_{22} =1 $ and $(\bSig)_{12} = (\bSig)_{21} = \rho $.
Similar to previous studies, we construct the independent sample and report the empirical size in Table \ref{table:sizematrix} at significance level $\alpha=0.05$. We  report the empirical powers over $\rho\in\{0,0.1,\ldots,0.9\}$ in Figure \ref{fig:powermatrix}.

\begin{table}[htbp!]
	\captionsetup{font = footnotesize}
	\caption{\footnotesize The empirical sizes of tests under three scenarios with  $(Z_{1}, Z_{2})$ from standard normal distribution ($z$) and $t$ distribution with 3 degrees of freedom ($t_3$). The tests include the generalized distance correlation test (GDC), the multivariate test of \cite{heller2013consistent} (HHG), the Ball covariance test (Ball), $\KAcov_{m}, m=1,2,3$. }
	\label{table:sizematrix}    
	\centering
	\begin{tabular}{r|r|cccccc}
		\hline
		\hline
		& & GDC & HHG  & Ball & $\KAcov_{1}$ &  $\KAcov_{2}$  &   $\KAcov_{3}$ \\
		\hline
		\multirow{2}{*}{(7)} & $z$ & 0.048 & 0.038 &0.056  & 0.068 & 0.068  & 0.064 \\
		& $t_3$ & 0.044 & 0.044 &0.050  & 0.042 & 0.050 & 0.052 \\
		\hline
		\multirow{2}{*}{(8)}  & $z$ & 0.044 & 0.052 & 0.044 & 0.058 & 0.056 & 0.060 \\
		& $t_3$ & 0.056 & 0.044 & 0.060 & 0.056  & 0.068 & 0.056 \\
		\hline
		\multirow{2}{*}{(9)}  & $z$ & 0.034 & 0.044 & 0.042 & 0.050 & 0.054 &0.054  \\
		& $t_3$ & 0.050 &0.058 &0.050& 0.050 & 0.050 & 0.048 \\
		\hline
		\hline
	\end{tabular}
\end{table}

Table \ref{table:sizematrix} confirms that all methods can control the empirical sizes of tests when data are random matrices. Figure \ref{fig:powermatrix}  illustrates our methods have better performance in detecting the dependence structures than other existing methods.

\begin{figure}[htbp!]
	%\addtocounter{subfigure}{0}
	\captionsetup{font = footnotesize}
	\centering
	\subfigure[]{
		\begin{minipage}[t]{0.24\linewidth}
			\centering
			\includegraphics[width=1.5in]{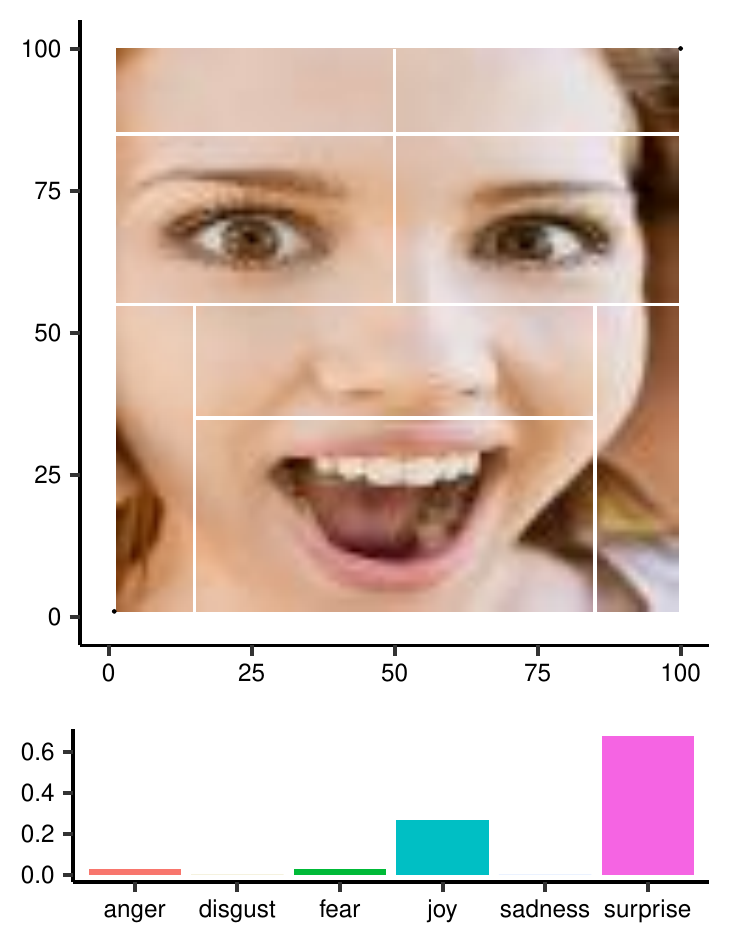}
			%\caption{fig1}
		\end{minipage}%
	}%
	\subfigure[]{
		\begin{minipage}[t]{0.24\linewidth}
			\centering
			\includegraphics[width=1.5in]{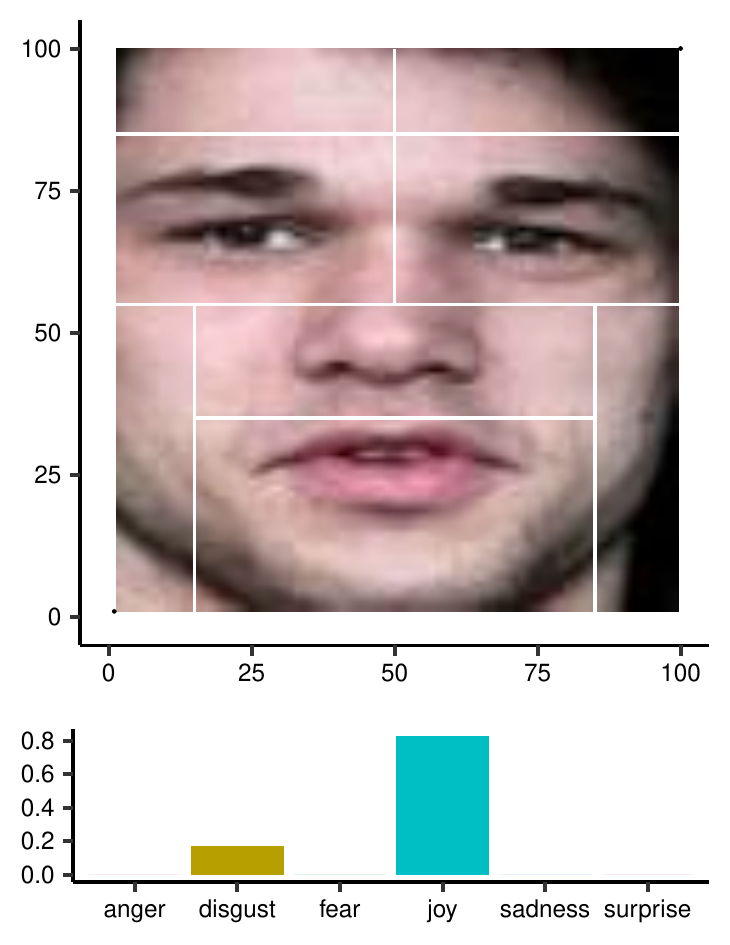}
			%\caption{fig1}
		\end{minipage}%
	}%
 	\subfigure[]{
		\begin{minipage}[t]{0.24\linewidth}
			\centering
			\includegraphics[width=1.5in]{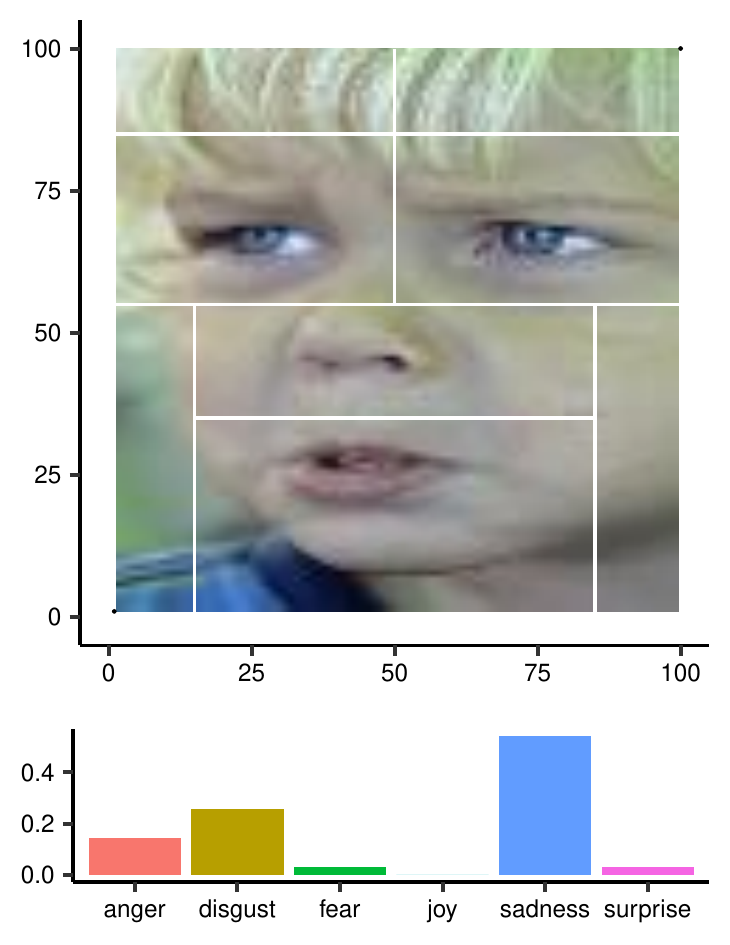}
			%\caption{fig1}
		\end{minipage}%
	}%
 	\subfigure[]{
		\begin{minipage}[t]{0.24\linewidth}
			\centering
			\includegraphics[width=1.5in]{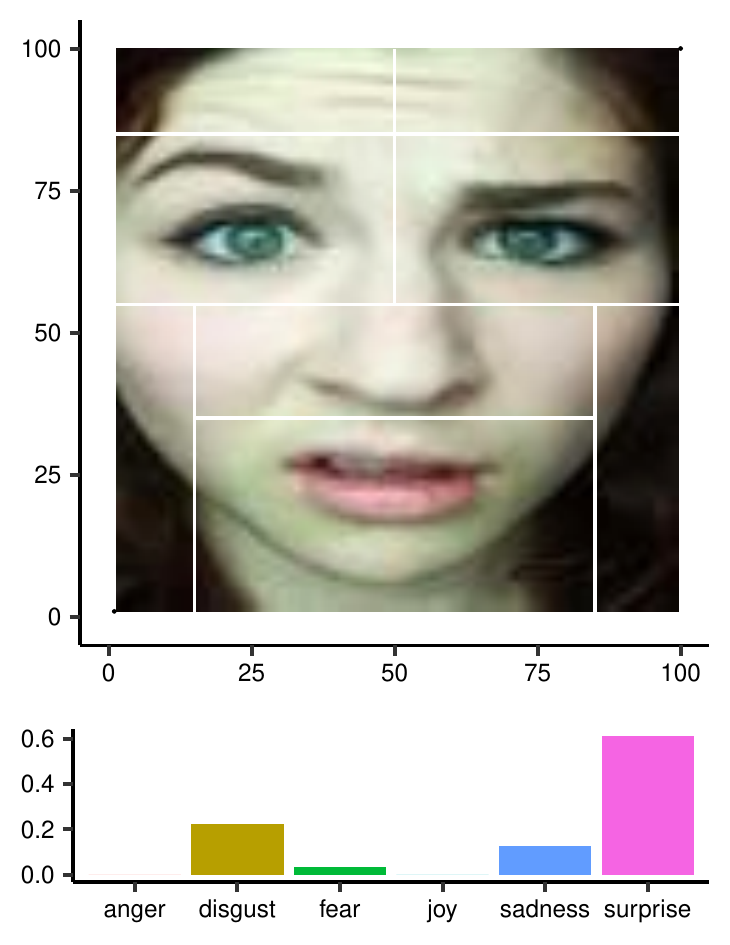}
			%\caption{fig1}
		\end{minipage}%
	}%
	\centering
	\caption{Segmented facial images with probabilities of $6$ emotions. The RGB images are of size $100\times 100$. Lower-left coordinates and top-right coordinates of $8$ areas are, left head: $(0,85),(50,100)$; right head: $(50,85),(100,100)$; left eye: $(0, 55),(50,85)$; right eye: $(50,55),(100,85)$; nose: $(15,35),(85,55)$; mouth: $(15,0),(85,35)$; left cheek: $(0,0),(15,55)$; right cheek: $(85,0),(100,55)$. Six emotions are anger, disgust, fear, joy, sadness, and surprise.
       }
	\label{fig:faces}
\end{figure}

\csection{Real Data Analysis}
\csubsection{Facial Expression Data}
Recognizing emotion through facial expressions has attracted broad attention due to its numerous potential applications in human-computer interaction \citep{rosenberg2020face}.
In recognition tasks, it is of great help to screen out unimportant areas in a large image to build features for further classification or prediction.
To address this issue, we apply our proposed methods to test the dependence between facial areas and emotions.
We choose the dataset, Realworld affective face multi-label (RAF-ML), which is collected by \cite{li2019blended}.
This dataset contains $4908$ aligned facial images, and every image is assigned by a probability vector of six emotions including anger, disgust, fear, joy, sadness, and surprise.
We segment each image into $8$ areas including left head, right head, left eye, right eye, nose, mouth, left cheek, and right cheek. The segmented facial images with probability vectors of emotions are displayed in Figure \ref{fig:faces}.

To construct symmetric semi-definite positive matrices to describe the facial images, we follow the commonly used procedures in face recognition \citep{wang2018low}.
Firstly, we convert the RGB image to the gray image by averaging  the RGB values. Then we calculate the covariance matrices of the feature vector $(I_{(u,v)}, u, v, G_{(u,v)}^{0,0},\ldots,$\\$ G_{(u,v)}^{4,7})$ in each subfigure, where $u,v$ are the coordinates within the subfigure, $I_{(u,v)}$ represents the average illumination values at position $(u,v)$, and $G_{(u,v)}^{0,0},\ldots, G_{(u,v)}^{4,7}$ are the 2D Gabor wavelet features at position $(u,v)$.
The orientation and direction of 2D Gabor wavelet transformation are $5$ and $8$, respectively.
We  randomly select $n=200$ images and conduct independence tests between $8$ covariance matrices of different facial areas and the vector of emotions.
We compare $6$ methods that can be applied to matrix-valued data. They are the generalized distance correlation test \citep{sejdinovic2013equivalence}, the multivariate test of \cite{heller2013consistent},  the Ball covariance test \citep{pan2020ball} and our methods using $\KAcov_m$, $m=1,2,3$. The distances between matrices are calculated using the log-Euclidean kernel, and the distances between vectors are calculated using distance kernels or Euclidean distances. 
We report $p$-values of these independence tests in Table \ref{table:realdata1}.

\begin{table}[htbp!]
	\captionsetup{font = footnotesize}
	\caption{\footnotesize $p$-values of independence tests between covariances matrices of $8$ facial areas and the vector of emotions. The tests include the generalized distance correlation test (GDC), the multivariate test of \cite{heller2013consistent} (HHG), the Ball covariance test (Ball), $\KAcov_{m}, m=1,2,3$. }
	\label{table:realdata1}    
	\centering
	\begin{tabular}{r|cccccc}
		\hline
		\hline
		& GDC & HHG  & Ball & $\KAcov_{1}$ &  $\KAcov_{2}$  &   $\KAcov_{3}$ \\
		\hline
	  Left head & 0.370  & 0.406     & 0.275   & 0.276 & 0.401  & 0.394 \\
		Right head & 0.730 & 0.770      & 0.785  & 0.701 & 0.753 & 0.748 \\
	Left eye & 0.010   & 0.005      & 0.005 & $<$0.001 & $<$0.001 &  $<$0.001\\
		Right eye & 0.015  & 0.020      & 0.060 & $<$0.001  & 0.011 & 0.012 \\
		Nose & 0.005       & 0.014      & 0.010 & $<$0.001 &  $<$0.001 & $<$0.001  \\
	  Mouth & 0.010      & 0.005      & 0.005 & $<$0.001 & 0.002 & 0.002 \\
   	Left cheek & 0.660 & 0.480      & 0.410 & 0.365 & 0.360 & 0.369  \\
	  Right cheek & 0.170 & 0.009     & 0.025 & 0.002 & 0.030 & 0.031 \\
		\hline
		\hline
	\end{tabular}
\end{table}
Table \ref{table:realdata1} demonstrates left eye, right eye, nose, and mouth are the four most related areas with emotions. In addition, $5$ of these $6$ methods also detect the dependence between the right cheek and emotions. 
This may be due to some faces turning to the left, e.g., (c) in Figure \ref{fig:faces}. By contrast, the generalized distance correlation test fails to detect such dependence.
These dependent areas can serve as  important features for further prediction or classification.

\csubsection{ Microbial Communities Data}

The microbial communities in the human intestinal have significant impacts on human health, and their states are associated with a series of host factors.
In this study, we  test the dependence between  age and intestinal microbiota by our proposed methods. 
The data are collected by \cite{lahti2014tipping} and can be downloaded from \url{https://datadryad.org/stash/dataset/doi:10.5061/dryad.pk75d}.
The dataset includes 130 genus-like phylogenetic groups that cover the majority of the known bacterial diversity of the human intestine. 
The Absolute Human Intestinal Tract Chip (HITChip) signal estimates of these 130 phylogenetic groups are provided.
The sample includes $1006$ adults in 15 countries, and the age, sex, and BMI groups of the subjects are also included in the dataset.

We remove the empty values in the dataset. 
According to BMI and Sex, we divide the data into eight groups, which are lean male ($n_{1}=189$), lean female ($n_{2}=304$), overweight male ($n_{3}=102$), overweight female ($n_{4}=102$), obese male ($n_{5}=91$), obese female ($n_{6}=133$), severeobese male ($n_{7}=30$), and severeobese female ($n_{8}=70$). 
In each group we test dependence between  age ($X$) and 130 phylogenetic groups ($\y$).  
We compare five independence tests including the distance correlation test \citep{szekely2007measuring}, the Ball covariance test \citep{pan2020ball}, Hilbert-Schmidt information criterion
\citep{gretton2007kernel} using Gaussian kernel, the multivariate test of \citep{heller2013consistent}
and our methods  $\KAcov_m(X, \y)$, $m=1,2,3$ with the distance kernel for $X$ and $L^1$-norm based kernel for $\y$.

In Table \ref{table:readdata2}, we report the $p$-values for these tests for eight groups. 
From the table, we can see that the $p$-values of our tests using $\KAcov_m(X, \y)$, $m=1,2,3$ are less than $0.1$, which indicates the strong dependence between age and phylogenetic groups.
This result is the same as \cite{zhang2020distance}.
By contrast, the $p$-values of the other four methods are larger than $0.2$ in overweight male group and severeobese male group, which shows that they  fail to detect dependence in these groups.

\begin{table}[htbp!]
	\captionsetup{font = footnotesize}
	\caption{\footnotesize The $p$-values of tests in eight groups. The tests include the distance correlation test (DC), the multivariate test of \cite{heller2013consistent} (HHG), the Hilbert-Schmidt independence criterion using Gaussian kernel (HSIC), Ball covariance test (Ball), $\KAcov_{m}(X,\y), m=1,2,3$ using distance kernel for $X$ and $L^1$-norm based kernel for $\y$.}
	\label{table:readdata2}    
	\centering
	\begin{tabular}{r|r|ccccccc}
		\hline
		\hline
		&	& DC & HHG  &  HSIC & Ball & $\KAcov_{1}$ &  $\KAcov_{2}$  &   $\KAcov_{3}$ \\
		\hline
		\multirow{2}{*}{lean} & male & 0.005 & 0.004 & 0.005 & 0.005 & $<$0.001& $<$0.001& $<$0.001 \\
		& female & 0.005 & 0.004 & 0.005 & 0.005 & $<$0.001& $<$0.001& $<$0.001 \\
		\hline
		\multirow{2}{*}{overweight}  & male & 0.225 & 0.210 & 0.335 & 0.225 & 0.002& $<$0.001& $<$0.001  \\
		& female & 0.020 & 0.156 & 0.010 & 0.155 &  $<$0.001& $<$0.001& $<$0.001  \\
		\hline
		\multirow{2}{*}{obese}  & male & 0.060 & 0.074 & 0.095 & 0.190 & $<$0.001 &0.073&0.073 \\
		& female & 0.030 & 0.121 & 0.035 & 0.080 &  $<$0.001& $<$0.001& $<$0.001  \\
		\hline
		\multirow{2}{*}{severeobese}  & male & 0.225 & 0.296 & 0.380 & 0.370 &  $<$0.001& $<$0.001& $<$0.001   \\
		& female & 0.055 & 0.015 & 0.005 & 0.010 &  $<$0.001& $<$0.001& $<$0.001  \\
		\hline
		\hline
	\end{tabular}
\end{table}

%%%%%%%%%%%%%%%%%%%%%%%%%%%%%%%%%%%%%%%%%%%%%%%%%%%%%%%%%%%
%%%%%%%%%%%%%%%%%%%%%%%%%%%%%%%%%%%%%%%%%%%%%%%%%%%%%%%%%%%
%
%\csection{EXTENSIONS USING ANGLE BASED DISTANCE}
%
%\csubsection{Angle based divergence
%
%mann-witney wilcoxon
%
%\cite{kim2020robust}
%
%\cite{li2020projective}

\csection{Extension: Connection with  Generalized Distance Covariance}
In this section, we provide a new integral derivation for 
generalized distance covariance \citep{sejdinovic2013equivalence}  by applying \eqref{equation:integration}.

{\defi{\label{definition:semimetric}} The function $\rho:\calZ\times\calZ\to[0,\infty)$ is called semimetric on $\calZ$, if it satisfies
\begin{itemize}
    \item[(i)] $\rho(z_{1},z_{2}) = \rho(z_{2},z_{1})$,
    \item[(ii)]  $\rho(z_{1},z_{2}) = 0$ if and only if $z_{1}=z_{2}$,   
\end{itemize}
for any $z_{1},z_{2}\in \calZ$. $(\calZ,\rho)$ is a semimetric space. 
In addition, the semimetric space $(\calZ, \rho)$ is of negative type, if for any $z_1,\ldots,z_{n}\in \calZ$ and $\alpha_1,\ldots,\alpha_{n}\in \calZ$, it satisfies
\beqrs
\sum_{i=1}^{n}\sum_{j=1}^{n}\alpha_{i}\alpha_{j}\rho(z_{i},z_{j})\leq 0, \textrm{ where } \sum_{j=1}^{n}\alpha_{j} = 0.
\eeqrs
}
Suppose  $\wt\rho_{1}:\calX\times\calX\to [0,+\infty)$ and $\wt\rho_{2}:\calY\times\calY\to [0,+\infty)$ are semimetric of negative type,
\cite{sejdinovic2013equivalence} put forward generalized distance correlation as 
$E\{\wt\rho_{1}(X_{1}, X_{2})^{q}\wt\rho_{2}(Y_{1}, Y_{2})^{q} - 
2 \wt\rho_{1}(X_{1}, X_{2})^{q}\wt\rho_{2}(Y_{1}, Y_{3})^{q} + 
\wt\rho_{1}(X_{1}, X_{2})^{q}\wt\rho_{2}(Y_{3}, Y_{4})^{q}\}$ with $0<q\leq 1$,.
$\wt\rho:\calZ\times\calZ\to [0,+\infty)$ is generated by $K:\calZ\times\calZ\to \mR$, if  $\wt\rho(Z_{1},Z_{2}) = K(Z_{1},Z_{1}) + K(Z_{2},Z_{2}) -2 K(Z_{1},Z_{2})$.
When $q = 1/2$, $\wt\rho_{1}$ and $\wt\rho_{2}$ are generated by $K_1$ and $K_{2}$ respectively. Then the generalized distance covariance is 
\beqrs
\GDcov(X,Y)\defby  E\{\rho_{1}(X_{1}, X_{2})\rho_{2}(Y_{1}, Y_{2}) - 2 \rho_{1}(X_{1}, X_{2})\rho_{2}(Y_{1}, Y_{3}) + \rho_{1}(X_{1}, X_{2})\rho_{2}(Y_{3}, Y_{4})\},
\eeqrs
where  $\rho_{1}(X_{1}, X_{2})\defby\|\phi_{1}(X_{1})-\phi_{1}(X_{2})\|_{\calH_1} = \{K_{1}(X_{1}, X_{1}) - 2K_{1}(X_{1}, X_{2}) + K_{1}(X_{2}, X_{2})\}^{1/2}$ and  $\rho_{2}(Y_{1},Y_{2})\defby\|\phi_{2}(Y_{1})-\phi_{2}(Y_{2})\|_{\calH_2}= \{K_{2}(Y_{1},Y_{1}) - 2K_{2}(Y_{1}, Y_{2}) + K_{2}(Y_{2}, Y_{2})\}^{1/2}$.
{\theo{\label{theorem:gdc_correlation}} 
Suppose the weight function $d\omega_{4}(u,v) \defby du\times v$ and the scale parameter $c=(\pi/32)^{1/2}$.
\beqrs
\GDcov(X,Y) =  c\int_{\calH_1}\int_{\calH_2}\int_{\mR^2}\cov^{2}\{1( f(X)\leq u), 1(g(Y)\leq v)\}
d \omega_{4}(u,v) \mu_1(df) \mu_2(dg), 
\eeqrs
}
Theorem \ref{theorem:gdc_correlation} provides the integration form for generalized distance covariance in the following theorem.

$\KAcov_{1}(X,Y)$ and  $\KAcov_{3}(X,Y)$ can also be seen as members of generalized distance covariance.
By defining $\rho_{angle,1}^{\prime}(x_1,x_{2}) \defby \ang_{1}^{\prime}(x_1, x_2)$, $\rho_{angle,2}^{\prime}(y_1,y_{2}) \defby \ang_{2}^{\prime}(y_1, y_2)$, $\rho_{angle,1}(x_1,x_{2}) \defby E\{\ang_{1}(x_1, x_2; X_{3})\}$ and $\rho_{angle,2}(y_1,y_{2}) \defby E\{\ang_{2}(y_1, y_2; Y_{3})\}$, we can represent 
$\KAcov_{1}(X,Y) = E\{\rho_{angle,1}^{\prime}(X_{1}, X_{2})\rho_{angle,2}^{\prime}(Y_{1}, Y_{2}) - 2 \rho_{angle,1}^{\prime}(X_{1}, X_{2})$\\$\rho_{angle,2}^{\prime}(Y_{1}, Y_{3})+ \rho_{angle,1}^{\prime}(X_{1}, X_{2})\rho_{angle,2}^{\prime}(Y_{3}, Y_{4})\}$ and $\KAcov_{3}(X,Y) = E\{\rho_{angle,1}(X_{1}, X_{2})$\\$\rho_{angle,2}(Y_{1}, Y_{2}) - 2 \rho_{angle,1}(X_{1}, X_{2})\rho_{angle,2}(Y_{1}, Y_{3}) + \rho_{angle,1}(X_{1}, X_{2})\rho_{angle,2}(Y_{3}, Y_{4})\}$.

{\prop{\label{proposition:semimetric}} $(\calX, \rho_{angle,1})$, $(\calY, \rho_{angle,2})$, $(\calX, \rho_{angle,1}^{\prime})$ and $(\calY, \rho_{angle,2}^{\prime})$ are semimetric spaces of negative type.
}

%{\color{red} Therefore, they belong to the member of generalized distance covariance. Still, we can not represent $\KAcov_{2}(X,Y)$ as the form of generalized distance covariance. Because we can not separate $E\{\ang_{2}(X_1,X_2;X_5)\ang_{2}(Y_3,Y_4;X_5)\mid X_1,X_2, Y_3,Y_4\}$ to the multiplication of two semimetric for the dependence between $X_5$ and $Y_5$.}

%%%%%%%%%%%%%%%%%%%%%%%%%%%%%%%%%%%%%%%%%%%%%%%%%%%%%%%%%%%
%%%%%%%%%%%%%%%%%%%%%%%%%%%%%%%%%%%%%%%%%%%%%%%%%%%%%%%%%%%

\csection{CONCLUDING REMARKS}
In this article, we introduce kernel angle dependence measures in metric spaces.  
By making use of the reproducing kernel Hilbert space equipped with Gaussian measure, we derive kernel angle covariances with simple and explicit forms via direct integration.
This group of dependence measures can be adapted to various complex objects, including low dimensional vectors, high dimensional vectors, and symmetric positive definite matrices.  It also incorporates several existing angle-based measures in Euclidean space.
We build estimates for kernel angle covariance upon $U$-statistics and adopt Gamma approximation in the testing procedure to accelerate the tests.
We conduct comprehensive simulations on three different complex objects, which demonstrate the remarkable performances of the proposed independence tests.
The framework can also be used to generalize other test statistics such as Mann–Whitney test statistics and Kendall's tau.

%%%%%%%%%%%%%%%%%%%%%%%%%%%%%%%%%%%%%%%%%%%%%%%%%%%%%%%%%%%
%%%%%%%%%%%%%%%%%%%%%%%%%%%%%%%%%%%%%%%%%%%%%%%%%%%%%%%%%%%
\begin{center}
\bibliography{reference}

\begin{thebibliography}{47}
\newcommand{\enquote}[1]{``#1''}
\providecommand{\natexlab}[1]{#1}
\expandafter\ifx\csname urlstyle\endcsname\relax
  \providecommand{\doi}[1]{doi:\discretionary{}{}{}#1}\else
  \providecommand{\doi}{doi:\discretionary{}{}{}\begingroup
  \urlstyle{rm}\Url}\fi

\bibitem[{Arsigny et~al.(2007)Arsigny, Fillard, Pennec, and
  Ayache}]{arsigny2007geometric}
Arsigny, V., Fillard, P., Pennec, X., and Ayache, N. (2007).
\newblock \enquote{Geometric means in a novel vector space structure on
  symmetric positive-definite matrices.}
\newblock \emph{SIAM Journal on Matrix Analysis and Applications},
  \textbf{29(1)}, 328--347.

\bibitem[{Berlinet and Thomas-Agnan(2011)}]{berlinet2011reproducing}
Berlinet, A. and Thomas-Agnan, C. (2011).
\newblock \emph{Reproducing kernel Hilbert spaces in probability and
  statistics}.
\newblock Springer Science \& Business Media.

\bibitem[{Bogomolny et~al.(2007)Bogomolny, Bohigas, and
  Schmit}]{bogomolny2007distance}
Bogomolny, E., Bohigas, O., and Schmit, C. (2007).
\newblock \enquote{Distance matrices and isometric embeddings.}
\newblock \emph{arXiv preprint arXiv:0710.2063}.

\bibitem[{Da~Prato(2014)}]{da2014introduction}
Da~Prato, G. (2014).
\newblock \emph{Introduction to stochastic analysis and Malliavin calculus},
  volume~13.
\newblock Springer.

\bibitem[{Damodaran et~al.(2017)Damodaran, Courty, and
  Lef{\`e}vre}]{damodaran2017sparse}
Damodaran, B.B., Courty, N., and Lef{\`e}vre, S. (2017).
\newblock \enquote{Sparse hilbert schmidt independence criterion and
  surrogate-kernel-based feature selection for hyperspectral image
  classification.}
\newblock \emph{IEEE Transactions on Geoscience and Remote Sensing},
  \textbf{55(4)}, 2385--2398.

\bibitem[{Deb et~al.(2020)Deb, Ghosal, and Sen}]{deb2020measuring}
Deb, N., Ghosal, P., and Sen, B. (2020).
\newblock \enquote{Measuring association on topological spaces using kernels
  and geometric graphs.}
\newblock \emph{arXiv preprint arXiv:2010.01768}.

\bibitem[{Deb and Sen(2021)}]{deb2021multivariate}
Deb, N. and Sen, B. (2021).
\newblock \enquote{Multivariate rank-based distribution-free nonparametric
  testing using measure transportation.}
\newblock \emph{Journal of the American Statistical Association}, pages 1--16.

\bibitem[{Greenfeld and Shalit(2020)}]{greenfeld2020robust}
Greenfeld, D. and Shalit, U. (2020).
\newblock \enquote{Robust learning with the hilbert-schmidt independence
  criterion.}
\newblock In \enquote{International Conference on Machine Learning,} pages
  3759--3768. PMLR.

\bibitem[{Gretton et~al.(2007)Gretton, Fukumizu, Teo, Song, Sch{\"o}lkopf, and
  Smola}]{gretton2007kernel}
Gretton, A., Fukumizu, K., Teo, C., Song, L., Sch{\"o}lkopf, B., and Smola, A.
  (2007).
\newblock \enquote{A kernel statistical test of independence.}
\newblock \emph{Advances in Neural Information Processing Systems},
  \textbf{20}.

\bibitem[{Gretton et~al.(2005)Gretton, Smola, Bousquet, Herbrich, Belitski,
  Augath, Murayama, Pauls, Sch{\"o}lkopf, and Logothetis}]{gretton2005kernel}
Gretton, A., Smola, A., Bousquet, O., Herbrich, R., Belitski, A., Augath, M.,
  Murayama, Y., Pauls, J., Sch{\"o}lkopf, B., and Logothetis, N. (2005).
\newblock \enquote{Kernel constrained covariance for dependence measurement.}
\newblock In \enquote{International Workshop on Artificial Intelligence and
  Statistics,} pages 112--119. PMLR.

\bibitem[{Gupta(1963)}]{gupta1963probability}
Gupta, S.S. (1963).
\newblock \enquote{Probability integrals of multivariate normal and
  multivariate t1.}
\newblock \emph{The Annals of Mathematical Statistics}, \textbf{34(3)},
  792--828.

\bibitem[{Heller et~al.(2013)Heller, Heller, and
  Gorfine}]{heller2013consistent}
Heller, R., Heller, Y., and Gorfine, M. (2013).
\newblock \enquote{A consistent multivariate test of association based on ranks
  of distances.}
\newblock \emph{Biometrika}, \textbf{100(2)}, 503--510.

\bibitem[{Jacod and Protter(2004)}]{jacod2004probability}
Jacod, J. and Protter, P. (2004).
\newblock \emph{Probability essentials}.
\newblock Springer Science \& Business Media.

\bibitem[{Ke and Yin(2020)}]{ke2019expected}
Ke, C. and Yin, X. (2020).
\newblock \enquote{Expected conditional characteristic function-based measures
  for testing independence.}
\newblock \emph{Journal of the American Statistical Association},
  \textbf{115(530)}, 985--996.

\bibitem[{Kim et~al.(2020)Kim, Balakrishnan, and Wasserman}]{kim2020robust}
Kim, I., Balakrishnan, S., and Wasserman, L. (2020).
\newblock \enquote{Robust multivariate nonparametric tests via projection
  averaging.}
\newblock \emph{The Annals of Statistics}, \textbf{48(6)}, 3417--3441.

\bibitem[{Lahti et~al.(2014)Lahti, Saloj{\"a}rvi, Salonen, Scheffer, and
  De~Vos}]{lahti2014tipping}
Lahti, L., Saloj{\"a}rvi, J., Salonen, A., Scheffer, M., and De~Vos, W.M.
  (2014).
\newblock \enquote{Tipping elements in the human intestinal ecosystem.}
\newblock \emph{Nature Communications}, \textbf{5(1)}, 1--10.

\bibitem[{Lai et~al.(2021)Lai, Zhang, Wang, and Kong}]{lai2021testing}
Lai, T., Zhang, Z., Wang, Y., and Kong, L. (2021).
\newblock \enquote{Testing independence of functional variables by angle
  covariance.}
\newblock \emph{Journal of Multivariate Analysis}, \textbf{182}, 104711.

\bibitem[{Li and Deng(2019)}]{li2019blended}
Li, S. and Deng, W. (2019).
\newblock \enquote{Blended emotion in-the-wild: Multi-label facial expression
  recognition using crowdsourced annotations and deep locality feature
  learning.}
\newblock \emph{International Journal of Computer Vision}, \textbf{127(6-7)},
  884--906.

\bibitem[{Li and Zhang(2020)}]{li2020projective}
Li, Z. and Zhang, Y. (2020).
\newblock \enquote{On a projective ensemble approach to two sample test for
  equality of distributions.}
\newblock In \enquote{International Conference on Machine Learning,} pages
  6020--6027. PMLR.

\bibitem[{Liu et~al.(2022)Liu, Si, Xu, and Zhang}]{liu2022new}
Liu, J., Si, Y., Xu, W., and Zhang, R. (2022).
\newblock \enquote{A new nonparametric extension of anova via a projection mean
  variance measure.}
\newblock \emph{Statistica Sinica}, \textbf{32(1)}, 367--390.

\bibitem[{Moon and Chen(2022)}]{moon2022interpoint}
Moon, H. and Chen, K. (2022).
\newblock \enquote{Interpoint-ranking sign covariance for the test of
  independence.}
\newblock \emph{Biometrika}, \textbf{109(1)}, 165--179.

\bibitem[{Pan et~al.(2020)Pan, Wang, Zhang, Zhu, and Zhu}]{pan2020ball}
Pan, W., Wang, X., Zhang, H., Zhu, H., and Zhu, J. (2020).
\newblock \enquote{Ball covariance: A generic measure of dependence in banach
  space.}
\newblock \emph{Journal of the American Statistical Association},
  \textbf{115(529)}, 307--317.

\bibitem[{Pfister et~al.(2018)Pfister, B{\"u}hlmann, Sch{\"o}lkopf, and
  Peters}]{pfister2018kernel}
Pfister, N., B{\"u}hlmann, P., Sch{\"o}lkopf, B., and Peters, J. (2018).
\newblock \enquote{Kernel-based tests for joint independence.}
\newblock \emph{Journal of the Royal Statistical Society: Series B (Statistical
  Methodology)}, \textbf{80(1)}, 5--31.

\bibitem[{Rosenberg and Ekman(2020)}]{rosenberg2020face}
Rosenberg, E.L. and Ekman, P. (2020).
\newblock \emph{What the face reveals: Basic and applied studies of spontaneous
  expression using the Facial Action Coding System (FACS)}.
\newblock Oxford University Press.

\bibitem[{Rudra et~al.(2022)Rudra, Baxter, Hsieh, and
  Ghosh}]{rudra2022compositional}
Rudra, P., Baxter, R., Hsieh, E.W., and Ghosh, D. (2022).
\newblock \enquote{Compositional data analysis using kernels in mass cytometry
  data.}
\newblock \emph{Bioinformatics Advances}, \textbf{2(1)}.
\newblock Vbac003.

\bibitem[{Sarkar et~al.(2020)Sarkar, Biswas, and Ghosh}]{sarkar2020some}
Sarkar, S., Biswas, R., and Ghosh, A.K. (2020).
\newblock \enquote{On some graph-based two-sample tests for high dimension, low
  sample size data.}
\newblock \emph{Machine Learning}, \textbf{109(2)}, 279--306.

\bibitem[{Sarkar and Ghosh(2018)}]{sarkar2018some}
Sarkar, S. and Ghosh, A.K. (2018).
\newblock \enquote{On some high-dimensional two-sample tests based on averages
  of inter-point distances.}
\newblock \emph{Stat}, \textbf{7(1)}, e187.

\bibitem[{Satterthwaite(1946)}]{satterthwaite1946approximate}
Satterthwaite, F.E. (1946).
\newblock \enquote{An approximate distribution of estimates of variance
  components.}
\newblock \emph{Biometrics Bulletin}, \textbf{2(6)}, 110--114.

\bibitem[{Sch{\"o}lkopf et~al.(2002)Sch{\"o}lkopf, Smola, Bach
  et~al.}]{scholkopf2002learning}
Sch{\"o}lkopf, B., Smola, A.J., Bach, F., et~al. (2002).
\newblock \emph{Learning with kernels: support vector machines, regularization,
  optimization, and beyond}.
\newblock MIT press.

\bibitem[{Sejdinovic et~al.(2013)Sejdinovic, Sriperumbudur, Gretton, and
  Fukumizu}]{sejdinovic2013equivalence}
Sejdinovic, D., Sriperumbudur, B., Gretton, A., and Fukumizu, K. (2013).
\newblock \enquote{Equivalence of distance-based and rkhs-based statistics in
  hypothesis testing.}
\newblock \emph{The Annals of Statistics}, \textbf{41(5)}, 2263--2291.

\bibitem[{Serfling(1980)}]{serfling1980approximation}
Serfling, R.L. (1980).
\newblock \emph{Approximation Theorems in Mathematical Statistics}.
\newblock New York: Wiley.

\bibitem[{Shi et~al.(2022)Shi, Drton, and Han}]{shi2020distribution}
Shi, H., Drton, M., and Han, F. (2022).
\newblock \enquote{Distribution-free consistent independence tests via
  center-outward ranks and signs.}
\newblock \emph{Journal of the American Statistical Association},
  \textbf{117(537)}, 395--410.

\bibitem[{Sriperumbudur et~al.(2011)Sriperumbudur, Fukumizu, and
  Lanckriet}]{sriperumbudur2011universality}
Sriperumbudur, B.K., Fukumizu, K., and Lanckriet, G.R. (2011).
\newblock \enquote{Universality, characteristic kernels and rkhs embedding of
  measures.}
\newblock \emph{Journal of Machine Learning Research}, \textbf{12(7)}.

\bibitem[{Sz{\'e}kely and Rizzo(2014)}]{szekely2014partial}
Sz{\'e}kely, G.J. and Rizzo, M.L. (2014).
\newblock \enquote{Partial distance correlation with methods for
  dissimilarities.}
\newblock \emph{The Annals of Statistics}, \textbf{42(6)}, 2382--2412.

\bibitem[{Sz{\'e}kely et~al.(2007)Sz{\'e}kely, Rizzo, and
  Bakirov}]{szekely2007measuring}
Sz{\'e}kely, G.J., Rizzo, M.L., and Bakirov, N.K. (2007).
\newblock \enquote{Measuring and testing dependence by correlation of
  distances.}
\newblock \emph{The Annals of Statistics}, \textbf{35(6)}, 2769--2794.

\bibitem[{van Zanten and van~der Vaart(2008)}]{van2008reproducing}
van Zanten, J. and van~der Vaart, A. (2008).
\newblock \enquote{Reproducing kernel hilbert spaces of gaussian priors.}
\newblock In \enquote{Pushing the limits of contemporary statistics:
  contributions in honor of Jayanta K. Ghosh,} pages 200--222. Institute of
  Mathematical Statistics.

\bibitem[{Wainwright(2019)}]{wainwright2019high}
Wainwright, M.J. (2019).
\newblock \emph{High-dimensional statistics: A non-asymptotic viewpoint},
  volume~48.
\newblock Cambridge University Press.

\bibitem[{Wang et~al.(2018)Wang, Hu, Gao, Ali, Tien, Sun, and
  Yin}]{wang2018low}
Wang, B., Hu, Y., Gao, J., Ali, M., Tien, D., Sun, Y., and Yin, B. (2018).
\newblock \enquote{Low rank representation on spd matrices with log-euclidean
  metric.}
\newblock \emph{Pattern Recognition}, \textbf{76}, 623--634.

\bibitem[{Welch(1938)}]{welch1938significance}
Welch, B.L. (1938).
\newblock \enquote{The significance of the difference between two means when
  the population variances are unequal.}
\newblock \emph{Biometrika}, \textbf{29(3/4)}, 350--362.

\bibitem[{Xu and Zhu(2022)}]{xu2022power}
Xu, K. and Zhu, L. (2022).
\newblock \enquote{Power analysis of projection-pursuit independence tests.}
\newblock \emph{Statistica Sinica}, \textbf{32}, 417--33.

\bibitem[{Yan and Zhang(2021)}]{yan2021kernel}
Yan, J. and Zhang, X. (2021).
\newblock \enquote{Kernel two-sample tests in high dimension: Interplay between
  moment discrepancy and dimension-and-sample orders.}
\newblock \emph{arXiv preprint arXiv:2201.00073}.

\bibitem[{Yao et~al.(2018)Yao, Zhang, and Shao}]{yao2018testing}
Yao, S., Zhang, X., and Shao, X. (2018).
\newblock \enquote{Testing mutual independence in high dimension via distance
  covariance.}
\newblock \emph{Journal of the Royal Statistical Society: Series B (Statistical
  Methodology)}, \textbf{80(3)}, 455--480.

\bibitem[{Ying and Yu(2022)}]{ying2022frechet}
Ying, C. and Yu, Z. (2022).
\newblock \enquote{Fr{\'e}chet sufficient dimension reduction for random
  objects.}
\newblock \emph{Biometrika}, \textbf{109(4)}, 975--992.

\bibitem[{Zhang and Dao(2020)}]{zhang2020distance}
Zhang, Q. and Dao, T. (2020).
\newblock \enquote{A distance based multisample test for high-dimensional
  compositional data with applications to the human microbiome.}
\newblock \emph{BMC Bioinformatics}, \textbf{21(9)}, 1--17.

\bibitem[{Zhang and Zhu(2022)}]{zhang2022projective}
Zhang, Y. and Zhu, L. (2022).
\newblock \enquote{Projective independence tests in high dimensions: the curses
  and the cures.}
\newblock \emph{Journal of Machine Learning Research}, \textbf{(under review)}.

\bibitem[{Zhu et~al.(2020)Zhu, Zhang, Yao, and Shao}]{zhu2020distance}
Zhu, C., Zhang, X., Yao, S., and Shao, X. (2020).
\newblock \enquote{Distance-based and rkhs-based dependence metrics in high
  dimension.}
\newblock \emph{The Annals of Statistics}, \textbf{48(6)}, 3366--3394.

\bibitem[{Zhu et~al.(2017)Zhu, Xu, Li, and Zhong}]{zhu2017projection}
Zhu, L., Xu, K., Li, R., and Zhong, W. (2017).
\newblock \enquote{Projection correlation between two random vectors.}
\newblock \emph{Biometrika}, \textbf{104(4)}, 829--843.

\end{thebibliography}
\end{center}

%%%%%%%%%%%%%%%%%%%%%%%%%%%%%%%%%%%%%%%%%%%%%%%%%%%%%%%%%%%
%%%%%%%%%%%%%%%%%%%%%%%%%%%%%%%%%%%%%%%%%%%%%%%%%%%%%%%%%%%

\setcounter{section}{0} % reset section counter
\renewcommand{\theequation}{\Alph{section}.\arabic{equation}} % prefix equation counter with s
\renewcommand{\thetable}{\Alph{section}.\arabic{table}} % prefix equation counter with 
\renewcommand{\thesection}{\Alph{section}}% Adjust section printing (from here onward)
\renewcommand{\thetheo}{\Alph{section}.\arabic{theo}}
\renewcommand{\theprop}{\Alph{section}.\arabic{prop}}
\renewcommand{\thecoll}{\Alph{section}.\arabic{coll}}
\renewcommand{\thelemm}{\Alph{section}.\arabic{lemm}}
% \section{APPENDIX: PROOFS OF THEOREMS}

%\section{Proof of Theorem \ref{theorem:equivalence}}
%% \renewcommand{\theequation}{A.\arabic{equation}}
%% \setcounter{equation}{0}
%\noindent
%To prove this equation, we only need to exchange the order of summations.
%\beqrs
%\wh\calS(X,Y) &=& 1 - 6\{n^2(c-1)\}^{-1}\sum\limits_{i=1}^{n}\sum_{h=1}^{H}\sum_{j<l}^{c}\{1(Y_{(h,j)}\geq Y_i)-1(Y_{(h,l)}\geq Y_i)\}^2\\
%&=& 1 - 6\{n^2(c-1)\}^{-1}\sum_{h=1}^{H}\sum_{j<l}^{c}\sum\limits_{i=1}^{n}\{1(Y_{(h,j)}\geq Y_i)-1(Y_{(h,l)}\geq Y_i)\}^2\\
%&=&1- 6\{n^2(c-1)\}^{-1}\sum_{h=1}^{H}\sum_{j<l}^{c} |r_{(h,j)}-r_{(h,l)}|.
%\eeqrs
%The equation can be verified using the definition of rank.
%\hfill$\fbox{}$

\newpage
\begin{center}
	{\bf\large  Supplement to ``Kernel Angle Dependence Measures for Complex Objects"}
\end{center}
In this Supplement Material, we provide all the proofs in the main context.  For the notation clarity, 
we denote $B(\calX)$ as the subset of $C(\calX)$, which are continuous functions bounded by 1 with respect to infinite norm. $B(\calY)$ is denoted in an analogous manner.

\section{Technical Lemmas}\label{section:technical_lemmas}

In the following context, we provide a spectral view of reproducing kernel Hilbert space and introduce the important series representation for the inner product.
Then, we introduce Gaussian measures and consider a special case, the Gaussian measure with  covariance identity operator.
By making use of the series representation, we derive the integration results for the reproducing kernel Hilbert space equipped with Gaussian measure using identity covariance operator in Lemma \ref{lemma:normal_hilbertspace} and \ref{lemma:abs_hilbertspace}.

Given real-valued reproducing kernel Hilbert space $\calH_{K}$ with reproducing kernel $K:\calZ\times\calZ\to\mR$, if $K$ is continuous,
we define the integral operator $T_{K}:L^{2}(\calZ, \mu)\to L^{2}(\calZ, \mu)$ by
\beqrs
(T_{K}f)(\cdot) = \int_{z\in\calZ} K (\cdot,z) f(z)d\mu (z),
\eeqrs
According to Mercer's Theorem \citep[Theorem 12.20]{wainwright2019high}, there is an orthonormal basis $\{\psi_{j}\}$ of $L^{2}(\calZ, \mu)$ consisting of eigenfunctions $T_{K}$ such that the corresponding sequence of eigenvalues $\{\lambda_{j}\}$ are non-negative. 
$K(z_1, z_2)$ has the representation
\beqrs
K(z_1, z_2) = \sum_{j=1}^{\infty}\lambda_{j} \psi_{j}(z_1)\psi_{j}(z_2),
\eeqrs
where the convergence is absolute and uniform. 
And this series representation refers to Mercer's representation of $K$. This representation gives an explicitly characterization of reproducing kernel Hilbert space \citep[Corollary 12.26]{wainwright2019high}. For any $g_{1},g_{2}\in \calH_{K}$,
\beqr\label{equation:seriesrepresentation}
\langle g_{1}, g_{2}\rangle_{
\calH_{K}} = \sum_{j=1}^{\infty}\lambda_{j}^{-1}
\langle g_{1},\psi_{j} \rangle_{L^{2}}\langle g_{2},\psi_{j} \rangle_{L^2}.
\eeqr

Suppose $\calH$ is separable Hilbert space, $h,f\in\calH$ and $Q:\calH\to \calH$ is continuous, symmetric and positive linear operator.
According to \citet[Theorem 1.11]{da2014introduction}, a Gaussian measure $\mu$ on $\calH$ with mean zero and covariance operator $Q$, has the characteristic function as
\beqrs
\int_{h\in\calH}\exp(i\langle h, f\rangle_{\calH}) \mu(dh) = \exp( - 2^{-1}\langle Qf,f\rangle_{\calH}).
\eeqrs
Let $\{e_{j}\}$ be orthonormal eigenfunctions and  $\{\sigma_{j}\}$  be corresponding eigenvalues, satisfying that $Qe_{j} = \sigma_{j}e_{j}$. The Gaussian measure $\mu$ has the following representation,
\beqr\label{equation:museries}
\mu(dh) = \prod_{m=1}^{\infty} \mu_{m}(dh), 
\textrm{ where } \mu_{m}(dh) = (2\pi \sigma_{m})^{-1/2}\exp\{-\langle h,e_{m}\rangle_{\calH}^2/(2\sigma_{m})\}d\langle h,e_{m}\rangle_{\calH}.
\eeqr

For the reproducing kernel Hilbert space equipped with Gaussian measures using mean zero and  covariance identity operator, we choose a special orthonormal basis in  $\calH_{K}$, $\{\lambda_{j}^{1/2}\psi_j\}$. For $h\in\calH_{K}$, denote $h_j\defby\langle h, \lambda_{j}^{1/2}\psi_j\rangle_{\calH_{K}}$.
Using the representation of inner product in reproducing kernel Hilbert space \eqref{equation:seriesrepresentation}, we have $h_j = \lambda_{j}^{-1/2} \langle h, \psi_j\rangle_{L^2}$.
By applying \eqref{equation:museries}, we get
\beqrs
\mu(dh) = 
\prod_{j=1}^{\infty} (2\pi)^{-1/2}\exp(-h_{j}^2/2)dh_{j}.
\eeqrs
With this preparation knowledge, we will prove Lemma \ref{lemma:normal_hilbertspace} and \ref{lemma:abs_hilbertspace}.

{\lemm{\label{lemma:normal_uni}\citep[Page 801]{gupta1963probability}} Let $(Z_{1},Z_{2})\trans$ be the bivariate normally distribution with mean $0$ and correlation $\rho$. Then,
	$\pr(Z_{1}\leq 0, Z_{2}\leq 0) = 4^{-1} + (2\pi)^{-1}\arcsin \rho$.	
}

{\lemm{\label{lemma:normal_vector}} Let $\z_{1}$ and $\z_{2}$ be two vectors in $\mR^{p}$. Suppose $\x$ is a random vector in $\mR^{p}$ and follows the multivariate standard normal distribution. Then,
	$E\{1(\langle\z_{1},\x\rangle\leq 0) 1(\langle\z_{2},\x\rangle\leq 0)  = 2^{-1} - (2\pi)^{-1}\arccos\{ \z_{1}\trans\z_{2}/(\|\z_1\|\|\z_2\|)\}$.}

\noindent\textbf{Proof of Lemma \ref{lemma:normal_vector}:} $\langle\z_{1},\x\rangle$ and $\langle\z_{2},\x\rangle$ follow normal distribution with mean $0$ and correlation $\z_{1}\trans\z_{2}/(\|\z_1\|\|\z_2\|)$. From Lemma \ref{lemma:normal_uni}, we can straightly derive that $E\{1(\langle\z_{1},\x\rangle\leq 0) 1(\langle\z_{2},\x\rangle\leq 0) \} = 2^{-1} - (2\pi)^{-1}\arccos\{ \z_{1}\trans\z_{2}/(\|\z_1\|\|\z_2\|)\}$.
\hfill$\fbox{}$

\noindent\textbf{Proof of Lemma \ref{lemma:normal_hilbertspace}:}
We prove the first equation.
As previously discussed, we choose orthonormal basis $\{\lambda_{j}^{1/2}\psi_j\}$ to represent Gaussian measure  with zero mean identity covariance operator.
We derive the integral that
\beqrs
&&\int_{h\in\calH_{K}}1(\langle s_{1},h\rangle_{\calH_{K}}\leq 0) 1(\langle s_{2}, h\rangle_{\calH_{K}}\leq 0) \mu(dh)\\
&=& \int_{h\in\calH_{K}}
1\bigg(\sum_{j=1}^\infty\lambda_{j}^{-1}
\langle s_{1},\psi_{j}\rangle_{L^2} 
\langle h,\psi_{j}\rangle_{L^2}\leq 0\bigg) 
1\bigg(\sum_{j=1}^\infty\lambda_{j}^{-1}
\langle s_{2},\psi_{j}\rangle_{L^2} 
\langle h,\psi_{j}\rangle_{L^2}\leq 0\bigg) \mu(dh).
\eeqrs
Using the uniformity of the series representation and Fubini's Theorem, the above formula equals the following equation as $M$
diverges to infinity.
\beqrs
\int_{h\in\calH_{K}}1\bigg(\sum_{j=1}^{M}\lambda_{j}^{-1}
\langle s_{1},\psi_{j}\rangle_{L^2} 
\langle h,\psi_{j}\rangle_{L^2}\leq 0\bigg)1\bigg(\sum_{j=1}^{M}\lambda_{j}^{-1}
\langle s_{2},\psi_{j}\rangle_{L^2} 
\langle h,\psi_{j}\rangle_{L^2}\leq 0\bigg) \mu(dh).
\eeqrs
This integral term equals
\beqrs
&&\int_{h\in\calH_{K}}
1\bigg(\sum_{j=1}^{M}\lambda_{j}^{-1/2}
\langle s_{1},\psi_{j}\rangle_{L^2} 
h_{j}\leq 0\bigg)
1\bigg(\sum_{j=1}^{M}\lambda_{j}^{-1/2}
\langle s_{2},\psi_{j}\rangle_{L^2} 
h_{j}\leq 0\bigg) \mu(dh)\\
&=& \int_{h\in\calH_{K}}
1\bigg(\sum_{j=1}^{M}\lambda_{j}^{-1/2}
\langle s_{1},\psi_{j}\rangle_{L^2} 
h_{j}\leq 0\bigg)
1\bigg(\sum_{j=1}^{M}\lambda_{j}^{-1/2}
\langle s_{2},\psi_{j}\rangle_{L^2} 
h_{j}\leq 0\bigg) \mu_{1}(dh)\ldots\mu_{M}(dh)\\
&=& 
2^{-1} - (2\pi)^{-1}\arccos\bigg\{\\
&& \Big(\sum_{j=1}^{M}\lambda_{j}^{-1}\langle s_{1},\psi_{j}\rangle_{L^2}\langle s_{2},\psi_{j}\rangle_{L^2}\Big)
\Big(\sum_{j=1}^{M}\lambda_{j}^{-1}\langle s_{1},\psi_{j}\rangle_{L^2}^2\Big)^{-1/2}
\Big(\sum_{j=1}^{M}\lambda_{j}^{-1}\langle s_{2},\psi_{j}\rangle_{L^2}^2\Big)^{-1/2}\bigg\}.
\eeqrs
The second equality can be established by applying Lemma \ref{lemma:normal_vector}.
As $M$ goes to infinity, we apply the series representation \eqref{equation:seriesrepresentation} and derive the final result as,
\beqrs
&&\int_{h\in\calH_{K}}1(\langle s_{1},h\rangle_{\calH_{K}}\leq 0) 1(\langle s_{2}, h\rangle_{\calH_{K}}\leq 0) \mu(dh)\\
&=& 2^{-1} - (2\pi)^{-1}\arccos\{ \langle s_1,s_2\rangle_{\calH_{K}}/(\|s_1\|_{\calH_{K}}\|s_2\|_{\calH_{K}})\}.
\eeqrs
Following the above paradigm of the proof, we prove the second equation.
We derive the integral that
\beqrs
&&\int_{h\in\calH_{K}}\int_{u\in\mR}1(\langle s_{1},h\rangle_{\calH_{K}}\leq u) 1(\langle s_{2}, h\rangle_{\calH_{K}}\leq u) d\Phi_{1}(u) \mu(dh)\\
&=& \int_{h\in\calH_{K}}\int_{u\in\mR}
1\bigg(\sum_{j=1}^\infty\lambda_{j}^{-1}
\langle s_{1},\psi_{j}\rangle_{L^2} 
\langle h,\psi_{j}\rangle_{L^2}\leq u\bigg) \\
&&\hspace{6cm}
1\bigg(\sum_{j=1}^\infty\lambda_{j}^{-1}
\langle s_{2},\psi_{j}\rangle_{L^2} 
\langle h,\psi_{j}\rangle_{L^2}\leq u\bigg) d\Phi_{1}(u)\mu(dh).
\eeqrs
Using the uniformity of the series representation and Fubini's Theorem, the above formula equals the following equation as $M$
diverges to infinity.
\beqrs
\int_{h\in\calH_{K}}\int_{u\in\mR}
1\bigg(\sum_{j=1}^{M}\lambda_{j}^{-1/2}
\langle s_{1},\psi_{j}\rangle_{L^2} h_{j}\leq u\bigg)
1\bigg(\sum_{j=1}^{M}\lambda_{m}^{-1/2}
\langle s_{2},\psi_{j}\rangle_{L^2} h_{j}\leq u\bigg) d\Phi_{1}(u) \mu(dh).
\eeqrs
Denote $\h\defby(h_{1},\ldots,h_{M},u)\trans$, we know $\h$ follows $M+1$ dimensional standard multi-variate normal distribution. Let $\Phi_{M+1}(\h)$ be the $M+1$ dimensional standard multi-variate normal distribution function. The above integration equals
\beqrs
&&\int_{\h\in\mR^{M+1}}
1\bigg\{ (-1) u+\sum_{j=1}^{M}\lambda_{j}^{-1/2}
\langle s_{1},\psi_{j}\rangle_{L^2} h_{j}\leq 0\bigg\}\\
&&\hspace{5cm}1\bigg\{(-1) u+\sum_{j=1}^{M}\lambda_{j}^{-1/2}
\langle s_{2},\psi_{j}\rangle_{L^2} h_{j} \leq 0\bigg\} d\Phi_{M+1}(\h)\\
&=& 2^{-1} - (2\pi)^{-1}\arccos\bigg\{\Big(1+\sum_{j=1}^{M}\lambda_{j}^{-1}\langle s_{1},\psi_{j}\rangle_{L^2}\langle s_{2},\psi_{j}\rangle_{L^2}\Big)\\
&& \hspace{4cm}
\Big(1+\sum_{j=1}^{M}\lambda_{j}^{-1}\langle s_{1},\psi_{j}\rangle_{L^2}^2\Big)^{-1/2}
\Big(1+\sum_{j=1}^{M}\lambda_{j}^{-1}\langle s_{2},\psi_{j}\rangle_{L^2}^2\Big)^{-1/2}\bigg\}.
\eeqrs
The above equality can be established by applying Lemma \ref{lemma:normal_vector}.
As $M$ goes to infinity, we complete the proof for the second equation.
	\beqrs
&&\int_{h\in\calH_{K}}\int_{u\in\mR}1(\langle s_{1},h\rangle_{\calH_{K}}\leq u) 1(\langle s_{2}, h\rangle_{\calH_{K}}\leq u) d\Phi_{1}(u) \mu(dh)\\
&=& 2^{-1} - (2\pi)^{-1}\arccos\{ (1+\langle s_1,s_2\rangle_{\calH_{K}})(1+\|s_1\|_{\calH_{K}}^{2})^{-1/2}(1+\|s_2\|_{\calH_{K}}^{2})^{-1/2}\}.
\eeqrs
\hfill$\fbox{}$

{\lemm{\label{lemma:abs_integral}} Let $\z,\x\in\mR^p$. $\z$ follows multivariate standard normal distribution. We have
	$E\{\abs{\z\trans\x}\} = 2(2\pi)^{-1/2}\|\x\|$.}

\noindent\textbf{Proof of Lemma \ref{lemma:abs_integral}:}
Without loss of generality, we assume $\x = (X_1, 0,\ldots,0)\trans$.
Else, we can do orthogonal transformation and inverse of orthogonal transformation for $\x$ and $\z$ respectively. 
And $\z$ still follows normal distribution.
Let $\z = (Z_1,\ldots,Z_p)$.
\beqrs
E\{\abs{\z\trans\x}\}
&=& \int_{\z\in\mR^m}\abs{\z\trans\x}\exp\{-\|\z\|^2/2\}d\z \\
&=& \|\x\|\int_{\z\in\mR^m}\abs{Z_1}\exp\{-\|\z\|^2/2\}d\z= 2(2\pi)^{-1/2}\|\x\|.
\eeqrs
The last equality is derived via direct integration using spherical coordinates.
\hfill$\fbox{}$

{\lemm\label{lemma:abs_hilbertspace}
Suppose $s$ and $h$ are in separable reproducing kernel  Hilbert space $\calH_{K}$, which has reproducing kernel $K:\calX\times\calX\rightarrow\mR$. 
Let $\mu$ be Gaussian measure on $\calH_{K}$ with mean zero and identity covariance operator. Then,
	\beqrs
	\int_{h\in\calH_{K}}\abs{\langle s,h \rangle_{\calH_{K}}}\mu(dh) = 2(2\pi)^{-1/2}\|s\|_{\calH_{K}}.
	\eeqrs
} 
\noindent\textbf{Proof of Lemma \ref{lemma:abs_hilbertspace}:}
We choose orthonormal basis $\{\lambda_{m}^{1/2}\psi_m\}$ to represent Gaussian measure  with zero mean identity covariance operator.
Then, we derive the integral that
\beqr\label{equation:integral_innerproduct}
&&\int_{h\in\calH_{K}}\abs{\langle s,h \rangle_{\calH_{K}}}\mu(dh)\nonumber\\
&=& \int_{h\in\calH_{K}}
\abs{\sum_{m=1}^\infty\lambda_{j}^{-1}
	\langle s,\psi_{j}\rangle_{L^2} 
	\langle h,\psi_{j}\rangle_{L^2}}\mu(dh)\nonumber\\
&=& \lim_{M\to\infty}\int_{h\in\calH_{K}}
\abs{\sum_{j=1}^{M}\lambda_{j}^{-1}
	\langle s,\psi_{j}\rangle_{L^2} 
	\langle h,\psi_{j}\rangle_{L^2}}\mu_{1}(dh)\ldots\mu_{M}(dh).
\eeqr
Then we calculate the integral term .
\beqr\label{equation:finite_dimension_integral}
&&\int_{h\in\calH_{K}}
\abs{\sum_{m=1}^{M}\lambda_{m}^{-1}
	\langle s,\psi_{m}\rangle_{L^2} 
	\langle h,\psi_{m}\rangle_{L^2}}\mu_{1}(dh)\ldots\mu_{M}(dh)\nonumber\\
&=& \int_{h\in\calH_{K}}
\abs{\sum_{m=1}^{M}\lambda_{m}^{-1/2}
	\langle s,\psi_{m}\rangle_{L^2} 
	h_{m}}\mu_{1}(dh)\ldots\mu_{M}(dh)\nonumber\\
&=& \left\{\sum_{m=1}^{M}\lambda_{m}^{-1}\langle s,\psi_{m}\rangle_{L^2}^2 \right\}^{1/2}.
\eeqr
By plugging the above integral result into \eqref{equation:integral_innerproduct}, we can derive that
\beqrs
&&\int_{h\in\calH_{K}}\abs{\langle s,h \rangle_{\calH_{K}}}\mu(dh)\\
&=& \lim_{M\to\infty} \left\{\sum_{m=1}^{M}\lambda_{m}^{-1}\langle s,\phi_{m}\rangle_{L^2}^2 \right\}^{1/2}
=  \left\{\lim_{M\to\infty}\sum_{m=1}^{M}\lambda_{m}^{-1}\langle s,\phi_{m}\rangle_{L^2}^2 \right\}^{1/2}=\|s\|_{\calH_{K}}.
\eeqrs
\hfill$\fbox{}$

\section{Proof of Lemma \ref{lemma:independence_equivalence}}
\setcounter{equation}{0}

\noindent We first provide the definition of universal kernel and then prove the this theorem. 

{\lemm{\label{theorem:independent}\citep[Theorem 10.1]{jacod2004probability}} 
Let $X$ and $Y$ be random variables on metric spaces.
$X$ and $Y$ are independent if and only if $\cov\{f(X), g(Y)\} = 0$ for any pair $(f,g)$ of bounded, continuous functions.}

{\defi{\label{definition:universal}}
A continuous kernel $K:\calZ\times\calZ\to\mR$ on a compact metric space $(\calZ, d)$ is called universal if and only if the reproducing kernel Hilbert space $\calH$ induced by the kernel $K$ is dense in the space of continuous functions on $\calZ$, with respect to the infinity norm.
}

When $X$ and $Y$ are independent, it is obvious that $\cov\{1(f(X)\leq u), 1(g(Y)\leq v)\} = 0$ for any $f\in \calH_1$, $g\in \calH_2$, $u\in \mR$ and $v\in\mR$. We complete the ``if"  part using the proof by contradiction.

Suppose that $X$ and $Y$ are not independent, given $\cov\{1(f(X)\leq u), 1(g(Y)\leq v)\} = 0$ for any $f\in \calH_1$, $g\in \calH_2$, $u\in \mR$ and $v\in\mR$. From Theorem \ref{theorem:independent}, there must exist $f_{0}\in B(\calX)$ and $g_{0}\in B(\calY)$ that $\cov\{f_{0}(X), g_{0}(Y)\} = c$, where $1\geq c>0$.

Given that $\calH_1$ and $\calH_2$ are induced by continuous universal kernel, from the Definition \ref{definition:universal}, we know that $\calH_1$ and $\calH_2$ are dense in $C(\calX)$ and $C(\calY)$ with respect to infinite norm respectively.
Therefore, we can find $f^{\ast}\in \calH_1$ and  $g^{\ast}\in \calH_2$ that $\|f^{\ast} - f_{0}\|_{\infty}\leq \varepsilon$ and $\|g^{\ast} - g_{0}\|_{\infty}\leq \varepsilon$, where $c/3>\varepsilon>0$
\beqrs
\cov\{f^{\ast}(X),g^{\ast}(Y)\} &=& \cov\{f^{\ast}(X) - f_{0}(X) + f_{0}(X),g^{\ast}(Y)- g_{0}(Y) + g_{0}(Y)\} \\
& = & \cov\{f_{0}(X),g_{0}(Y)\} + \cov\{f^{\ast}(X) - f_{0}(X), g^{\ast}(Y)- g_{0}(Y)\} \\
&& + \cov\{f_{0}(X),  g^{\ast}(Y)- g_{0}(Y)\} + \cov\{g_{0}(Y),  f^{\ast}(X)- f_{0}(X)\}\\
& \geq & \cov\{f^{\ast}(X),g^{\ast}(Y)\} - \varepsilon^2 - \varepsilon[\abs{E\{f_{0}(X)\}} + \abs{E\{g_{0}(Y)\}}]\\
&\geq& c- \varepsilon^2 - 2\varepsilon > c/9.
\eeqrs
Given the above inequality,  the fact $\cov\{1(f^{\ast}(X)\leq u), 1(g^{\ast}(Y)\leq v)\} = 0$ for any $u\in \mR$ and $v\in\mR$ doesn't hold up. The proof is complete for this contradiction.
\hfill$\fbox{}$

\section{Proof of Theorem \ref{theorem:angel_correlation}}

Firstly, we prove statement (i).
We rewrite the covariance term as
\beqrs
&&\cov\{1( \langle\phi_{1}(X), f \rangle\leq u), 1( \langle\phi_{2}(Y), g \rangle\leq v)\} \\
&=& 
E\{1( \langle\phi_{1}(X_{1}), f \rangle\leq u)1( \langle\phi_{2}(Y_{1}), g \rangle\leq v) - 1( \langle\phi_{1}(X_{1}), f \rangle\leq u)1( \langle\phi_{2}(Y_{3}), g \rangle\leq v)\}.
\eeqrs
The square of the covariance can be represented as the expectation form.
\beqr\label{equation:covariance_square}
&&\cov^{2}\{1( \langle\phi_{1}(X), f \rangle\leq u), 1( \langle\phi_{2}(Y), g \rangle\leq v)\} \\
&=& 
E\Big[\{1( \langle\phi_{1}(X_{1}), f \rangle\leq u)1( \langle\phi_{2}(Y_{1}), g \rangle\leq v) - 1( \langle\phi_{1}(X_{1}), f \rangle\leq u)1( \langle\phi_{2}(Y_{3}), g \rangle\leq v)\}\nonumber\\
&&\{1( \langle\phi_{1}(X_{2}), f \rangle\leq u)1( \langle\phi_{2}(Y_{2}), g \rangle\leq v) - 1( \langle\phi_{1}(X_{2}), f \rangle\leq u)1( \langle\phi_{2}(Y_{4}), g \rangle\leq v)\}\Big].\nonumber
\eeqr
Using Fubini's Theorem, we can exchange the order of integrals.
\beqr\label{equation:fubini}
&&\int_{\calH_1}\int_{\calH_2}\int_{\mR^2}\cov^{2}\{1( \langle\phi_{1}(X), f \rangle\leq u), 1( \langle\phi_{2}(Y), g \rangle\leq v)\}
d \Phi(u,v)  \mu_1(df) \mu_2(dg)\nonumber\\
&=&E\bigg[\int_{\calH_1}\int_{\calH_2}\int_{\mR^2}\{1( \langle\phi_{1}(X_{1}), f \rangle\leq u)1( \langle\phi_{2}(Y_{1}), g \rangle\leq v) - 1( \langle\phi_{1}(X_{1}), f \rangle\leq u)\nonumber\\
&&1( \langle\phi_{2}(Y_{3}), g \rangle\leq v)\}
\{1( \langle\phi_{1}(X_{2}), f \rangle\leq u)1( \langle\phi_{2}(Y_{2}), g \rangle\leq v) - 1( \langle\phi_{1}(X_{2}), f \rangle\leq u)\nonumber\\
&&1( \langle\phi_{2}(Y_{4}), g \rangle\leq v)\}d \Phi(u,v)  \mu_1(df) \mu_2(dg)\bigg].
\eeqr
The integral term of equation \eqref{equation:fubini} can be rewritten as
\beqr\label{equation:angel_intgeral}
&&\int_{\calH_1^{\prime}}\int_{\calH_2^{\prime}}\{
1( \langle(\phi_{1}(X_{1}),-1), (f,u) \rangle\leq 0)
1( \langle(\phi_{2}(Y_{1}),-1), (g,v) \rangle\leq 0) - \\
&&\hspace{1.6cm}1( \langle(\phi_{1}(X_{1}),-1),  (f,u) \rangle\leq 0)
1( \langle(\phi_{2}(Y_{3}),-1), (g,v)  \rangle\leq 0)\}\nonumber\\
&&\hspace{1.4cm}\{1( \langle(\phi_{1}(X_{2}),-1),  (f,u) \rangle\leq 0)
1( \langle(\phi_{2}(Y_{2}),-1), (g,v)  \rangle\leq 0) - \nonumber\\
&&\hspace{1.6cm}1( \langle(\phi_{1}(X_{2}),-1),  (f,u) \rangle\leq 0)1( \langle(\phi_{2}(Y_{4}),-1), (g,v)  \rangle\leq 0)\}  
\mu_1^{\prime}(df) \mu_2^{\prime}(dg).\nonumber
\eeqr
By applying Lemma \ref{lemma:normal_hilbertspace} repeatedly, we can simplify the  equation \eqref{equation:angel_intgeral} as
$(2\pi)^{-2}\big\{$\\$\ang_{1}(X_{1}, X_{2})\ang_{1}(Y_{1}, Y_{2}) - 2 \ang_{1}(X_{1}, X_{2})\ang_{1}(Y_{1}, Y_{3}) 
+ \ang_{1}(X_{1}, X_{2})\ang_{1}(Y_{3}, Y_{4})\big\}$.
Taking expectation, we complete the proof for statement (i).

Next, we simplify the result in (ii). Following the proof for (i), we can similarly derive equation \eqref{equation:covariance_square}.
Given $f$ and $g$, $F_{U,V}(u,v)$ is the distribution function for $(\langle\phi_{1}(X_5), f \rangle,
\langle\phi_{2}(Y_5), g \rangle)\trans$. 
Combined with equation \eqref{equation:covariance_square}, the innermost integral can be rewritten as
\beqrs
&&\int_{\mR^2}\cov^{2}\{1( \langle\phi_{1}(X), f \rangle\leq u), 1( \langle\phi_{2}(Y), g \rangle\leq v)\}
d F_{U,V}(u,v) \\
&=&E\Big[\big\{1( \langle\phi_{1}(X_{1})-\phi_{1}(X_{5}), f \rangle\leq 0)1( \langle\phi_{2}(Y_{1})-\phi_{2}(Y_{5}), g \rangle\leq 0) \\
&&- 1( \langle\phi_{1}(X_{1})-\phi_{1}(X_{5}), f \rangle\leq 0)1( \langle\phi_{2}(Y_{3})-\phi_{2}(Y_{5}), g \rangle\leq 0)\big\}\\
&&\big\{1( \langle\phi_{1}(X_{2})-\phi_{1}(X_{5}), f \rangle\leq 0)1( \langle\phi_{2}(Y_{2})-\phi_{2}(Y_{5}), g \rangle\leq 0) \\
&&- 1( \langle\phi_{1}(X_{2})-\phi_{1}(X_{5}), f \rangle\leq 0)1( \langle\phi_{2}(Y_{4})-\phi_{2}(Y_{5}), g \rangle\leq 0)\big\}\Big].
\eeqrs
By Fubini's Theorem, we can exchange the order of the two integrals with the expectation. Applying the integral result of Lemma \ref{lemma:normal_hilbertspace} repeatedly, we can derive 
\beqrs
&&4\pi^2\int_{\calH_1}\int_{\calH_2}\int_{\mR^2}\cov^{2}\{1( \langle\phi_{1}(X), f \rangle\leq u), 1( \langle\phi_{2}(Y), g \rangle\leq v)\}
d F_{U,V}(u,v)  \mu_1(df) \mu_2(dg)\\
&=&  E\{\ang_{2}(X_{1}, X_{2};X_{5})\ang_{2}(Y_{1}, Y_{2};Y_{5}) -  \ang_{2}(X_{1}, X_{2};X_{5})\ang_{2}(Y_{2}, Y_{3};Y_{5}) \\
&&-  \ang_{2}(X_{1}, X_{2};X_{5})\ang_{2}(Y_{1}, Y_{4};Y_{5}) 
+ \ang_{2}(X_{1}, X_{2};X_{5})\ang_{2}(Y_{3}, Y_{4};Y_{5})\}\\
&=&  E\{\ang_{2}(X_{1}, X_{2};X_{5})\ang_{2}(Y_{1}, Y_{2};Y_{5}) - 2 \ang_{2}(X_{1}, X_{2};X_{5})\ang_{2}(Y_{1}, Y_{3};Y_{5}) \\
&&+ \ang_{2}(X_{1}, X_{2};X_{5})\ang_{2}(Y_{3}, Y_{4};Y_{5})\}.
\eeqrs
We complete the proof for Theorem \ref{theorem:angel_correlation} (ii). 
Given $f$ and $g$, $F_{U}(u)F_V(v)$ is the distribution function for $(\langle\phi_{1}(X_5), f \rangle,
\langle\phi_{2}(Y_6), g \rangle)\trans$. 
Using this fact, Theorem \ref{theorem:angel_correlation} (iii) can be proved following the same paradigm.
We omit the details here.
\hfill$\fbox{}$

\section{Proof of Theorem \ref{theorem:angel_correlation_ustat}}

Following the equation \eqref{equation:covariance_square}, we can further rewrite the square of covariance term $4\cov^{2}\{1( \langle\phi_{1}(X), f \rangle\leq u), 1( \langle\phi_{2}(Y), g \rangle\leq v)\}$ via direct calculation as
\beqrs
&& E\Big[
\{1( \langle\phi_{1}(X_{1}), f \rangle\leq u)-
1( \langle\phi_{1}(X_{2}), g \rangle\leq u)\}^{2}
\{1( \langle\phi_{2}(Y_{1}), f \rangle\leq v) -
1( \langle\phi_{2}(Y_{2}), g \rangle\leq v)\}^{2} \\
&&- \{1( \langle\phi_{1}(X_{1}), f \rangle\leq u)-
1( \langle\phi_{1}(X_{2}), g \rangle\leq u)\}^{2}
\{1( \langle\phi_{2}(Y_{1}), f \rangle\leq v) -
1( \langle\phi_{2}(Y_{4}), g \rangle\leq v)\}^{2} \\
&&- \{1( \langle\phi_{1}(X_{1}), f \rangle\leq u)-
1( \langle\phi_{1}(X_{2}), g \rangle\leq u)\}^{2}
\{1( \langle\phi_{2}(Y_{2}), f \rangle\leq v) -
1( \langle\phi_{2}(Y_{3}), g \rangle\leq v)\}^{2} \\
&&+\{1( \langle\phi_{1}(X_{1}), f \rangle\leq u)-
1( \langle\phi_{1}(X_{3}), g \rangle\leq u)\}^{2}
\{1( \langle\phi_{2}(Y_{3}), f \rangle\leq v) -
1( \langle\phi_{2}(Y_{4}), g \rangle\leq v)\}^{2}  \Big].
\eeqrs
Given the integral result that
\beqrs
\int_{u\in\mR}\{1( \langle\phi_{1}(X_{1}), f \rangle\leq u)-
1( \langle\phi_{1}(X_{2}), g \rangle\leq u)\}^{2}du 
= \abs{ \langle\phi_{1}(X_{1}), f \rangle-
	\langle\phi_{1}(X_{2}), f \rangle}
\eeqrs
We can apply Funibi's Theorem and get
\beqrs
&&4^{-1}\int_{\mR^2}\cov^{2}\{1( \langle\phi_{1}(X), f \rangle\leq u), 1( \langle\phi_{2}(Y), g \rangle\leq v)\}
dudv \\
&=& E\Big[
\abs{ \langle\phi_{1}(X_{1}), f \rangle-
 \langle\phi_{1}(X_{2}), f \rangle}
\abs{\langle\phi_{2}(Y_{1}), g \rangle -
\langle\phi_{2}(Y_{2}), g \rangle} \\
&&- \abs{ \langle\phi_{1}(X_{1}), f \rangle-
 \langle\phi_{1}(X_{2}), f \rangle}
\abs{ \langle\phi_{2}(Y_{1}), g \rangle) -
 \langle\phi_{2}(Y_{4}), g \rangle)} \\
&&- \abs{\langle\phi_{1}(X_{1}), f \rangle-
 \langle\phi_{1}(X_{2}), f \rangle}
\abs{ \langle\phi_{2}(Y_{2}), g \rangle -
 \langle\phi_{2}(Y_{3}), g \rangle} \\
&&+\abs{ \langle\phi_{1}(X_{1}), f \rangle-
 \langle\phi_{1}(X_{3}), f \rangle}
\abs{ \langle\phi_{2}(Y_{3}), g \rangle -
 \langle\phi_{2}(Y_{4}), g \rangle}  \Big].
\eeqrs
Based on the above equation, we apply Fubini's Theorem again. 
Using the result of Lemma \ref{lemma:abs_hilbertspace} repeatedly, we have
\beqrs
&&(\pi/32)^{1/2}\int_{\calH_1}\int_{\calH_2}\int_{\mR^2}\cov^{2}\{1( \langle\phi_{1}(X), f \rangle\leq u), 1( \langle\phi_{2}(Y), g \rangle\leq v)\}
d u dv  \mu_1(df) \mu_2(dg)\\
&=& E\Big[
\|\phi_{1}(X_{1})-\phi_{1}(X_{2})\|_{\calH_1} \|\phi_{2}(Y_{1})-\phi_{2}(Y_{2})\|_{\calH_2}
-2\|\phi_{1}(X_{1})-\phi_{1}(X_{2})\|_{\calH_1} \|\phi_{2}(Y_{1})-\\
&& \hspace{0.8cm}
\phi_{2}(Y_{3})\|_{\calH_2}
+\|\phi_{1}(X_{1})-\phi_{1}(X_{2})\|_{\calH_1} \|\phi_{2}(Y_{3})-\phi_{2}(Y_{4})\|_{\calH_2}\Big]\\
&=& E\big[ \rho_{1}(X_{1},X_2)\rho_{2}(Y_{1},Y_{2})
-2\rho_{1}(X_{1},X_2)\rho_{2}(Y_{1},Y_{3})
+\rho_{1}(X_{1},X_2)\rho_{2}(Y_{3},Y_{4})
\big].
\eeqrs
\hfill$\fbox{}$

\section{Proof of Theorem \ref{theorem:angel_correlation_ustat}}

We shall prove this theorem following two steps.

\noindent\textbf{Step 1.} We consider the case when $X$ and $Y$ are independent. We prove the asymptotic properties for $\wh\KAcov_{1}(X,Y)$ first.

Denote $d_{1,1}(X_1, X_2, X_3, X_4)\defby \ang_{1}^{\prime}(X_{1}, X_{2}) + \ang_{1}^{\prime}(X_{3}, X_{4}) - \ang_{1}^{\prime}(X_{1}, X_{3}) - \ang_{1}^{\prime}(X_{2}, X_{4})$, $d_{1,2}(Y_1, Y_2, Y_3, Y_4)\defby \ang_{2}^{\prime}(Y_{1}, Y_{2}) +  \ang_{2}^{\prime}(Y_{3}, Y_{4}) - \ang_{2}^{\prime}(Y_{1}, Y_{3}) - \ang_{2}^{\prime}(Y_{2}, Y_{4})$ and 
\beqrs
&&h_{1}((X_{1}, Y_{1}),(X_{2}, Y_{2}),(X_{3}, Y_{3}),(X_{4}, Y_{4}))\\
&\defby& \{4(4!)\}^{-1}\sum_{(i_1,i_2,i_3,i_4)}^{4} d_{1,1}(X_{i_1}, X_{i_2}, X_{i_3}, X_{i_4})d_{1,2}(Y_{i_1}, Y_{i_2}, Y_{i_3}, Y_{i_4}).
\eeqrs
With straight calculation, we can rewrite $\wh\KAcov_{1}(X,Y)$ as
\beqrs
\wh\KAcov_{1}(X,Y) = \{(n)_{4}\}^{-1}\sum_{(i,j,k,l)}^{n} h_{1}((X_{i}, Y_{i}),(X_{j}, Y_{j}),(X_{k}, Y_{k}),(X_{l}, Y_{l})).
\eeqrs
 When $X$ and $Y$ are independent, the kernel $h_{1}(\cdot)$ is a degenerate kernel. We define
$h_{1,2}((X_{1},Y_{1}),(X_{2},Y_{2}))\defby 6E\{h_{1}((X_{1}, Y_{1}),(X_{2}, Y_{2}),(X_{3}, Y_{3}),(X_{4}, Y_{4}))\mid (X_{1},Y_{1}),(X_{2},Y_{2})\}$.
We can simplify this expectation term and get $h_{1,2}((X_{1},Y_{1}),(X_{2},Y_{2})) = \ang_{1,\textrm{center}}^{\prime}(X_1,X_2)$\\$\ang_{2,\textrm{center}}^{\prime}(Y_1,Y_2)$, where $\ang_{1,\textrm{center}}^{\prime}(X_1,X_2)\defby
E\{d_{1,1}(X_1, X_2, X_3, X_4)\mid X_1, X_2\}$ and $\ang_{2,\textrm{center}}^{\prime}(Y_1,Y_2)\defby
E\{d_{1,2}(y_1, y_2, Y_3, Y_4)\mid Y_{1}, Y_{2}\}$.
According to \citet[Theorem 5.5.2]{serfling1980approximation}, as $n\to\infty$,
\beqrs
n\wh\KAcov_{1}(X,Y) \stackrel{d}{\longrightarrow} \sum_{j=1}^{\infty}\eta_{1,j}(\zeta^{2}_{1,1,j}-1),
\eeqrs
where $\{\eta_{1,j}, j=1,\ldots,\infty\}$ are eigenvalues of the integral operator $T_{1}:L^{2}(\calX\times\calY, \mu_{X}\times\mu_{Y})\to L^{2}(\calX\times\calY, \mu_{X}\times\mu_{Y})$.
\beqrs
(T_{1}f)(x_2,y_2)\defby\int_{\calX\times\calY} h_{1,2}((x_{1},y_{1}),(x_{2},y_{2})) f(x_1,y_1)d \mu_{X}(x_1) d\mu_{Y}(y_1).
\eeqrs
We approximate the distribution for the following term using gamma distribution.
\beqrs
\sum_{j=1}^{\infty}\eta_{1,j}\zeta^{2}_{1,1,j}.
\eeqrs
The first and second moment can be calculated as
\beqrs
E\bigg\{\sum_{j=1}^{\infty}\eta_{1,j}\zeta^{2}_{1,1,j}\Big\}&=&\sum_{j=1}^{\infty}\eta_{1,j} =E\big\{h_{1,2}((X_{1},Y_{1}),(X_{1},Y_{1}))\big\}\\
&=&E\big\{\ang_{1,\textrm{center}}^{\prime}(X_1,X_1)\big\}E\big\{\ang_{2,\textrm{center}}^{\prime}(Y_1,Y_1)\big\}\\
&=&E\big\{\ang_{1}^{\prime}(X_1,X_2)\big\}E\big\{\ang_{2}^{\prime}(Y_1,Y_2)\big\}.
\eeqrs
And the second moment can be calculated as
\beqrs
\var\bigg\{\sum_{j=1}^{\infty}\eta_{1,j}\zeta^{2}_{1,1,j}\Big\}
&=&\sum_{j=1}^{\infty}2\eta_{1,j}^2 
=2E\big\{h_{1,2}((X_{1},Y_{1}),(X_{2},Y_{2}))^{2}\big\}\\
&=& 2E\big\{\ang_{1,\textrm{center}}^{\prime}(X_1,X_2)^{2}\big\}E\big\{\ang_{2,\textrm{center}}^{\prime}(Y_1,Y_2)^{2}\big\} \\
&=& 2E\big\{\KAcov_{1}(X_1,X_2)\big\}E\big\{\KAcov_{1}(Y_1,Y_2)\big\}
\eeqrs
Therefore, the shape and rate parameter $\alpha_{1}$ and $\beta_{1}$ are 
\beqrs
\alpha_{1} &=& \Big[E\big\{\ang_{1}^{\prime}(X_1,X_2)\big\}E\big\{\ang_{2}^{\prime}(Y_1,Y_2)\big\} \Big]^{2}  \Big[2E\big\{\KAcov_{1}(X_1,X_2)\big\}E\big\{\KAcov_{1}(Y_1,Y_2)\big\}\Big]^{-1},\\
\beta_{1} &=&  \Big[E\big\{\ang_{1}^{\prime}(X_1,X_2)\big\}E\big\{\ang_{2}^{\prime}(Y_1,Y_2)\big\} \Big]  \Big[2E\big\{\KAcov_{1}(X_1,X_2)\big\}E\big\{\KAcov_{1}(Y_1,Y_2)\big\}\Big]^{-1}.
\eeqrs

Next, we prove the asymptotic properties for $\wh\KAcov_{2}(X,Y)$.

Denote $d_{2,1}(X_1, X_2, X_3, X_4, X_5)\defby \ang_{1}(X_{1}, X_{2};X_{5}) + \ang_{1}(X_{3}, X_{4};X_{5}) - $\\$\ang_{1}(X_{1}, X_{3};X_{5}) - \ang_{1}(X_{2}, X_{4};X_{5})$, $d_{2,2}(Y_1, Y_2, Y_3, Y_4, Y_5)\defby \ang_{2}(Y_{1}, Y_{2};Y_{5}) + $\\$ \ang_{2}(Y_{3}, Y_{4};Y_{5}) - \ang_{2}(Y_{1}, Y_{3};Y_{5}) - \ang_{2}(Y_{2}, Y_{4};Y_{5})$ and 
\beqrs
&&h_{2}((X_{1}, Y_{1}),(X_{2}, Y_{2}),(X_{3}, Y_{3}),(X_{4}, Y_{4}),(X_{5}, Y_{5}) )\\
&\defby& \{4(5!)\}^{-1}\sum_{(i_1,i_2,i_3,i_4,i_5)}^{5} d_{2,1}(X_{i_1}, X_{i_2}, X_{i_3}, X_{i_4}, X_{i_5})d_{2,2}(Y_{i_1}, Y_{i_2}, Y_{i_3}, Y_{i_4}, Y_{i_5}),
\eeqrs
With straight calculation, we can rewrite $\wh\KAcov_{2}(X,Y)$ as
\beqrs
\wh\KAcov_{2}(X,Y) = \{(n)_{5}\}^{-1}\sum_{(i,j,k,l,r)}^{n} h_{2}((X_{i}, Y_{i}),(X_{j}, Y_{j}),(X_{k}, Y_{k}),(X_{l}, Y_{l}),(X_{r}, Y_{r})).
\eeqrs
When $X$ and $Y$ are independent, the kernel $h$ is a degenerate kernel. We define
$h_{2,2}((X_{1},Y_{1}),(X_{2},Y_{2}))\defby 10 E\{h_{2}((X_{1}, Y_{1}),(X_{2}, Y_{2}),(X_{3}, Y_{3}),(X_{4}, Y_{4}),(X_{5}, Y_{5}))\mid (X_{1},Y_{1}),(X_{2},Y_{2})\}$.
We can simplify this expectation term and get $h_{2,2}((X_{1},Y_{1}),(X_{2},Y_{2})) $\\$=\ang_{1,\textrm{center}}(X_1,X_2)\ang_{2,\textrm{center}}(Y_1,Y_2)$,
where $\ang_{1,\textrm{center}}(x_1,x_2)\defby
E\{d_{2,1}(x_1, x_2, X_3,$\\$ X_4, X_{5})\}$ and $\ang_{2,\textrm{center}}(y_1,y_2)\defby
E\{d_{2,2}(y_1, y_2, Y_3, Y_4, Y_{5})\}$.
According to \citet[Theorem 5.5.2]{serfling1980approximation}, as $n\to\infty$,
\beqrs
n\wh\KAcov_{2}(X,Y) \stackrel{d}{\longrightarrow} \sum_{j=1}^{\infty}\eta_{2,j}(\zeta^{2}_{1,2,j}-1),
\eeqrs
where $\{\eta_{2,j}, j=1,\ldots,\infty\}$ are eigenvalues of the integral operator $T_{2}:L^{2}(\calX\times\calY, \mu_{X}\times\mu_{Y})\to L^{2}(\calX\times\calY, \mu_{X}\times\mu_{Y})$.
\beqrs
(T_{2}f)(x_2,y_2)\defby\int_{\calX\times\calY} h_{2,2}((x_{1},y_{1}),(x_{2},y_{2})) f(x_1,y_1)d \mu_{X}(x_1) d\mu_{Y}(y_1).
\eeqrs
We approximate the distribution for the following term using gamma distribution.
\beqrs
\sum_{j=1}^{\infty}\eta_{2,j}\zeta^{2}_{1,2,j}.
\eeqrs
The first and second moment can be calculated as
\beqrs
E\bigg\{\sum_{j=1}^{\infty}\eta_{2,j}\zeta^{2}_{1,2,j}\Big\}&=&\sum_{j=1}^{\infty}\eta_{2,j} =E\big\{h_{2,2}((X_{1},Y_{1}),(X_{1},Y_{1}))\big\}\\
&=&E\big\{\ang_{1,\textrm{center}}(X_1,X_1)\big\}E\big\{\ang_{2,\textrm{center}}(Y_1,Y_1)\big\}\\
&=&E\big\{\ang_{1}(X_1,X_2;X_3)\big\}E\big\{\ang_{2}(Y_1,Y_2;Y_3)\big\}.
\eeqrs
And the second moment can be calculated as
\beqrs
\var\bigg\{\sum_{j=1}^{\infty}\eta_{2,j}\zeta^{2}_{1,2,j}\Big\}&=&\sum_{j=1}^{\infty}2\eta_{2,j}^2 
=2E\big\{h_{2,2}((X_{1},Y_{1}),(X_{2},Y_{2}))^{2}\big\}\\
&=& 2E\big\{\ang_{1,\textrm{center}}(X_1,X_2)^{2}\big\}E\big\{\ang_{2,\textrm{center}}(Y_1,Y_2)^{2}\big\} \\
&=& 2E\big\{\KAcov_{3}(X_1,X_2)\big\}E\big\{\KAcov_{3}(Y_1,Y_2)\big\}.
\eeqrs
Therefore, the shape and rate parameter $\alpha_{1}$ and $\beta_{1}$ are 
\beqrs
\alpha_{2} &=& \Big[E\big\{\ang_{1}(X_1,X_2;X_3)\big\}E\big\{\ang_{1}(Y_1,Y_2;Y_3)\big\} \Big]^{2}  \Big[2E\big\{\KAcov_{3}(X_1,X_2)\big\}E\big\{\KAcov_{3}(Y_1,Y_2)\big\}\Big]^{-1},\\
\beta_{2} &=&  \Big[E\big\{\ang_{1}(X_1,X_2;X_3)\big\}E\big\{\ang_{1}(Y_1,Y_2;Y_3)\big\} \Big]  \Big[2E\big\{\KAcov_{3}(X_1,X_2)\big\}E\big\{\KAcov_{3}(Y_1,Y_2)\big\}\Big]^{-1}.
\eeqrs
Following the similar proof for $\wh\KAcov_{2}(X,Y)$, denote 
\beqrs
&&h_{3}((X_{1}, Y_{1}),(X_{2}, Y_{2}),(X_{3}, Y_{3}),(X_{4}, Y_{4}),(X_{5}, Y_{5}),(X_{6}, Y_{6}) )\\
&\defby& \{4(6!)\}^{-1}\sum_{(i_1,i_2,i_3,i_4,i_5,i_6)}^{6} d_{2,1}(X_{i_1}, X_{i_2}, X_{i_3}, X_{i_4}, X_{i_5})d_{2,2}(Y_{i_1}, Y_{i_2}, Y_{i_3}, Y_{i_4},  Y_{i_6}).
\eeqrs
We can rewrite $\wh\KAcov_{3}(X,Y)$ as
\beqrs
\wh\KAcov_{3}(X,Y) = \{(n)_{6}\}^{-1}\sum_{(i,j,k,l,r,t)}^{n} h_{2}((X_{i}, Y_{i}),(X_{j}, Y_{j}),(X_{k}, Y_{k}),(X_{l}, Y_{l}),(X_{r}, Y_{r}),(X_{t}, Y_{t})).
\eeqrs
When $X$ and $Y$ are independent, the kernel $h$ is a degenerate kernel. We define
$h_{3,2}((X_{1},Y_{1}),(X_{2},Y_{2}))\defby 15 E\{h_{3}((X_{1}, Y_{1}),(X_{2}, Y_{2}),(X_{3}, Y_{3}),(X_{4}, Y_{4}),(X_{5}, Y_{5}),(X_{6}, Y_{6}))\mid (X_{1},Y_{1}),(X_{2},Y_{2})\}$.
We can simplify this expectation term and get $h_{3,2}((X_{1},Y_{1}),(X_{2},Y_{2})) $\\$=\ang_{1,\textrm{center}}(X_1,X_2)\ang_{2,\textrm{center}}(Y_1,Y_2)$.
According to \citet[Theorem 5.5.2]{serfling1980approximation}, as $n\to\infty$,
\beqrs
n\wh\KAcov_{3}(X,Y) \stackrel{d}{\longrightarrow} \sum_{j=1}^{\infty}\eta_{3,j}(\zeta^{2}_{1,3,j}-1),
\eeqrs
where $\{\eta_{3,j}, j=1,\ldots,\infty\}$ are eigenvalues of the integral operator $T_{3}:L^{2}(\calX\times\calY, \mu_{X}\times\mu_{Y})\to L^{2}(\calX\times\calY, \mu_{X}\times\mu_{Y})$.
\beqrs
(T_{3}f)(x_2,y_2)\defby\int_{\calX\times\calY} h_{3,2}((x_{1},y_{1}),(x_{2},y_{2})) f(x_1,y_1)d \mu_{X}(x_1) d\mu_{Y}(y_1).
\eeqrs
We approximate the distribution for the following term using gamma distribution.
\beqrs
\sum_{j=1}^{\infty}\eta_{3,j}\zeta^{2}_{1,3,j}.
\eeqrs
The shape and scale parameter can be calculated as
\beqrs
\alpha_{3} &=& \Big[E\big\{\ang_{1}(X_1,X_2;X_3)\big\}E\big\{\ang_{1}(Y_1,Y_2;Y_3)\big\} \Big]^{2}  \Big[2E\big\{\KAcov_{3}(X_1,X_2)\big\}E\big\{\KAcov_{3}(Y_1,Y_2)\big\}\Big]^{-1},\\
\beta_{3} &=&  \Big[E\big\{\ang_{1}(X_1,X_2;X_1)\big\}E\big\{\ang_{1}(Y_1,Y_2;Y_3)\big\} \Big]  \Big[2E\big\{\KAcov_{3}(X_1,X_2)\big\}E\big\{\KAcov_{3}(Y_1,Y_2)\big\}\Big]^{-1}.
\eeqrs

\noindent\textbf{Step 2.} We consider the case when $X$ and $Y$ are dependent. 

We prove the asymptotic properties for $\wh\KAcov_{1}(X,Y)$ first.
We define
$h_{1,1}(X_{1},Y_{1})\defby 4 E\{h_{1}((X_{1}, Y_{1}),(X_{2}, Y_{2}),(X_{3}, Y_{3}),(X_{4}, Y_{4}))\mid (X_{1},Y_{1})\}$, which can be simplified as 
$h_{1,1}(X_{1},Y_{1})\defby E\big\{d_{1,1}(X_1, X_2, X_3, X_4)d_{1,2}(Y_1, Y_2, Y_3, Y_4)\mid (X_{1},Y_{1})\big\}$. Denote
\beqr\label{equation:sigma1}
\sigma_{1}^{2}\defby E\big\{h_{1,1}(X_{1},Y_{1})^2\big\}.
\eeqr
According to \citet[Theorem 5.5.1A]{serfling1980approximation}, as $n\to\infty$,
\beqrs
n^{1/2}\{\wh\KAcov_{1}(X,Y)-\KAcov_{1}(X,Y)\} \stackrel{d}{\longrightarrow} \calN(0, \sigma_{1}^{2}).
\eeqrs
Next, we prove the asymptotic properties for $\wh\KAcov_{2}(X,Y)$.
We define
$h_{2,1}(X_{1},Y_{1})\defby 5 E\{h_{2}((X_{1}, Y_{1}),(X_{2}, Y_{2}),(X_{3}, Y_{3}),(X_{4}, Y_{4}),(X_{5}, Y_{5}))\mid (X_{1},Y_{1})\}$. Denote 
\beqr\label{equation:sigma2}
\sigma_{2}^{2}\defby E\big\{h_{2,1}(X_{1},Y_{1})^2\big\}.
\eeqr
According to \citet[Theorem 5.5.1A]{serfling1980approximation}, as $n\to\infty$,
\beqrs
n^{1/2}\{\wh\KAcov_{2}(X,Y)-\KAcov_{2}(X,Y)\} \stackrel{d}{\longrightarrow} \calN(0, \sigma_{2}^{2}).
\eeqrs
Following the similar paradigm of the prove for $\wh\KAcov_{2}(X,Y)$, we can complete the proof for $\wh\KAcov_{3}(X,Y)$. Denote $h_{3,1}(X_{1},Y_{1})\defby 6 E\{h_{3}((X_{1}, Y_{1}),(X_{2}, Y_{2}),(X_{3}, Y_{3}),(X_{4}, Y_{4}),$\\$(X_{5}, Y_{5}),(X_{6}, Y_{6}))\mid (X_{1},Y_{1})\}$, the corresponding variance can be represented as
\beqr\label{equation:sigma3}
\sigma_{3}\defby E\big\{h_{3,1}(X_{1},Y_{1})^2\big\}.
\eeqr
\hfill$\fbox{}$

\section{Proof of Proposition \ref{proposition:semimetric}}
We prove $(\calX, \rho_{angle,1})$ is semimetric space of negative type and omit the proof for others which can be shown similarly.

Given $x_1, x_2\in \calX$, it is easy to see that $\rho_{angle,1}(x_{1}, x_{2}) = \rho_{angle,1}(x_{2}, x_{1})$.

If $x_1 = x_2$, we have $\rho_{angle,1}(x_{1}, x_{2}) = 0$. And if  $\rho_{angle,1}(x_{1}, x_{2}) = 0$, it implies that
$\langle \phi_{1}(x_{1})-\phi_{1}(X), \phi_{1}(x_{2})-\phi_{1}(X)\rangle_{\calH_{1}} = \|\phi_{1}(x_{1})-\phi_{1}(X)\|_{\calH_{1}}\|\phi_{1}(x_{2})-\phi_{1}(X)\|_{\calH_{1}}$ almost surely. By Cauchy-Schwarz inequality, this equality  holds if and only if $\phi_{1}(x_{1})-\phi_{1}(X) = \phi_{1}(x_{1})-\phi_{1}(X)$ almost surely. Given $\phi_{1}(\cdot)$ is injection, we know $x_{1} = x_{2}$.

Next, we show the negative type. Given $x_{3}\in\calX$, from \cite{bogomolny2007distance}, we know for fixed $x_{3}$,
\beqrs
\sum_{i=1}^{n}\sum_{j=1}^{n}\alpha_{i}\alpha_{j}\ang_{1}(x_1,x_{2};x_{3})\leq 0,
\eeqrs
By taking expectation of $X_{3}$, this inequality also holds. Thus, $\rho_{angle,1}$ is of negative type.
\hfill$\fbox{}$

\end{document}